# Ultracold Bose and Fermi dipolar gases: a Quantum Monte Carlo study

**Phd Program in Computational and Applied Physics**

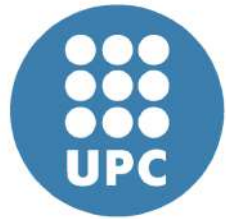

**Raúl Bombín Escudero**

**Supervisors:** Jordi Boronat Medico
Ferran Mazzanti Castrillejo

Department of Physics
Universitat Politècnica de Catalunya

This dissertation is submitted for the degree of
*Doctor of Physics*

October 2019

# Acknowledgements

**English:**  First of all, I would like to give special thanks to my advisors, Professors Jordi Boronat and Ferran Mazzanti, for giving me the opportunity of getting started in the scientific career. In the moment of depositing this thesis I can not avoid remembering that first Skype call that we had, after which, they decided to accept me in the Barcelona Quantum Monte Carlo group; the assessment of the consequences of this decisions is left to the judgment of the reader. I am delighted for having work under their guidance in the good atmosphere that characterizes the group and having learned from them all that I know about ultracold gases and the quantum Monte Carlo. For the contribution to the later, I am also grateful to all the staff of the Department of Physics of Universitat Politècnica de Catalunya with special mention to the other predoctoral researchers with whom I have coincide during most of the time that I have spent in the department and with whom I have shared some hundreds of lunches and a few dinners: Joan Sanchez, Grecia Guijarro and Huixia Lu. Also to Dr. Guillem Ferré, thanks to whom I was able to take the first steps in the Path Integral Monte Carlo, Dr. Adrián Macía whose Phd. Thesis was a source of knowledge during my first months here. Also to my office mates: Jordi Ortiz who introduced me to climbing (to the regret of my mother) and Viktor Cikojević who definitely contributed to dynamize the group during the months that he was here.

I fondly remember the months that I have expended in Trento, for which I have to be grateful to Tommaso Comparin, who invited me there for the first time. There is no doubt that his mark is printed in this thesis, fruit of the collaboration that he has promoted between the Barcelona and the Trento group. Also special thanks for Prof. Stefano Giorgini, for accepting me in his group during that months and for the familiar way in which he used to show up in the office, not only to talk about physics but to remain me that it was time to climb. I am also indebted to all the people in the BEC center at Trento with whom I had useful discussions and definitely made my stay enjoyable, specially the ones with whom I shared office during most of that time: Albert Gallemí and Russel Bisset.

Here I would like to acknowledge all the other collaborators that I have not mentioned up to now and with whom I have had useful discussions: Markus Holzmann for sharing his expertise on fermionic systems, Gianluca Bertaina for his patience and for finding time to work in our shared project and Fabian Böttcher and Matthias

Wenzel for translating the setups in a language that a computational physicist can understand.

Now that I am concluding this stage, I have also to thanks to my parents, who never gave up on pushing me to continue studying, despite they had to hear my comments that first day of school: "every day? and what if I get tired?". Also to my sisters, Ana and Belén, whose initial expectations, when I moved to Barcelona, about how much would they visit a city with beach probably have not been fulfilled. And, of course, the rest of my family: I am sure that finally this Christmas (almost) all of them would have it clear whether I am a physicist or a chemist. Especial mention to my grandparents Rufina and Agustín and also the ones that are not with us anymore: Balbino and Carmen. Finally, I cannot forget my uncle Pablo who used to say that I would be "one of those quantum physicist".

I would like to specially devote this thesis to Beatriz for being my partner over these years, giving support when needed, and for all the good moments that we have had together.

Although it is not possible to make an individual acknowledge of all of them, it is of justice to mention here all the professionals of the public educational system that have dedicated some time to my formation, both in the school and university. Thanks to their labor, people form the surrounding neighborhoods, as is my case, can access to the higher education.

I can not forget the partners from the assembly D-Recerca and from CC.OO.,together with whom we have obtained improvements in the labor conditions of predoctoral researchers.

To finish, thanks to all the friends that have spent some time with me these years, all in all, to all the people who have ever asked me: and then what?

**Spanish** En primer lugar, quiero agradecer a mis tutores, Jordi Boronat y Ferran Mazzanti, por haberme dado la oportunidad de comenzar la carrera investigadora. Ahora que estoy apunto de depositar esta Tesis, no puedo evitar recordar aquella primera llamada de Skype tras la cual decidieron aceptarme en el grupo Barcelona Quantum Monte Carlo: la evaluación de las consecuencias de esta decisión se deja a criterio del lector. Ha sido un placer haber trabajado bajo su guía y haber podido aprender de ellos todo lo que se sobre los gases ultrafríos y el Monte Carlo Cuántico, rodeado del buen ambiente que caracteriza al grupo. Me gustaría agradecer también a todo el personal del Departamento de Física de la Universitat Politècnica de Catalunya por su contribución en este último aspecto. Especial mención merecen los otros investigadores predoctorales con los que he coincidido en el departamento durante buena parte de estos años y con quienes he compartido cientos de almuerzos y algunas cenas: Joan Sánchez, Grecia Guijarro y Huixia Lu. Sin olvidar a los que pasaron antes por aquí: al Dr. Guillem Ferré, gracias a quien pude dar los primeros pasos con el *Path*

*Integral Monte Carlo*, y al Dr. Adrián Macía cuya tesis fue una fuente de conocimiento durante mis primeros meses aquí. También a mis compañeros de despacho: Jordi Ortiz, quien además me introdujo en el mundo de la escalada (para disgusto de mi madre) y Viktor Cikojević, que sin duda ha dinamizado el grupo durante los meses que ha estado con nosotros.

Recuerdo con especial cariño los meses de mi estancia en Trento. Debo agradecer esta experiencia al Dr. Tommaso Comparin, por haber sido quien me invitó allí por primera vez. No hay duda de que su marca está impresa en esta tesis, fruto de la colaboración que el mismo ha promovido entre el grupo de Barcelona y Trento. Por mi estancia en Trento, tengo que agradecer también de forma especial a Prof. Stefano Giorgini, por aceptarme en su grupo durante esos meses y por su forma familiar de aparecer en el despacho, unas veces para hablar de física y otras para recordarme que era hora de ir a escalar. También etoy en deuda con todas las personas del BEC Center en Trento con quienes pude compartir discusiones sobre física y ratos agradables. Y especialmente a quienes me acogieron en su despacho durante esos meses: Dr. Albert Gallemí y Dr. Russel Bisset.

Quiero mencionar también al resto de colaboradores que no he mencionado hasta ahora: Markus Holzmann por compartir su experiencia en sistemas fermiónicos, Gianluca Bertaina por su paciencia y por encontrar tiempo para trabajar en nuestro proyecto compartido pese a no ser su ocupación principal y finalmente a Fabian Böttcher y Matthias Wenzel por traducir los montajes experimentales a un lenguaje que puede entender un físico computacional.

Ahora que esta etapa concluye he de agradecer a mis padres, que nunca dejaron de animarme a estudiar pese a haber tenido que oir aquel primer día de escuela aquello de: "¿todos los días? ¿Y qué pasa si me canso?". También a mis hermanas, Ana y Belén, cuyas expectativas iniciales sobre cuánto me visitarían en una ciudad con playa probablemente no han sido satisfechas. Y, por supuesto, al resto de mi familia: estoy seguro de que finalmente esta Navidad (casi) todos tendrán claro si soy físico o químico. Mención especial a mis abuelos Rufina y Agustín y también a los que ya no están con nosotros: Balbino y Carmen. Finalmente, no puedo dejar de mencionar a mi tío Pablo, que decía que llegaría a ser "uno de esos físicos cuánticos".

Me gustaría dedicar esta tesis a Beatriz por ser mi compañera durante estos años, brindarme apoyo cuando ha sido necesario, y por todos los buenos momentos que hemos pasado juntos.

Aunque no es posible hacer un reconocimiento individual de todos ellos, creo que es de justicia mencionar aquí a todos los profesionales del sistema educativo público que han dedicado algún tiempo a mi formación, tanto en la escuela como en la universidad. Es gracias a su trabajo que las personas de barrios periféricos como el mío pueden acceder a la educación superior.

No puedo acabar sin felicitar a los compañeros de la asamblea de doctorandos D-Recerca y de CC.OO. junto a quienes hemos conseguido algunas mejoras sustanciales en las condiciones de los investigadores predoctorales.

Para acabar, gracias a todos los amigos que habéis pasado algún tiempo conmigo estos años y en general a todas las personas que alguna vez me han preguntado: ¿y ahora qué?

# Table of contents

# List of figures

# List of tables

# Chapter 1

# Introduction

> ''*Muchos años después, frente al pelotón de fusilamiento, el coronel Aureliano Buendía había de recordar aquella tarde remota en que su padre lo llevó a conocer el hielo.[...] El mundo era tan reciente, que muchas cosas carecían de nombre, y para mencionarlas había que señalarlas con el dedo.*"

> — Gabriel García Márquez, Cien años de soledad.[1]

Experiments with ultracold gases are performed nowadays in many laboratories around the world and techniques to achieve the quantum degenerate regime are becoming standard. However, as it usually happens with scientific developments, both the technical and theoretical tools that are used to explore this systems in the present, have arisen in many different contexts and throughout various decades. Just to highlight some of them: the development of the laser, and, related to it, the spectacular success in trapping and cooling techniques on the experimental side, and, from the theoretical point of view, the unquestionable role that thermodynamics, statistical physics and, certainly, quantum mechanics have played. Not only for the understanding of the experimental results, but also motivating new measurements. Precisely, the blow up that has taken place in the community is closely related to the possibilities that ultracold gases offer for testing theoretical models in experiments. Thanks to the incredible tunability that current state-of-the-art experimental setups have achieved, it has been possible to use them as a testbed for many-body theories in a wide range of interaction parameters. Before introducing the newest discoveries and the particular problems that have been tackled in this Thesis, it is worth to perform a brief historical overview.

Hundreds of years before the phenomenology of quantum gases was even imagined, some experiments with gases that have had a direct relevance in the field of low temperature physics were done. In 1662, Robert Boyle experimentally established a relation between the pressure and the volume of a gas inside a chamber. Almost at the same time, E. Mariotte complemented this study establishing that the law reported by Boyle was only true when the temperature is fixed. Despite of this, a systematic study of the effects of temperature in gases had to wait more than one hundred years. Around that time, James Watt patented its steam engine, giving rise to the first industrial revolution. The implications of the thermodynamics of gases on the incipient industry, explains the renewed interest on the topic at that time. In this context, J. Charles (1787) and Gay-Lussac (1808) studied and established laws about the effects of temperature on the volume and pressure of gases. In 1811, the study of chemical reactions between gases lead A. Avogadro to establish that gases that occupy the same volume, under the same conditions of pressure and temperature, contain the same number of elementary constituents: atoms or molecules. Avogadro's law, together with the ideas of Dalton and others, set the basis of the *atomic-molecular theory*. Finally, in 1834, E. Clapeyron combined the previous ideas to obtain the *ideal gas law*, whose extrapolation to zero temperature gave the first insight into the existence of a universal absolute minimum of temperature. This discovery sown the seed for the interest in low temperature physics, whose first steps where related to the development of techniques for cooling down different gaseous systems to their liquefaction point. Those first attempts ended up with the achievement of liquid helium at a temperature of about 4 K in 1908, which led Heike Kamerlingh Onnes to obtain the Nobel prize. As Helium has played a central role in low temperature physics, we will comment about it a bit later.

---

[1] *Many years later, as he faced the firing squad, Colonel Aureliano Buendía was to remember that distant afternoon when his father took him to discover ice. [...] The world was so recent that many things lacked names, and in order to indicate them it was necessary to point.*' — Gabriel García Márquez, One Hundred Years of Solitude.

The third principle of thermodynamics is closely related to existence of an absolute minimum of temperature. In 1926, Nerst enunciated his theorem [1], whose modern interpretation reveals that as temperature approaches zero, the change in the entropy of a system also tends to zero. Nerst theorem was soon applied to enunciate the third law of thermodynamics, that states that it is not possible to access the absolute minimum of temperature by any procedure that requires a finite number of steps [2], although it does not include any further limitation to access arbitrarily small temperatures.

Also in the down of the 20th century, the development of quantum mechanics, renewed the interest in the behavior of systems near the absolute zero of temperature. Classical mechanics states that the kinetic contribution to the energy of a system would vanish as its temperature approaches zero. However, the development of quantum physics revealed that the state of a system at zero temperature would have a contribution to the energy coming from the *zero point motion*. Regarding many particle systems, the concepts of statistical mechanics had also to be adapted to correctly describe quantum systems. In the quantum regime, the underlying symmetry between indistinguishable particles (bosons or fermions) have to be taken into account, and so the Boltzmann distribution, that describes the occupations of states at a given temperature, works only in the classical limit. The occupied states of a system composed of indistinguishable bosons follow Bose-Einstein statistics, while the equivalent fermionic system is described in terms of Fermi-Dirac statistics [3, 4]. Quantum statistics are crucial to correctly describe the properties of systems in the low temperature regime.

Following the ideas that were previously exposed by Satyendra Nath Bose, in 1926 Einstein predicted a new state of matter, that nowadays we know with the name of *Bose-Einstein condensate* (BEC). When a bosonic system is cooled down to a low enough temperature, particles (or at least a large fraction of them) would condensate to the minimum energy state [5]. This theoretical prediction appeared just six years before the *lambda transition* was observed in liquid $^4$He. The specific heat of liquid helium was measured at low temperature showing a divergence around the lambda temperature $T_\lambda = 2.17K$ [6], which suggested the existence of helium in two liquid phases, named Helium-I and Helium-II, above and below this temperature respectively. Nowadays we know that Helium-II has a condensate fraction of about 8% [7], that doubtless influences its properties. These two phases of helium were experimentally studied during the decade of 1930's and a novel phenomena was reported by two experimental teams [8, 9]: Helium-II flows without viscosity. This effect, that was called *superfluidity*, constituted the first observed macroscopic effect that had to be explained purely in terms of the underlying quantum nature of its microscopic constituents.

The situation is different when dealing with fermionic systems, where the Pauli exclusion principle prevents indistinguishable particles from occupying the same microscopic sate. A theoretical understanding of these systems came by the hand of Lev Landau [10, 11], who studied the concept of *Fermi liquid* in the decade of 1950 in order to explain systems such as $^3$He. Superfluidity in fermionic systems is more subtle than in their bosonic counterparts because its existence requires the mechanism of pairing: the appearance of pairs of fermions that form quasi-particles with bosonic properties, allowing the existence of a condensate of pairs. As this mechanism is highly suppressed by thermal fluctuations, the discovery of superfluidity in $^3$He came more than 30 years later than in $^4$He. Finally, in 1970, a superfluid phase of $^3$He was observed below a temperature of 3 mK [12, 13]. For decades, liquid helium, whether on its bosonic or fermionic isotope, caught much of the attention of the condensed matter community, being the best suitable candidate to experimentally test quantum many-body theories.

The invention of the laser in 1960, opened a window for performing new experiments, not only in optics, but in many other fields of physics. Concretely, after three decades of innovations, new techniques that are of interest for the atomic physics community were developed. In 1978, two groups, almost simultaneously [14, 15], and following the theoretical prescriptions of Ashkin [16], demonstrated that it is possible to trap and cool down atomic systems employing laser beams. After some years of continuum developments, finally, in 1995, it was possible to obtain the first gaseous Bose-Einstein

Condensate [17, 18]. Some years later, this achievement was complemented by the observation of a Fermi degenerated gas in a trap containing $^{40}$K atoms [19]. Although, in the beginning, experiments where only accessible in the weakly interacting regime, the scenario changed with the observation of Feshbach resonances [20]. These resonances allow to tune the scattering length, and thus, to perform experiments in a wide range of interaction strengths. A new era in Atomic, Molecular and Optical physics (AMO) was opened by this achievements because it made experimentally accessible phenomena that are characteristic of the strongly coupling regime, such as the BEC-BCS crossover (*cf.* [21–25] for experiments and [26, 27] for Monte Carlo studies), or the recent claim for the observation of itinerant ferromagnetism [28] which has been a long-standing topic in the field in the past decade [29–36] .

Usually, ultracold gases, being extremely dilute systems, are to some extent well described by an isotropic, short-ranged, contact interaction model. Recently, the scenario has become richer with the achievement of quantum degenerate systems with intrinsic dipolar interaction – see Ref. [37–39] for reviews. The realization of quantum degenerated systems composed of atoms with large magnetic or electric moment, gives access to experiments in which the long-ranged and anisotropic character of the interaction plays a crucial role. Experiments with bosonic and fermionic dipolar atoms have been performed in several laboratories: initially by employing Chromium atoms [40, 41] and more recently, with Dysprosium [42, 43] and Erbium [44, 45]. One of the first examples of the effect of anisotropy in such systems is the observation of the Fermi surface deformation [45]. Furthermore, the anisotropy of the dipolar interaction makes it possible to break a continuous translational symmetry, what gives rise to new phases of matter in the BEC field such as the stripe phase in two-dimensional geometries [46, 47] and the formation of ultra-dilute quantum droplets in the free space [48–53]. A lot of attention has been put recently on the superfluid properties of these phases with broken translational symmetry because they resemble the *supersolid phase* predicted for $^4$He decades ago [54]. With this in mind intense work is being developed, related to the study of the superfluid properties of the stripe phase [55, 56], and also to arrays of droplets [57–61] trapped in a cylindrical symmetry.

Another remarkable experimental improvement in the recent years has been the achievement of ultra-dilute liquid-like droplets in Bose-Bose mixtures. D. S. Petrov [62] put forward the idea of employing mixtures with attractive inter-species and repulsive intra-species interaction to obtain self-bound systems. Another remarkable property of these systems is that their density is usually higher than the usual dilute BEC systems, making their theoretical description more challenging. After that, additional work has been done in this direction [63–67] studying miscibility properties and determining the regime of universality in terms of the gas parameter. For a certain regime of parameters, the formation of self-bound droplets has also been reported [68–70]. Still, in relation with mixtures, a topic that has caught a lot of attention is the ultra-dilute concentration of impurities, that is usually referred as the *polaron* problem. It offers a first insight into some of the physical phenomena that are of interest in different correlated systems: the pairing mechanism that gives rise to the BEC-BCS crossover [71–73], the possible itinerant ferromagnetism in two-component systems [29–36] or the Kondo effect in systems containing magnetic impurities [74]. Regarding system with dipolar interaction, mixtures of isotopes of Dysprosium and Erbium have also been realized [75, 76], including the singular case in which a low concentration of impurities are embedded in a dipolar dysprosium droplet [77].

Finally, we would like to remark the exceptional paradigm that ultracold gases offer to study systems in reduced geometry. To achieve them, it is enough to impose a tight confinement over one or more of the three dimensions of space. For example, a one-dimensional (1D) configuration can be obtained by imposing a tight confinement along two directions. This has allowed, for example, to study the properties of one-dimensional systems in the Tonks-Girardeu regime, where repulsion between atoms together with the impossibility of crossing each other, gives place to a fermionization scenario (*cf.* [78, 79]). A peculiarity of 1D systems, that make them different from higher dimensional ones, is that superfluidity can appear only as a finite size effect. The scaling of the superfluidity in a 1D system as a function of the temperature and the system size, has been numerically studied at low temperatures [80], where the Luttinger Liquid theory stands. Additionally, interesting theoretical work has also been done to study the beyond-Luttinger-Liquid behavior [81] as a function of the temperature. The formation of droplets in one dimensional mixtures of bosons has been also predicted, both for mixtures of bosons [82] and for dipolar atoms [83], similarly to what has been observed in three dimensional systems. In what concerns to two-dimensional systems, some of their particularities have also been studied in ultracold gases.

Correlations are enhanced in two-dimensional systems as it is reflected in the appearance of deviations from the mean-field theory even in the very low density regime [84]. Another important characteristic of two-dimensional systems is the absence of off-diagonal long-range order for any finite temperature, what makes the superfluid to normal phase transition to follow the Berenzinskii, Kosterlitz and Thouless scenario [85, 86]. Two-dimensional systems, and properties related to the absence of off-diagonal long-range order, have been studied in ultracold gases confined in pancake geometries [87–90]. Of particular interest is the case of pure dipolar systems, whose stability against collapse is guaranteed once all the dipolar moments are polarized along a certain direction in space (below a maximum polarization angle). The ground-state diagram of such system has been studied considering both bosonic [46, 47, 91, 92] and fermionic gases [93, 94].

## Objectives and outline

The object of study of this Thesis are quantum dipolar systems. The properties of these systems have been analyzed using different Quantum Monte Carlo (QMC) methods, that allow to perform calculations both at zero and at finite temperature. The Thesis is structured as follows:

- In chapter 2 we present the Quantum Monte Carlo methods that we have employed in this Thesis. The simplest one is *Variational Monte Carlo* (VMC) that, as its name states, allows us to obtain a variational solution. An improved approximation to the quantum many-body problem is obtained with the *Diffusion Monte Carlo* (DMC) method. With it, the exact ground-state of a bosonic system is found by performing a propagation in imaginary time. However, when dealing with Fermions this method becomes variational. The third method that we use is based on the Feynman path integral formalism of Quantum mechanics: *Path Integral Monte Carlo* (PIMC). The main advantage with respect to other methods is that it gives us access to thermal properties of the quantum system. Finally the *Path Integral Ground State*, an adapted PIMC algorithm to evaluate properties in the limit of zero temperature, has also been used. These two path integral methods, when applied to evaluate properties of bosonic systems, provide us with exact results.

After introducing the method, we present the research that we have done regarding dipolar systems restricted to a two-dimensional geometry. We have split the discussion into two chapters, in chapter 3 we present our studies regarding the bosonic dipolar system while in chapter 4 we present that for dipolar fermions.

- In chapter 3 we study a system of indistinguishable bosonic dipoles that are restricted to move in the plane. They are all polarized along a direction that forms a certain angle (tilt angle) with respect to the normal vector of the plane containing them. In a previous work, the phase diagram of this system was studied as a function of the density and the tilt angle, revealing the existence of three different phases: gas, stripe, and solid [47]. In this chapter we present a characterization of the superfluid properties of these phases. We start analyzing the system at zero temperature by means of DMC and PIGS methods. Then this study is extended to finite temperature with the help of the PIMC method, which allows to characterize the thermal transition that exists in the gas and stripe phases between a superfluid at low temperature and a normal system above a critical temperature. In two-dimensions this transition follows the BKT scenario, as it will be discussed in the text. The full characterization of this transition leads us to propose the stripe phase as a good candidate for the supersolid state of matter. Finally, we discard the possibility of treating the stripe phase as an ensemble of one dimensional systems by comparing our Monte Carlo data with the predictions of the 1D Luttinger liquid model.

The result of this work has been published in the following two works:

- R. Bombín, J. Boronat, and F. Mazzanti Dipolar Bose Supersolid Stripes. *Physical Review Letters*, **119**, 250402 (2017).

- R. Bombín, F. Mazzanti, and J. Boronat Berezinskii-Kosterlitz-Thouless Transition in Two-Dimensional Dipolar Stripes *Physical Review A*, **100**, 063614 (2019).

- In chapter 4, we consider different dipolar systems in strictly two-dimensions in which fermionic species are present. In the DMC algorithm, the inclusion of Fermi statistics is carried out by implementing the *Fixed-Node technique*, that allows us to tackle with the sign problem, although in an approximate way, making the method variational. In this chapter, we have restricted our analysis to the particular case in which the dipoles are polarized in the perpendicular direction to the plane. We start by evaluating the low-density equation of state of a two component dipolar Fermi mixture. At high density, near the crystallization point, we discuss the possible existence of a ferromagnetic ground-state, which would constitute an example of *itinerant ferromagnetism*. In the second part of this chapter we investigate the properties of the Fermi polaron, consisting on a single atomic impurity embedded in a pure fermionic bath. In particular, we compare the properties of the dipolar polaron to those of a hard-disk model, what allows us to determine the regime of universality in terms of the gas parameter $na_s^2$, with $n$ the density of the system and $a_s$ the s-wave scattering length of the impurity-to-bath interaction.

This work has appeared in the following publications:

- T. Comparin., R. Bombin, M.Holzmann, F. Mazzanti, J. Boronat, and S. Giorgini Two-dimensional Mixture of Dipolar Fermions: Equation of State and Magnetic Phases. *Physical Review A*, **99**, 043609 (2018).

- R. Bombín, T. Comparin, G. Bertaina, F. Mazzanti, S. Giorgini, J. Boronat Two-dimensional repulsive Fermi polarons with short- and long-range interactions. *Physical Review A*, **100**, 023608 (2019).

In recent years, the study of self-bound dipolar droplets have caught much attention. Their formation is possible due to the competition between quantum correlations and the attractive and repulsive contributions to the energy of the inter-particle potential. The anisotropy of dipolar interaction, makes this droplets to be elongated along the dipole polarization direction, in contrast with the spherical droplets that are obtained in Bose-Bose mixtures described by contact interactions.

- In chapter 5, we study a system of dipolar atoms confined in a trap. The model potential that we use to describe this system includes a short-range repulsive part that prevents the system from collapse. We restrict our simulations to parameters that allow to compare with recent experiments of $^{162}$Dy and $^{164}$Dy. The employment of different model potentials allows to evaluate the deviations from the extended Gross-Pitaevskii equation (e-GPE) prediction, due to non-universal effects. The most direct comparison with experiments comes from the evaluation of the critical atom number, that is, the minimum number of atoms needed to form a self-bound droplet. In this same chapter other observables that are of interest in order to better understand the differences between PIGS and e-GPE are discussed.

This work has been published in the following work:

- F. Böttcher, M. Wenzel, J. N. Schmidt, M. Guo, T. Langen, I. Ferrier-Barbut, T. Pfau, R. Bombín, J. Sánchez-Baena, J. Boronat, and F. Mazzanti Quantum correlations in dilute dipolar quantum droplets beyond the extended Gross-Pitaevskii equation. *Physical Review Research* **1**, 033088 (2019).

# Chapter 2

# Quantum Monte Carlo Methods

*"Chi vuole il fine vuole i mezzi idonei a raggiungerlo"*

— Antonio Gramsci, Quaderni del carcere.[1]

The object of study of this Thesis are quantum many-body systems, in particular dipolar ones. Evaluating their properties requires us to deal with the many-body Schrödinger equation. Finding an analytical solution is hardly possible, specially when dealing with interacting systems. To tackle this problem some approximations to reduce its complexity such as mean-field or density functional theory can be applied. In our case, we employ Monte Carlo methods, that allow us to find numerical solutions by employing a set of stochastic techniques. The many-body Schrödinger equation, that describes the state of a quantum system, is written as the following eigenvalue differential equation:

$$\hat{H}\Psi = E\Psi, \tag{2.1}$$

where $\hat{H}$ is the Hamiltonian of the system, $\Psi$ is the many-body wave function describing its state, and $E$ the energy corresponding to that state. In what follows, we consider the particles of the system interact via a two-body potential. With this assumption, the Hamiltonian for the N particle system reads:

$$\hat{H} = -\frac{\hbar^2}{2m}\sum_i^N \nabla_i^2 + \sum_{i<j}^N V^{2B}(\mathbf{r}_{ij}) + \sum_i^N V^{Ext.}(\mathbf{r}_i) \tag{2.2}$$

where $V^{2B}$ is the two body potential between particles and $V^{Ext.}$ an external potential acting on the particles. In the notation that we use, we introduce $\mathbf{R} = \{\mathbf{r}_1, ..., \mathbf{r}_N\}$ to refer to the complete set of coordinates of the system.

When working with many-body physics, multidimensional integrals appear in the calculations, and their integration usually constitutes a challenge. Under the label of Monte Carlo (MC), there exist a set of standard methods that allow to evaluate these integrals by employing stochastic methods. In this chapter, we will discuss the Quantum Monte Carlo (QMC) methods that have been used in this Thesis. We start in section 2.1, we introduce some of the basis that are common between the techniques employed hereafter.

---

[1] *"Who wants the end, wants the means to achieving it"* —Antonio Gramsci, Prison notebooks.

In sec. 2.2 we introduce the Variational Monte Carlo (VMC) method. It takes advantage of the variational principle for obtaining an upper bound solution to the energy of the ground state of the system. This method was introduced by McMillan [95] in 1965 to study liquid $^4$He and it turned out to be an acceptable approximation for evaluating the structural properties, although it was not accurate enough to reproduce the experimental data quantitatively.

These results were improved some years later with the introduction of the Diffusion Monte Carlo (DMC) algorithm [96]. This method finds the ground state of the system by performing imaginary time propagation, which allows to obtain exact result for the ground state of bosonic systems. On the other hand, when dealing with fermionic species, the antisymmetry of the wave function makes us deal with the well known *sign problem*. One of the approximations that allow to go around this problem is the *Fixed-Node* technique, that gives a variational solution to the problem [97]. The DMC method will be deeply discussed in sec. 2.3.

Although DMC is a powerful tool for obtaining ground state properties, it does not allow to obtain results beyond the limit of $T = 0$. To understand thermal processes such as the superfluid-to-non-superfluid transition that we study in chapter 3, finite temperature calculations are needed. This can be achieved by taking advantage of the Feynman formalism of Quantum Mechanics [98]. Although Feynman proposed its formulation in the 60's of the past century, its application to calculate properties of a quantum many-body system employing MC techniques had to wait until the development of modern computers allowed it. In 1986, Ceperley and Pollock [99, 100] presented the *Path Intergral Monte Carlo Method* (PIMC) to solve the quantum many-body problem including the correct symmetry between particles. To do so, the method takes advantage of the classical isomorphism (established also by Feynman [98]), between the quantum many-body problem and a classical system of polymers.

In section 2.4 we introduce the PIMC method. There, it is shown how to exploit the classical isomorphism to map the quantum many-body system into a classical one of polymers [101, 102]. The first implementations of PIMC were not very efficient in sampling the permutation space, however, more recently, the introduction of the *worm algorithm* [103] has solved this problem in a more efficient way. The Path Integral Monte Carlo method is able to produce exact results for bosonic systems, allowing to study them in the low temperature regime, where quantum effects are more important. However, regarding fermionic systems, the efforts to adapt the method have not been as successful. A similar approach to the Fixed-Node technique can be implemented, usually referred in literature as *Restricted Path*, however, its application has not been as fruitful as its DMC counterpart.

Although the PIMC method allows to perform calculations of quantum systems in the low-temperature regime, the number of intermediate time steps needed for the method to converge increase proportionally to the inverse of the temperature. This means that, in practice, we cannot perform calculations at arbitrarily low temperatures. Even more, calculations at zero temperature are not allowed with this method. Fortunately, an adaptation of PIMC to zero temperature is possible by a combination of the variational principle and imaginary time propagation. This is known as the *Path Integral Ground State* method, and is discussed in 2.5.

## 2.1 Common basis

In this section we introduce the basis of Monte Carlo techniques. We also introduce the imaginary time propagation, which is a standard quantum many-body technique that will be common to most of the methods that we discuss in this chapter.

### 2.1.1 Monte Carlo integration and Importance Sampling

Dealing with multidimensional integrals is one of the difficulties that one finds when working with many-body problems. For the majority of systems of interest, they cannot be solved analytically, and numerical techniques have to be used. One of the options is to integrate them by employing stochastic methods, that are usually known as Monte Carlo techniques. Let us consider the integral:

$$\langle g \rangle = \int_V g(\vec{x})d\vec{x} = V \int_V \frac{1}{V} g(\vec{x})d\vec{x} \tag{2.3}$$

where the vector $\vec{x}$ represents the N variables that we have to integrate over. The subscript $V$ on the integration symbol indicates that we have to integrate over a multidimensional volume. The last equality in the above equation has been written for convenience. The reason is that it allows us to interpret the integral as the expectation value of $\frac{1}{V} g(\vec{x})$ over the uniform probability distribution in the integration volume $V$. This result can be generalized to any probability distribution. To evaluate the integral, a set of N uniformly distributed random points is sampled inside the integration volume, $\{\vec{x}_i\}$ and the values $\{g(\vec{x}_i)\}$ are calculated. Then, the definite integral and its variance can be computed as:

$$\langle g \rangle_V = \lim_{N \to \infty} \frac{V}{N} \sum_i^N g(\vec{x_i}) \tag{2.4}$$

$$\sigma^2 = \lim_{N \to \infty} \left[ \frac{V}{N} \sum_i^N g^2(\vec{x_i}) - \left( \frac{V}{N} \sum_i^N g(\vec{x_i}) \right)^2 \right] \tag{2.5}$$

In general, in many-body physics, one is interested in the computation of expectation values over a given probability distribution. The expectation value of a given function $g(\vec{x})$ over a given probability distribution $f(\vec{x})$ reads

$$\langle g \rangle = \int g(\vec{x}) f(\vec{x}) d\vec{x}. \tag{2.6}$$

with $f(\vec{x}) > 0$ in all the integration domain. Similarly to the Monte Carlo integration that we have already commented, the expectation value of $g(\vec{x})$ can be computed by sampling over the probability distribution $f(\vec{x})$ instead of from the uniform probability distribution employed in (2.5). Moreover, in general, any integral of the form of (2.3) can be expressed in the form of an expectation value over a probability distribution as it follows:

$$\langle g \rangle = \int_V \frac{g(\vec{x})}{f(\vec{x})} f(\vec{x}) d\vec{x} \equiv \langle f \rangle_f. \tag{2.7}$$

The above is well justified as long as the function that is used as a probability distribution is positive definite $f(\vec{x}) > 0$. It is clear that both choices will give the same result, so that $\langle g \rangle = \langle g \rangle_f$ but in general one can improve the variance by choosing a proper probability distribution $f(\vec{x})$. This is known as *Importance Sampling* technique and will be used in section 2.3 to improve the efficiency of DMC method.

The main difficulty that we have to deal with, in order to obtain the value of the integral of Eq. (2.7), is to sample random numbers distributed according to $f(\vec{x})$. Although some probability distributions can be sampled easily, a general purpose method is needed of other cases. In the following we introduce a general method, that will allow us to sample from any probability distribution.

In general, finding empirically a probability distribution corresponding to a given finite set of points is possible up to some statistical uncertainty[2]: this is called the direct problem. However, the inverse problem, that is, sampling numbers from a given probability distribution, presents some additional difficulties. A solution to this problem, was proposed in the context of Markov chains that we briefly introduce in the next section.

---

[2]This is guaranteed by the central limit theorem and the law of large numbers.

### 2.1.2 Markov Chains

A Markovian chain is defined as a collection of states $\{s_1, s_2 ..., s_n\}$ and a collection of rules or transition probabilities $T_{ij}$ from the ith state to the jth state. These probabilities are defined in such a way that the succession of states do not depend on the previous history. The probabilities are normalized so the transition to any final state is guaranteed,

$$\sum_i T_{ij} = 1, \tag{2.8}$$

with $0 \le T_{ij} \le 1$. The solution of the direct problem for a Markov chain process is simple, the probability of arriving to a given state i reads:

$$P_i = \sum_j T_{ij} P_j. \tag{2.9}$$

The above expression, constitutes a set of linear equations that, together with the normalization condition $\sum_i P_i = 1$, yields an unique solution as long as the equations (2.9) are linearly independent.

The inverse problem, being able to obtain expectation values from the probability distribution, is more difficult, but also more interesting, because it can be used to evaluate physical observables.

### 2.1.3 The Metropolis Algorithm

In 1953, Nicholas Metropolis, Arianna W. Rosenbluth, Marshall N. Rosenbluth, Augusta.H. Teller and Edwar Teller, produced an algorithm with a solution to this problem [104]. This method is usually referred in literature as the $MR^2T^2$, of simply as the Metropolis method.

Let $q_{ij}$ be the elements of an auxiliary symmetric matrix so that the probability of going from the ith to jth state is the same as that of the reverse process. One defines a transition matrix as

$$
\begin{aligned}
T_{ij} &= q_{ij} & if \ P_i < P_j \quad i \ne j \\
T_{ij} &= q_{ij} P_j / P_i & if \ P_i > P_j \quad i \ne j \\
T_{ii} &= q_{ii} + \sum_k q_{ki}(1 - P_k/P_i) & for \ P_k < P_i \quad .
\end{aligned}
\tag{2.10}
$$

Notice that the last relation is set to fulfill the normalization condition (2.8). It is easy to check that the matrix of probabilities defined in (2.10) satisfies the detailed balance condition $T_{ij} P_j = T_{ji} P_i$, so that any process is compensated by its reverse one. This guarantees that the asymptotic probability distribution is stationary. Besides and more important, if the system is ergodic this probability distribution is unique. To check this, we have to distinguish the two cases $P_i > P_j$, and $P_i < P_j$. We focus on the second one, as the later can be proved in a similar way, while the case $i = j$ is straight forward. Following the Metropolis prescription:

$$P_i T_{ij} = P_j T_{ji}$$
$$P_i q_{ij} = P_j q_{ji} \frac{P_i}{P_j} = q_{ij} P_j \tag{2.11}$$

where in the second line we have make use of the definitions in Eq. (2.10) and of the fact that $q_{ij} = q_{ji}$. The last step for proving the validity of the Metropolis prescription is to show that it fulfills the condition (2.9) which can be shown straightforwardly by summing on j on the detailed balance condition and using the normalization condition (2.8).

The Metropolis algorithm gives a simple solution to the problem of sampling over a probability distribution that can be easily implemented in Monte Carlo calculations. The flow of the algorithm can be easily established:

1. Start from an initial state $s_i$ and propose a movement to a state $s_j$

2. If the probability $P_j$ is greater that that of the initial state $P_i$, accept the movement.

3. Otherwise accept the movement with probability $P_j/P_i$.

4. If the movement has been accepted, take the proposed state as the initial state for the next step.

5. Repeat from point 1.

For a given probability distribution with a continuous set of states $P(x)$, the analysis would be the same but replacing $T_{ij} \rightarrow P(x, x')$ and summations by integrals. In this case, instead of Markov chains, one has a Markov processes. An example of this would be a quantum (or classical) particle traveling through a medium with whom it can be scattered or absorbed.

### 2.1.4 Imaginary time propagation

Before introducing the QMC methods that we use in this Thesis, we schematically describe the imaginary time propagation technique. It constitutes a standard method (also known as projector method) that will be in the basis of both the DMC and the Path Integral methods that we describe in the rest of the chapter.

The time dependent Schrödinger equation of a system described by a Hamiltonian $\hat{H}$ reads

$$i\hbar \frac{\partial \Psi}{\partial t} = \hat{H} \Psi. \tag{2.12}$$

To find the ground state of the system, we can solve this equation in imaginary time. With this purpose, we define the imaginary time as $\tau = it/\hbar$, so that the previous equation reads:

$$-\frac{\partial \Psi}{\partial \tau} = \hat{H}\Psi \tag{2.13}$$

In Dirac's notation, the formal solution of the equation (2.13) for a time independent $\hat{H}$ can be written as

$$|\Psi(\tau)\rangle = e^{-\hat{H}\tau} |\Psi(\tau = 0)\rangle, \tag{2.14}$$

where $|\Psi(\tau = 0)\rangle$ represents the initial state of the system, which can be expanded into the basis of eigenstates of the Hamiltonian $\{\phi_i\}$,

$$|\Psi(\tau = 0)\rangle = \sum_{i=0}^{\infty} a_i |\phi_i\rangle. \tag{2.15}$$

In the following, we assume that the eigenstates of the Hamiltonian are ordered so that $E_0 < E_1 < E_2....$ Introducing (2.15) into (2.14), one obtains:

$$|\Psi(\tau)\rangle = \sum_{i=0}^{\infty} e^{-E_i\tau} a_i |\phi_i\rangle. \tag{2.16}$$

The exponential suppression of the excited estates, guarantees that, for long imaginary times, the only relevant contribution (in relative terms) to the normalized wave function will come from the ground state of the system.

$$\lim_{\tau \to \infty} \frac{\Psi(\tau)}{\sqrt{\langle \Psi(\tau)|\Psi(\tau)\rangle}} \sim |\phi_0\rangle. \tag{2.17}$$

## 2.2 Variational Monte Carlo

Variational Monte Carlo (VMC) is the simplest method that we use in this Thesis. Although it only offers a variational solution to the problem, among its advantages we can mention its simpler implementation and lower computational cost compared with other QMC methods.

### 2.2.1 The Variational Principle

As is indicated by its name, VMC takes advantage of the variational principle, which states that, for any trial wave function $\Psi_T$, the quantity:

$$E_V = \frac{\langle \Psi_T | \hat{H} | \Psi_T \rangle}{\langle \Psi_T | \Psi_T \rangle}, \tag{2.18}$$

constitutes an upper bound to the energy of the exact ground state of the system, $E_0$. To show this, we can express the trial wave function in the orthonormal basis of eigenstates of the Hamiltonian, $\{\phi_n\}$ , $\langle\phi_n|\phi_m\rangle = \delta_{n,m}$. As before

$$\Psi_T = \sum_{n=0} a_n|\phi_n\rangle, \tag{2.19}$$

with $\phi_n$ the eigenfunctions satisfying Schrodinger equation

$$\hat{H}|\phi_n\rangle = E_n|\phi_n\rangle, \tag{2.20}$$

Using this, Eq. (2.18) reads:

$$
\begin{aligned}
E_V &= \frac{(\sum_{n=0} a_n^*\langle\phi_n|)\hat{H}(\sum_{m=0} a_m|\phi_m\rangle)}{(\sum_{n=0} a_n^*\langle\phi_n|)(\sum_{m=0} a_m|\phi_m\rangle)} \\
&= \frac{\sum_{n=0}|a_n|^2 E_n}{\sum_{n=0}|a_n|^2} = E_0 + \frac{\sum_{n=1}|a_n|^2 E_n}{\sum_{n=0}|a_n|^2} \geq E_0.
\end{aligned}
\tag{2.21}
$$

In general, the trial wave function is made to depend on a set of parameters, $\Psi_T = \Psi_T(R; \lambda_1...\lambda_M)$, so that an improved upper bound can be found from the variational principle imposing the conditions:

$$
\begin{aligned}
\frac{\partial E_T(R; \lambda_1...\lambda_M)}{\partial \lambda_i} &= 0 \qquad i = 1, N \\
\frac{\partial^2 E_T(R; \lambda_1...\lambda_M)}{\partial^2 \lambda_i} &> 0 \qquad i = 1, N
\end{aligned}
\tag{2.22}
$$

The variational principle together with the sampling procedure explained in section 2.1, offer us a powerful tool to obtain suitable upper bounds to the ground state energy. Finally it is worth noticing that, if the trial wave function coincides with an exact eigenstate $\phi_i$ of the Hamiltonian with energy $E_i$ the variance of the Monte Carlo estimation is exactly zero [105, 106]:

$$\sigma_T^2 = \frac{\langle\Psi_T|(\hat{H} - E_V)^2|\Psi_T\rangle}{\langle\Psi_T|\Psi_T\rangle} = 0 \qquad if \ \Psi_T = \phi_i \ (E_V = E_i). \tag{2.23}$$

### 2.2.2 Computation of Observables

The objective of Variational Monte Carlo is to evaluate expectation values of physical quantities using the variational principle. In coordinate representation, the energy estimator in Eq.(2.18) reads

$$E_V = \frac{\int \Psi_T^*(\mathbf{R})H\Psi_T(\mathbf{R})d\mathbf{R}}{\int \Psi_T^*(\mathbf{R})\Psi_T(\mathbf{R})d\mathbf{R}}. \tag{2.24}$$

As our purpose is to evaluate this quantity with Monte Carlo, it is convenient to remember that the density of probability of finding the system in the state $\mathbf{R}$ is given by the square modulus of the wave function. With this in mind, we can define the following probability distribution:

$$P(\mathbf{R}) = \frac{\Psi_T^*(\mathbf{R})\Psi_T(\mathbf{R})}{\int \Psi_T^*(\mathbf{R})\Psi_T(\mathbf{R})d\mathbf{R}}, \tag{2.25}$$

so that the expectation value of the variational energy, can be written as

$$E_V = \int d\mathbf{R} P(\mathbf{R}) E_L(\mathbf{R}), \tag{2.26}$$

where we have introduced the so called local energy:

$$E_L = \frac{1}{\Psi_T(\mathbf{R})} H(\mathbf{R})\Psi_T(\mathbf{R}). \tag{2.27}$$

In this way, the local energy of the system is computed by sampling the local energy over the probability distribution defined by the squared wave function. To do so, we take advantage of the Metropolis algorithm described in 2.1. In general, not only the local energy, but any other observable $\hat{O}$ can be computed at the variational level following this expression

$$\langle \hat{O} \rangle_{VMC} = \int d\mathbf{R} P(\mathbf{R}) O_L(\mathbf{R}) \tag{2.28}$$

by sampling its associated local quantity, $O_L$, over the probability distribution $P(\mathbf{R})$

$$O_L(\mathbf{R}) = \frac{1}{\Psi_T(\mathbf{R})} O(\mathbf{R})\Psi_T(\mathbf{R}). \tag{2.29}$$

## 2.3 Diffusion Monte Carlo

An improvement over the Variational Monte Carlo method is given by the Diffusion Monte Carlo (DMC) technique. This method finds the real ground state of a bosonic system by scholastically solving the time dependent Schrödinger equation in imaginary time (see subsection 2.1.4). When dealing with Fermions, the method becomes variational as will be commented in section 2.3.5.

### 2.3.1 DMC principles

Our starting point is, once again, the time dependent Schrödinger equation (2.13) in which we introduce an energy shift $E_T$, which is equivalent to replacing $\hat{H} \rightarrow \hat{H} - E_T$,

$$-\frac{\partial \Psi}{\partial \tau} = \left( \hat{H} - E_T \right) \Psi, \tag{2.30}$$

the justification for introducing the constant shift $E_T$ will be clear later. This equation has a formal solution that can be expanded in the form of (2.16),

$$|\Psi(\tau)\rangle = \sum_{i=0}^{\infty} e^{-(E_i - E_T)\tau} a_i |\phi_i\rangle. \tag{2.31}$$

It is clear, as it was explained in subsection 2.1.4, that the only long-imaginary time contribution will be the one coming from the ground state. For a Hamiltonian of the form given by Eq. (2.2), the time dependent Schrödinger equation of (2.30) reads:

$$-\frac{\partial \Psi}{\partial \tau} = -D\nabla^2 \Psi + (V - E_T)\Psi, \tag{2.32}$$

where we have introduced the diffusion coefficient $D = \frac{\hbar^2}{2m}$ and the notation $\nabla^2 = \sum_{i=1}^{N} \nabla_i^2$ and $V = \sum_{i<j}^{N} V^{2B}(\mathbf{r}_{ij}) + \sum_i^N V^{Ext.}(\mathbf{r}_i)$. The first term in this equation describes a *diffusion process* in imaginary time, that actually gives its name to the method. The last one, is a *branching term*, that affects the norm of the system as it evolves in imaginary time. It is worth noticing that the shift $E_T$ introduced in the Hamiltonian (and appearing in the branching term), allows us to keep the norm under control: if $E_T \approx E_0$ the final state ($\tau \to \infty$) will be the ground state of the Hamiltonian (*cf.* Eq. (2.31)). Notice that if the value $E_T$ is very different from the ground state energy, the solution would diverge (or vanish) if $E_T > (<)E_0$.

Before considering how to integrate the previous differential equation, some remarks have to be done:

1. In order to find the real ground state of the system $\phi_0$, the initial state $\Psi(\tau = 0)$ needs to have a finite overlap with $\phi_0$. The larger this overlap is, the faster the method will converge.

2. The reference energy $E_T$ introduced in equation 2.30 has to be chosen in a smart way to improve the convergence.

In order to solve the time dependent Schrödinger equation, we need to determine how to perform the propagation from an initial state at imaginary time $\tau = 0$, to the state at time $\tau$. This can done by introducing the Green's function formalism,

$$\Psi(\mathbf{R}, \tau) = \int G(\mathbf{R}, \mathbf{R}', \tau)\Psi(\mathbf{R}', 0)d\mathbf{R}'. \tag{2.33}$$

Considering the formal solution of the Schrödinger equation:

$$|\Psi(\tau)\rangle = e^{-(\hat{H} - E_T)\tau}\Psi(0)\rangle, \tag{2.34}$$

projecting into the basis of position coordinates, and introducing the completeness relation, one obtains

$$\langle \mathbf{R}|\Psi(\tau)\rangle = \int \langle \mathbf{R}|e^{-(\hat{H}-E_T)\tau}|\mathbf{R}'\rangle \langle \mathbf{R}'|\Psi(0)\rangle d\mathbf{R}'. \tag{2.35}$$

By comparison with Eq. (2.33), we can identify the Green's function or propagator

$$G(\mathbf{R}, \mathbf{R}', \tau) = \langle \mathbf{R}|e^{-(\hat{H}-E_T)\tau}|\mathbf{R}'\rangle. \tag{2.36}$$

To obtain an exact solution of (2.35) is not possible in general. To tackle this problem one can use short-time expansions leading to approximations that work for small $\tau$. With this in mind, the propagation to longer imaginary times can be done by taking advantage of the convolution property of the Green's function: The Green's function $G(\mathbf{R}, \mathbf{R}', \tau)$ propagating the state of the system a time interval $\tau$ can be obtained as the integral of the product of two Green's function propagating over an interval $\tau/2$,

$$G(\mathbf{R}, \mathbf{R}', \tau) = \int G(\mathbf{R}, \mathbf{R}'', \tau/2)G(\mathbf{R}'', \mathbf{R}', \tau/2)d\mathbf{R}''. \tag{2.37}$$

Using this relation, we can consider the propagation in an imaginary time interval $\tau$ as the sum of many imaginary time propagation with a smaller time step $\delta\tau$, which can be made arbitrarily small just by introducing intermediate time steps. This allows us to use any short time expansion like Suzuki Trotter's expansion [107, 108] or more elaborated ones (see, for example, [109])

Working with these short-time expansions of the propagator allows us to integrate the Schrödinger equation in imaginary time. But, before explaining the details of the integration method itself, let us introduce the *Importance Sampling* technique. This technique supposes a key improvement in the efficiency of DMC simulations. It consists on introducing a trial wave function $\Psi_T$ in the method, that is expected to have a significant overlap with the exact ground state of the system in order to reduce the variance. The crude case that we have been discussing so far, can be obtained as a particular case of the DMC method with importance sampling technique just by setting $\Psi_T = 1$

### 2.3.2 Importance Sampling

In real many-body problems, potentials with some divergences are usually involved. Due to this, the variance in the estimation of observables can be very large, making the DMC method to have a poor convergence. A way to get around this problem is to use the Importance Sampling technique, which employs a trial wave function, that normally has been previously optimized employing the VMC method. The trial wave function acts as a guiding function in order to enhance the exploration of regions of the phase space where the wave function is larger, and suppress the sampling in regions where it tends to zero, as happens in the vicinity of a divergent repulsive potential.

The first step consists on writing the Schrödinger equation in imaginary time not for the wave function of the system, but for the product of the real wave function times a trial one, independent of the imaginary time

$$f(\mathbf{R}, \tau) = \Psi_T(\mathbf{R})\phi(R, \tau). \qquad (2.38)$$

The above is usually referred as a mixed probability distribution. With this definition, the time dependent Schrödinger equation reads

$$-\frac{\partial f(\mathbf{R}, \tau)}{\partial \tau} = -D\left(\nabla^2 f(\mathbf{R}, \tau) + \frac{2}{\Psi_T^2(\mathbf{R})}(\nabla\Psi_T(\mathbf{R}))^2 f(\mathbf{R}, \tau) - \frac{2}{\Psi_T(\mathbf{R})}\nabla\Psi_T(\mathbf{R})f(\mathbf{R}, \tau)\right.$$
$$\left. -\frac{1}{\Psi_T(\mathbf{R})}\nabla^2\Psi_T(\mathbf{R})f(\mathbf{R}, \tau)\right) + (V(\mathbf{R}) - E_T)f(\mathbf{R}, \tau). \qquad (2.39)$$

Now, we define the quantum force $f(\mathbf{R}) = \frac{2}{\Psi_T(\mathbf{R})}\nabla\Psi_T(\mathbf{R})$, whose physical meaning will be clear later, and also take into account the definition of the local energy in Eq. (2.27). Then, the above equation is written in the compact form:

$$-\frac{\partial f(\mathbf{R}, \tau)}{\partial \tau} = -D\nabla^2 f(\mathbf{R}, \tau) + D\nabla\left(F(\mathbf{R})f(\mathbf{R}, \tau)\right) + (E_L(\mathbf{R}) - E_T)f(\mathbf{R}, \tau). \quad (2.40)$$

It is worth noticing that Eq. (2.32) can be recovered by setting $\Psi_T(\mathbf{R}) = 1$, so that $F(\mathbf{R}) = 0$ and $E_L(\mathbf{R}) = V(R)$. The evolution described by the previous equation can be though as the sum of three different operators acting on $f(\mathbf{R}, \tau)$:

$$-\frac{\partial f(\mathbf{R}, \tau)}{\partial \tau} = \left(\hat{O}_K + \hat{O}_D + \hat{O}_B\right)f(\mathbf{R}, \tau). \qquad (2.41)$$

where $\hat{O}_K$, $\hat{O}_D$ and $\hat{O}_B$ are the short-time Diffusion, drift and branching operators, that correspond respectively to the three terms on the right hand side of Eq. (2.40). Also in analogy with Eq. (2.35) the evolution of the mixed probability distribution $f(\mathbf{R})$ can be written in terms of a Green's function

$$\langle\mathbf{R}|f(\tau)\rangle = \int\langle\mathbf{R}|G(\tau)|\mathbf{R}'\rangle\langle\mathbf{R}'|f(0)\rangle d\mathbf{R}'. \qquad (2.42)$$

In the following we discuss how to integrate the imaginary time Schrödinger equation with importance sampling.

### 2.3.3 Monte Carlo integration

Up to now we have presented all the necessary ingredients for DMC method. The only remaining step, is how to integrate Eq. (2.41). Before giving an answer to this question, let us summarize the previous ideas:

1. We consider the Schrödinger equation (2.13) in imaginary time and its formal solution (2.35), that after long enough propagation in imaginary time gives the ground state of the system.

2. The wave function of the system is represented by a set of vectors of coordinates $\{\mathbf{R}\}$ (walkers) that evolve in imaginary time according to (2.13).

3. When considering a system of bosons, the ground state of the system is described by a positive definite wave function, so it can be used as a probability distribution. (The fermionic case will be considered in section 2.3.5)

4. In general it is not possible to solve the evolution given by Eq. (2.35) exactly. This problem can be sorted out by taking advantage of the convolution property of the Green's Function (2.37). In this way, the time evolution is computed as a product of Green's functions at small time steps $\delta\tau$:

$$|\Psi(\tau)\rangle = \prod_{i=1}^{n} e^{-(\hat{H}-E_T)\delta\tau} |\Psi(0)\rangle . \tag{2.43}$$

Now, we focus on how to treat the Green's function at each time step of our simulations. To first order in $\delta\tau$, the Green's function can be further decomposed in the form

$$e^{-(\hat{H}-E_T)\delta\tau} \approx e^{\hat{O}_K \delta\tau} e^{\hat{O}_D \delta\tau} e^{\hat{O}_B \delta\tau} + \mathcal{O}(\delta\tau^2). \tag{2.44}$$

The above expression makes sense when we define the short-time Green's function $\hat{G}_K = e^{\hat{O}_K \delta\tau}$, $\hat{G}_D = e^{\hat{O}_D \delta\tau}$ and $\hat{G}_B = e^{\hat{O}_B \delta\tau}$, following the notation introduced in equation (2.41) for the kinetic, drift and branching parts respectively. In coordinate representation, one has [92]

$$G_K(\mathbf{R}, \mathbf{R}', \tau) = (4\pi D\tau)^{-dN/2} \exp[-\frac{(\mathbf{R} - \mathbf{R}')^2}{4D\tau}] \tag{2.45}$$

$$G_D(\mathbf{R}, \mathbf{R}', \tau) = \delta\left(\mathbf{R} - \mathcal{R}'(\tau)\right) \tag{2.46}$$

$$G_B(\mathbf{R}, \mathbf{R}', \tau) = \exp\left[-(E_L(\mathbf{R}) - E_T)\tau\right] \delta\left(\mathbf{R} - \mathbf{R}'\right), \tag{2.47}$$

where $G_K(\mathbf{R}, \mathbf{R}', \tau)$ is the well known solution for the non-interacting problem, corresponding free propagation between $\mathbf{R}$ and $\mathbf{R}'$. $G_D(\mathbf{R}, \mathbf{R}', \tau)$ is usually called the drift term, and it represents a deterministic evolution given by the drift force coming from the introduction of a trial wave function. This evolution leads to a new set of coordinates $\mathcal{R}(\tau)$ defined by the equations

$$\frac{d\mathcal{R}(\tau)}{d\tau} = D\mathbf{F}(\mathcal{R}(\tau)),$$
$$\mathcal{R}(0) = \mathbf{R}, \tag{2.48}$$

which is equivalent to the equation describing the movement of a classical particle under the action of a force $\mathbf{F}$. Notice that actually, $\mathbf{F}(\mathcal{R}(\tau))$ is not a real force, as it does not have units of force. Finally the propagator $G_B(\mathbf{R}, \mathbf{R}', \tau)$ of Eq. (2.47) corresponds to the branching term: determined by the exponential of the difference between the local energy and the energy shift $E_T$. It acts as a reweighing term, giving higher weights to the walkers that have a lower energy. Later on in this section we will comment how is this implemented in the program.

If one wants to go beyond this first order approximation one of the possible, but not unique, solutions [109] is to use the following expansion:

$$e^{-\hat{H}\delta\tau} \approx e^{\hat{O}_B\delta\tau/2}e^{\hat{O}_D\delta\tau/2}e^{\hat{O}_K\delta\tau}e^{\hat{O}_D\delta\tau/2}e^{\hat{O}_B\delta\tau/2} + \mathcal{O}(\delta\tau^3), \tag{2.49}$$

which, indeed, is very convenient for its implementation in a DMC algorithm. Notice that, as the previous operator will be applied iteratively, in practice one can implement it so that any of the $\hat{G}_\alpha$ ($\alpha = K, D, B$) operators can be the first one at each time step. For example, reordering them one can construct a second order algorithm in which the first propagator would be the diffusion one, similar to the first order algorithm (see Eq. (2.44)). This, makes it rather easy to go from a first to a second order algorithm. Making use of the results of equations (2.45), (2.46) and (2.47) and inserting them in the short imaginary time expansion of (2.44), the total Green's function for the first order propagator in coordinate representation reads

$$\begin{aligned}
G(\mathbf{R}, \mathbf{R}', \tau) &= \int d\mathbf{R}_1 d\mathbf{R}_2 G_B(\mathbf{R}', \mathbf{R}_2, \tau) G_D(\mathbf{R}_1, \mathbf{R}_2, \tau) G_K(\mathbf{R}, \mathbf{R}_1, \tau) \\
&= \int d\mathbf{R}_1 d\mathbf{R}_2 e^{-(E_L(\mathbf{R}_2)-E_T)\tau} \delta(\mathbf{R}_2 - \mathbf{R}') \delta(\mathbf{R}_1 - \mathcal{R}_2(\tau)) \\
&\qquad \times (4\pi D\tau)^{-dN/2} e^{-\frac{(\mathbf{R}-\mathbf{R}_1)^2}{4D\tau}} = \\
&= (4\pi D\tau)^{-dN/2} e^{-(E_L(\mathbf{R})-E_T)\delta\tau} \\
&\qquad \times \int d\mathbf{R}_1 \delta(\mathcal{R}_1(\tau) - R) e^{-\frac{(\mathbf{R}-\mathbf{R}_1)^2}{4D\tau}} + \mathcal{O}(\tau^2)
\end{aligned} \tag{2.50}$$

where he symbol $\mathcal{R}_1(\tau)$ is used to refer to a solution of Eq. (2.48) with initial condition $\mathbf{R}_1(0)$. Similarly, from Eqs. (2.45), (2.46), (2.47) and (2.49), the second order propagator reads

$$
\begin{aligned}
G(\mathbf{R}, \mathbf{R}', \tau) &= \int d\mathbf{R}_1 d\mathbf{R}_2 d\mathbf{R}_3 d\mathbf{R}_4 G_B(\mathbf{R}_4, \mathbf{R}', \tau/2) G_D(\mathbf{R}_3, \mathbf{R}_4, \tau/2) G_K(\mathbf{R}_2, \mathbf{R}_3, \tau/2) \\
&\quad \times G_D(\mathbf{R}_1, \mathbf{R}_2, \tau/2) G_B(\mathbf{R}, \mathbf{R}_1, \tau/2) \\
&= \int d\mathbf{R}_1 d\mathbf{R}_2 d\mathbf{R}_3 d\mathbf{R}_4 e^{-(E_L(\mathbf{R}_4) - E_T)\tau/2} \delta(\mathbf{R}_4 - \mathbf{R}') \delta(\mathbf{R}_3 - \mathcal{R}_4(\tau/2)) \\
&\quad \times (4\pi D\tau)^{-dN/2} e^{-\frac{(\mathbf{R}_2 - \mathbf{R}_3(\tau))^2}{4D\tau}} \\
&\quad \times \delta(\mathbf{R}_1 - \mathcal{R}_2(\tau/2)) e^{-(E_L(\mathbf{R}) - E_T)\tau/2} \delta(\mathbf{R} - \mathbf{R}_1) = \\
&= (4\pi D\tau)^{-dN/2} e^{-\left(\frac{E_L(\mathbf{R}) + E_L(\mathbf{R}')}{2} - E_T\right)\tau} \\
&\quad \times \int d\mathbf{R}_2 d\mathbf{R}_3 \delta(\mathcal{R}_3(\tau/2) - \mathbf{R}') \delta(\mathbf{R} - \mathcal{R}_2(\tau/2)) e^{-\frac{(\mathbf{R}_2 - \mathbf{R}_3)^2}{4D\tau}} + \mathcal{O}(\tau^3)
\end{aligned}
\tag{2.51}
$$

where the symbol $\mathcal{R}_2(\tau/2)$ and $\mathcal{R}_3(\tau/2)$ are used to refer to a solution of Eq. (2.48) with initial condition $\mathbf{R}_2(0)$ and $\mathbf{R}_3(0)$ respectively. From the above expressions, it is clear that the diffusion processes in both schemes are similar. The same occurs with the modifications needed in order to implement the branching term in a second order algorithm from that of a first order one. Regarding the drift term, equation (2.48), has to be solved with the same precision in $\delta\tau$ as we are requiring to the Green's function. Details about the numerical implementation of these three propagators are given in the following section.

### 2.3.4  DMC algorithm

In DMC, the probability distribution is represented by a set of vectors of coordinates, $\{\mathbf{R}_1, ..., \mathbf{R}_{N_w}\}$ that we call walkers, with each vector representing a set of coordinates of the whole N-particle system, $\mathbf{R} = \{\mathbf{r}_1, ..., \mathbf{r}_N\}$, so that at each step the probability distribution is represented by:

$$
f(\mathbf{R}, \tau) = \frac{1}{\mathcal{N}} \sum_{i=1}^{N_w} \delta(\mathbf{R} - \mathbf{R}_i(\tau)).
\tag{2.52}
$$

with $\mathcal{N}$ a normalization constant. In general, the evolution of this probability distribution is given by (2.41). But as we have shown, in practice one uses a short imaginary time expansion of the Green's function (*cf.* Eq. (2.44) for a first order expansion or (2.49) for the second order version). It allows us to act on the probability distribution with three different short-time propagators, as described in the previous section an whose explicit form is exposed in Eqs. (2.50) and (2.51). In DMC, the diffusion term $\hat{G}_K$ implemented by sampling a displacement for each of the coordinates of the walker from the a normalized Gaussian probability distribution,

$$
\mathbf{R} = \mathbf{R}' + \boldsymbol{\xi},
\tag{2.53}
$$

with $\boldsymbol{\xi}$ a vector whose coordinates are sampled form the normalized Gaussian distribution. The $\hat{G}_D$ represents the drift contribution coming from the use of a trial wave function for Importance Sampling. To first order in $\delta\tau$ this is given by $\mathcal{R}'(\delta\tau) = \mathcal{R}'(0) + \mathbf{F}(\mathbf{R}(0))\delta\tau$. While for the second order implementation, one has to solve Eq. (2.48) with a second order algorithm. In our case we use the following Runge-Kutta (predictor-corrector) method [109] for the displacement from $\mathbf{R}'$ to $\mathbf{R}$:

1. $\mathbf{R}_1 = \mathbf{R}' + \mathbf{F}(\mathbf{R})\delta\tau/2$

2. $\mathbf{R}_2 = \mathbf{R}' + \left(\mathbf{F}(\mathbf{R}') + \mathbf{F}(\mathbf{R}_1)\right)\delta\tau/4$

3. $\mathbf{R}_3 = \mathbf{R}' + \mathbf{F}(\mathbf{R}_2)\delta\tau$

4. $\mathbf{R} = \mathbf{R}_3$.

Notice that, in order to have a second order algorithm, not only one has to solve (2.48) with a second order method but the drift propagator has to be included twice at each iteration of the simulation, as can be deduced from Eq. (2.49).

The $\hat{G}_K$ and $\hat{G}_D$ propagators do not change the norm of $f(\mathbf{R}, \tau)$. On the contrary, the branching Green's function $G_B$ reweighs the walkers representing the probability distribution and changes its norm. The branching propagator acting on $f$ reads:

$$
\begin{aligned}
f(\mathbf{R}, \tau + \delta\tau) &= \int d\mathbf{R}' e^{-(E_L(R) - E_T)\tau} f(\mathbf{R'}, \tau)\delta(\mathbf{R'} - \mathbf{R}) \\
&= e^{-(E_L(R) - E_T)\tau} f(\mathbf{R}, \tau),
\end{aligned}
\tag{2.54}
$$

which clearly do not preserve the norm of $f$. In our algorithm this step is implemented by replicating (or killing) the walkers that have high (low) values of $e^{-(E_L(R) - E_T)}$. Schematically:

1. For each walker, in the first (second) order algorithm we evaluate the quantity $n_{sons} = e^{-(E_L(R) - E_T)\tau} + \eta$ ($n_{sons} = e^{-(\frac{E_L(\mathbf{R}) + E_L(\mathbf{R}')}{2} - E_T)\tau} + \eta$), with $\eta$ sampled from the uniform probability distribution $[0, 1)$.

2. $n_{sons}$ is rounded to its integer part.

3. If $n_{sons} = 0$ the walker is removed. On the contrary, if $n_{sons} > 0$ we replicate the walker $n_{sons}$ times and include it in our representation of $f(\mathbf{R})$.

4. After repeating the procedure for each walker, we obtain a *new generation* of walkers describing the new mixed-probability distribution function $f(\mathbf{R}, \tau + \delta\tau)$, that will be evolved in imaginary time during the following iteration.

At this point it is worth noticing that the constant shift $E_T$, introduced at the very beginning in the Schrödinger equation, should be adjusted during the simulation in order to control the population of walkers around a desired value. Finally, it is importance to remark that the convergence of the method is only guaranteed when the following two limits are accomplished simultaneously: $\delta\tau \to 0$ and $N_w \to \infty$, and therefore the convergence in these two parameters has to be studied in order to obtain reliable results.

### 2.3.5  Fermions

Systems involving fermions are usually more challenging than those corresponding to bosonic system. The anti-symmetry of the wave function lead to the so called *sign problem*. In this subsection we show how the DMC method can be adapted to tackle this problem. We focus in the particular case considered in this Thesis, in which we consider particles with spin but under the action of spin independent interactions. An adaptation of the DMC method to work with systems involving spin-orbit coupling can be found in [110]. In the particular case studied in this Thesis, this means that only Hamiltonians of the form of Eq. (2.2) are considered.

When using quantum Monte Carlo methods, the sign problem makes the probability distribution $f(\mathbf{R}, \tau) = \Psi_T^F(\mathbf{R})\phi(\mathbf{R}, \tau)$ to not be positive definite, and therefore it can no longer be interpreted as a probability distribution from where sampling can be done. Going around the problem is possible in several ways, for example through the so called *Fixed-Node* approximation. The prize that we have to pay is that the method becomes variational, except for the particular case in which we know the exact *nodal surface* of the problem. By nodal surface we refer to the hypersurface that divides the phase space in regions where the wave function is either positive or negative definite.

#### 2.3.5.1  The Fixed-Node approximation

The presence of nodes in the wave function of a Fermionic system implies that, in general, the mixed probability distribution

$$f^F(\mathbf{R}, \tau) = \Psi_T^F(\mathbf{R})\phi^F(\mathbf{R}, \tau) \tag{2.55}$$

is not positive definite. The superscript $F$ in the previous equation, indicates that we are referring to a fermionic system. In the following, we avoid it for simplicity. One way of going around this problem is to restrict ourselves to the regions of the Hilbert space in which $\phi$ and $\Psi_T$ have the same nodal surface, which constitutes the basis of the standard Fixed-Node technique. Actually, the implementation of Fixed-Node into DMC is natural when using the importance sampling technique described in section 2.3.2: the only thing that one has to do is to include the nodal surface into the trial wave function. The choice of the trial wave function divides the phase space in regions in which the wave function has different sign. Once a nodal surface is chosen the drift force near the nodes diverges, pushing the walker away from it, so that a walker will never be able to cross it. When the algorithm is implemented, however, finite values of $\delta\tau$ are used, and some walkers may cross the nodes spuriously (due to the Gaussian movements). If this is the case, the walker should be removed.

In general, the trial wave function for a fermionic system is chosen as the product of an antisymmetric part $\Psi_A$, times a symmetric term $\Psi_S$ which, for translationally invariant systems as the ones considered in this Thesis, is usually chosen to be of the Jastrow form [111], similar to what is done for bosonic systems. This fermionic wave function, termed Jastrow-Slater wave function then reads

$$\Psi_T(\mathbf{R}) = \Psi_A(\mathbf{R})\Psi_S(\mathbf{R}). \tag{2.56}$$

Withing this approximation, DMC allows to find efficiently the lower energy state compatible with the nodal surface that is imposed onto the trial wave function. The solution obtained in this way corresponds to the exact ground state of the system only when trial wave function has the nodal surface of the real ground state. Otherwise it becomes variational, as it was shown by Ceperley, Moskowitz and collaborators [112, 113].

The most simple choice for the nodal surface is to construct a Slater determinant for each of the species present in the system. Following this prescription, the antisymmetric part of the trial wave function reads:

$$\Psi_A(R) = \prod_\alpha D_\alpha(\mathbf{R}_\alpha) \tag{2.57}$$

with the index $\alpha$ labeling each of the different fermionic species present in the system and $\mathbf{R}_\alpha$ referring to the subset of coordinates of the $\alpha$ specie. The natural question now is: which is the better choice for the orbitals $\{\phi^j\}$ inside the Slater matrix $D_\alpha^{ij} = \phi^j(\mathbf{r}_\alpha^i)$?

**Free particle orbitals**

The simplest choice for the orbitals $\phi^j$, as long as the system is translationally invariant, consists on using the solution of the free Fermi system. In this approximation the orbitals that we use are plane waves:

$$\phi_\alpha^j(\mathbf{r}_\alpha^i) = e^{i\mathbf{k}\mathbf{r}_\alpha^i}. \tag{2.58}$$

This solution is expected to be a good approximation in the weakly interacting regime. However, when correlations become more important, it is necessary to improve the solution by employing more elaborated orbitals.

**The Backflow correction**

An improvement over the free particle orbitals is obtained by introducing Backflow coordinates. This idea was first introduced by Feynman and Cohen to study the phonon-roton spectrum in $^4$He [114]. Later on, the same idea was applied to fermionic systems, where backflow-based wave functions were used for variational calculations on $^3$He [115]. More recently, it has been applied to study other fermionic systems in the correlated regime, such as two-dimensional $^3$He (see, for example, [116]). Essentially the method consists in replacing the set of particle coordinates $\{\mathbf{r}_i\}$ in the plane wave orbitals by a new set of coordinates $\{\mathbf{q}_i\}$ defined as:

$$\mathbf{r}_i \to \mathbf{q}_i = \mathbf{r}_i + \lambda_B \sum_{j \neq i}^{N} \eta_B(\mathbf{r}_{ij}) \mathbf{r}_{ij}. \tag{2.59}$$

with $\lambda_B$ a variational parameter and $\eta_B$ the backflow correlation function. In the work presented in this Thesis this function is chosen to be a Gaussian $\eta_B(\mathbf{r}_{ij}) = e^{(r_{ij}-r_B)^2/\alpha_B^2}$, so that we have the two additional variational backflow parameters: $\alpha_B$ and $r_B$.

To obtain the previous result, let us consider that the wave function of the fermionic system can be constructed as in Eq. 2.56, that is, as a product of a symmetric function $\Phi_S(\mathbf{R})$ and an antisymmetric part $\Psi_A(\mathbf{R})$ that we write as a phase:

$$\Psi(\mathbf{R}) = e^{i\Omega(\mathbf{R},\tau)}\Phi_S(\mathbf{R},\tau). \tag{2.60}$$

Introducing the previous ansatz into the time dependent Schrödinger equation and separating the real and imaginary parts, one obtains the following two coupled differential equations:

$$\frac{\partial \Phi_S}{\partial \tau} = D\left[(\vec{\nabla}\Omega)^2 - \nabla^2\Phi_S\right] - (V-E)\Phi_S \tag{2.61}$$

$$\frac{\partial \Omega}{\partial \tau} = D\left[\vec{\nabla}^2\Omega + 2\vec{\nabla}\Omega\frac{\vec{\nabla}\Phi_S}{\Phi_S}\right]. \tag{2.62}$$

The second of these expressions allows us to find an improved nodal surface from a trial one. To do so, let us assume that, to first order, we can write

$$\Omega = \Omega_0 + \frac{\partial \Omega_0}{\partial \tau}\Delta\tau, \tag{2.63}$$

for simplicity we consider $\Omega_0 = \sum_i \mathbf{k}_i \mathbf{r}_i$ which can be thought as one of the terms coming from the Slater determinant. The results that we obtain with this approximation will tell us how the orbitals in the Slater determinant should be changed at first order in imaginary time. Introducing (2.62) into (2.63), a solution for the first correction to

$\Omega_0$, comes out as:

$$\Omega = \Omega_0 + \Delta\tau D \sum_k \left( \mathbf{k}_k \frac{\vec{\nabla}\Phi_S}{\Phi_S} \right). \tag{2.64}$$

If now we consider $\Phi_S = e^{\sum_{i<j} u(r_{ij})}$ which corresponds to write it as a Jastrow factor, and rename the constants, one can write:

$$\Omega = \Omega_0 + \lambda_B \sum_{i \neq k} \mathbf{k}_k u'(r_{ij}) \frac{\mathbf{r}_{ki}}{r_{ki}} \tag{2.65}$$

$$\Omega = \sum_k \mathbf{k}_k \left( \mathbf{r}_k + \lambda_B \sum_{i \neq k} \eta(r_{ij})\mathbf{r}_{ki} \right) \tag{2.66}$$

where we have introduced the backflow potential $\eta(r_{ij}) = \mathbf{k}_k \frac{u'(r_{ij})}{r_{ki}}$ The last expression is really interesting because it tells us that, the first backflow correction can be implemented in the plane wave orbitals by a new set of coordinates $\{\mathbf{q}_i\}$ as defined in Eq. (2.59), into the plane wave orbitals. Notice that in principle the function $\eta$ is defined in terms of the derivative of the Jastrow factor. For divergent potentials, such as the Lennard-Jones or dipolar ones considered in this work, the behavior is pathological due to its divergence at the origin. To sort out this inconvenient, we treat $\eta(r)$ as a variational function, and generally is chosen to be Gaussian.

The backflow correction described above can be improved by inserting the solution of Eq. (2.64) again into Eq. (2.62) to obtain a new correction to the orbitals [117]. Another possible improvements consist in adding three-body backflow correlations [118] or iterative procedures to improve the correlated coordinates $\{\mathbf{q}_i\}$. The later has been used to determine the ground state of paramagnetic and ferromagnetic phases in $^3$He [119, 120] and in dipolar systems [94].

### 2.3.6 Computation of Observables

In this section, we explain how the computation of observables is done in the DMC framework. As we will show, only estimations of observables that commute with the Hamiltonian give place to exact results when sampled from the mixed-probability distribution introduced with the Importance Sampling technique. Here we discuss how, in some cases, this problem can be avoided by the employment of the *forward walking technique*. We also show that, by using information coming from the variational estimator, an indicator of the quality of the DMC biased expectation values can be obtained.

### 2.3.6.1 Mixed estimators

In general, the expectation value of a given observable in a quantum system is obtained from the following quantity:

$$\langle \hat{O} \rangle = \frac{\langle \Psi | \hat{O} | \Psi \rangle}{\langle \Psi | \Psi \rangle}. \tag{2.67}$$

with $\Psi$ the wave function describing the state of the system. However, when using DMC to evaluate properties of the ground state of a system, we sample from the mixed probability distribution $f = \phi_0 \Psi_T$. Thus, we only have access to the *mixed-estimator*:

$$\langle \hat{O} \rangle_f = \frac{\langle \Psi_T | \hat{O} | \phi_0 \rangle}{\langle \Psi_T | \phi_0 \rangle}. \tag{2.68}$$

The above provides exact results for observables that commute with the Hamiltonian. In this case:

$$\langle \hat{O} \rangle_{DMC} = \frac{\langle \Psi_T | \hat{O} | \phi_0 \rangle}{\langle \Psi_T | \phi_0 \rangle} = O_0 \frac{\langle \Psi_T | \phi_0 \rangle}{\langle \Psi_T | \phi_0 \rangle} = O_0 \tag{2.69}$$

with $O_0$ the eigenvalue of the operator $\hat{O}$ corresponding to $\phi_0$. In the case of operators that do not commute with $\hat{H}$, the result of the mixed estimator will be biased by $\Psi_T$ except in the particular case in which the trial wave function corresponds to the real ground state of the system. However, it is still possible to obtain a first order correction in $\Psi_T$, which is termed the *extrapolated estimator*. To show this, let us assume that the trial wave function can be formally expressed as $\Psi_T = \phi_0 + \delta \Psi_T + \mathcal{O}((\delta \Psi_T)^2)$. Then, the probability distributions employed in VMC ($P_{VMC}$) and DMC ($P_{DMC}$) can be expanded up to first order as:

$$P_{VMC} = \Psi_T^* \Psi_T = \phi_0^2 + 2\phi_0 \delta \Psi_T + \mathcal{O}((\delta \Psi_T)^2) \tag{2.70}$$

$$P_{DMC} = \Psi_T \phi_0 = \phi_0^2 + \phi_0 \delta \Psi_T + \mathcal{O}((\delta \Psi_T)^2), \tag{2.71}$$

from where is straightforward to obtain the following relation:

$$\langle \hat{O} \rangle_{ext_1} \simeq 2\langle \hat{O} \rangle_{\text{DMC}} - \langle \hat{O} \rangle_{\text{VMC}} + \mathcal{O}((\delta \Psi_T)^2). \tag{2.72}$$

An alternative correction to the mixed estimator can be obtained by considering the quantity $P_{DMC}^2 = \phi_0^4 + 2\phi_0^3 \delta \Psi_T + \mathcal{O}((\delta \Psi_T)^2)$. Which allows us to write an alternative first order extrapolation of the mixed estimator:

$$\langle \hat{O} \rangle_{ext_2} \simeq \frac{\langle \hat{O} \rangle_{\text{DMC}}^2}{\langle \hat{O} \rangle_{\text{VMC}}} + \mathcal{O}((\delta \Psi_T)^2). \tag{2.73}$$

### 2.3.6.2 Pure Estimators Techniques

The above extrapolations are reliable when the DMC correction to the VMC estimation is small. In this section we comment the *forward walking* technique [121], that allows us to obtain pure estimators for observables that do not commute with the Hamiltonian. The basic idea behind this method is to relate the mixed estimator expression, accessible in DMC, to one of the form of Eq. (2.67). Such a relation is obtained as follows,

$$\langle \hat{O} \rangle = \frac{\langle \phi_0 | \hat{O} | \phi_0 \rangle}{\langle \phi_0 | \phi_0 \rangle} = \frac{\langle \Psi_T | \frac{\phi_0}{\Psi_T} \hat{O} | \phi_0 \rangle}{\langle \Psi_T | \frac{\phi_0}{\Psi_T} | \phi_0 \rangle} \equiv \left\langle \frac{\phi_0}{\Psi_T} \hat{O} \right\rangle_{DMC}. \qquad (2.74)$$

The previous result relates the expectation values of the operator $\hat{O}$ and the quantity $\frac{\phi_0}{\Psi_T} \hat{O}$, where we can identify a weight $W(\mathbf{R}) = \frac{\phi_0}{\Psi_T}$, defined from the quotient of the trial and the ground state wave functions. The challenge of computing this quantity is performed following Liu *et. al.* analysis (*cf.* Ref [122]). They showed that $\frac{\phi_0}{\Psi_T}$ can be computed, inside the DMC algorithm, from the asymptotic number of descendants of each of the walkers:

$$W(\mathbf{R}) = \lim_{\tau \to \infty} n_{sons}(\mathbf{R}(\tau)). \qquad (2.75)$$

The implementation in the DMC method is as follows [123]: after each iteration, when a walker is replicated, we replicate not only its coordinates $\{\mathbf{R}_i\}$ but also the weight defined in the above equation and the computed observables associated with it. Usually in Monte Carlo, to accumulate statistics, one calculates observables averaging a certain number of blocks of $N_{it}$ iterations. Inside one of these blocks, for each iteration $i_\tau$, corresponding to imaginary time propagation with time step $\delta\tau$, the two following quantities can be computed:

$$\langle O \rangle_i^{i_\tau} = \langle \hat{0}(\mathbf{R}_i(i_\tau \delta\tau) \rangle_{DMC}$$
$$W_i^{i_\tau} = n_{sons}^i(\mathbf{R}_i(i_\tau \delta\tau)). \qquad (2.76)$$

Once the block is completed, an estimation of the observable is computed as

$$\langle \hat{O} \rangle^{Block} = \frac{1}{\mathcal{W}} \sum_{i=1}^{N_w} \sum_{i_\tau=0}^{N_{it}} W_i^{i_\tau} \langle O \rangle_i^{i_\tau} \qquad (2.77)$$

where the normalization is obtained directly from the sum of descendants $\mathcal{W} = \sum_{i=1}^{N_w} \sum_{i_\tau=0}^{N_{it}} W_i^{i_\tau}$ with $N_w$ the total number of walkers. And of course, estimation must be obtain through the average of multiple blocks to reduce the statistical uncertainty, so one evaluates

$$\langle \hat{O} \rangle_{pure} = \frac{1}{N_{block}} \sum_{j_{block}=1}^{N_{block}} \langle \hat{O} \rangle^{j_{block}}. \qquad (2.78)$$

It is worth to recall that, as follows from Eq. (2.75), that the above prescription gives exact results only in the limit of large number of iterations inside each of the blocks. In practice, when DMC calculations are done employing the pure estimator technique, convergence on the window defined by a given block size ($\tau_{window} = N_{it}\delta\tau$), has to be checked.

## 2.4 Path Integral Monte Carlo

In this section we present the Path Integral Monte Carlo method. As it has been commented on the introduction of this chapter, this method will be used for computing properties of bosonic systems at finite temperature.

### 2.4.1 Basis of the method

One of the main differences of Path Integral Monte Carlo, when compared with the methods commented in the previous sections, is that in PIMC the sampling is not performed over a probability distribution related to a wave function (or a product of them). Instead, the sampling is done over the thermal density matrix, that gives access to properties of the system at thermal equilibrium. The normalized *thermal density matrix* for a system described by a certain Hamiltonian $\hat{H}$ at temperature $T$ in the canonical ensemble reads

$$\hat{\rho} = \frac{e^{-\beta\hat{H}}}{Z},\tag{2.79}$$

with $\beta = 1/(k_B T)$ and $k_B$ the Boltzmann constant. The normalization in the previous expression is the partition function, defined as

$$Z = \mathsf{Tr}(e^{-\beta\hat{H}}).\tag{2.80}$$

In principle, a complete knowledge of $\hat{\rho}$ would allow to calculate expectation values for any operator $\hat{O}$ just by evaluating the trace:

$$\langle\hat{O}\rangle = \mathsf{Tr}(\hat{\rho}\hat{O}).\tag{2.81}$$

For Monte Carlo sampling, a suitable form of the above expression is obtained when the trace is taken in the coordinate representation,

$$\langle\hat{O}\rangle = \int d\mathbf{R}d\mathbf{R}'\rho(\mathbf{R},\mathbf{R}';\beta)\langle\mathbf{R}'|O(\mathbf{R})|R\rangle,\tag{2.82}$$

where we have made use of the notation $\langle\mathbf{R}_1|\hat{\rho}|\mathbf{R}_2\rangle = \rho(\mathbf{R}_1,\mathbf{R}_2;\beta)$. For the case in which the operator $\hat{O}$ is local, the above expression is reduced to

$$\langle\hat{O}\rangle = \int d\mathbf{R}\rho(\mathbf{R},\mathbf{R};\beta)O(\mathbf{R}).\tag{2.83}$$

And similarly, the partition function, reads

$$Z = \int d\mathbf{R} \rho(\mathbf{R}, \mathbf{R}; \beta).$$ (2.84)

From the definition in Eq. (2.79), an important property of the density matrices is deduced: the product of two density matrices constitutes another density matrix at lower temperature. In coordinate representation:

$$\rho(\mathbf{R}_1, \mathbf{R}_2; \beta_1 + \beta_2) = \int d\mathbf{R}_3 \rho(\mathbf{R}_1, \mathbf{R}_3; \beta_1) \rho(\mathbf{R}_3, \mathbf{R}_2; \beta_2).$$ (2.85)

The exact computation of the partition function of Eq. (2.84) in general is not possible even for classical systems due to the huge dimension of the space than have to be explored. Moreover, another complication arises in quantum systems due to the non-commutability of the terms appearing in the Hamiltonian $\hat{H} = \hat{T} + \hat{V}$. Using recursively the convolution property of Eq. (2.85), the density matrix can be expressed as the product of $M$ density matrices at temperature $MT$:

$$e^{-\beta(\hat{T}+\hat{V})} = e^{-\frac{\beta}{M}(\hat{T}+\hat{V})^M}.$$ (2.86)

The above expression allows to build the thermal density matrix of the quantum system at low temperature from the product of matrices at higher temperature. In coordinate representation, the above expression reads,

$$\rho(\mathbf{R}_1, \mathbf{R}_{M+1}; \beta) = \int d\mathbf{R}_2...d\mathbf{R}_M \prod_{\alpha=1}^{M} \rho(\mathbf{R}_\alpha, \mathbf{R}_{\alpha+1}; \beta/M).$$ (2.87)

An important remark is that the thermal density matrix can also be interpreted as propagator in imaginary time. This is clear when the inverse of the temperature is identified with the imaginary time, $i\tau = \beta$. Actually, the Path integral method maps the d-dimensional quantum problem into a (d+1)-dimensional one, with the imaginary time playing the role of the extra dimension. Commonly, this is refereed in literature as the *classical isomorphism* and, as we will show, it corresponds to mapping the quantum N-particle systems to a classical system containing $N$ polymers. In this picture, each of this polymers is constituted by $M$ coordinates, that we call *beads* corresponding to each of the imaginary time slices. It is important to notice that, for the evaluation of diagonal operators, the sets coordinates $\mathbf{R}_1$ and $\mathbf{R}_{M+1}$ have to coincide (see Eq. (2.81)), and for this reason each of the polymers would constitute a close chain (ring-polymer). The computational cost that we have to pay, for solving the quantum problem taking advantage of this classical mapping is that the number of coordinates in our system passes to be $d \times M \times N$ instead of $d \times N$. This isomorphism will appear in a clearer way in the next section.

The previous means that, a the set of coordinates $\mathbf{R}_1$ is obtained after M imaginary time propagation, each of them with time step $\tau = \beta/M$, from the set $\mathbf{R}_{M+1}$. It can be thought that the sets of coordinates $\{\mathbf{R}_j\}$ play the same role as the walkers in the DMC method, in the sense that they reflect the delocalization of quantum particles in space. Finally, it is worth to say that in the following we usually refer to the product of the imaginary time and the Hamiltonian as the action of the system $\hat{S}$, so Eq. (2.79) reads:

$$\hat{\rho} = e^{-\hat{S}} \tag{2.88}$$

### 2.4.2 The primitive approximation

Although Eq. (2.86) allows us to obtain the thermal density matrix at low temperature from its equivalent at high temperature, one still has to deal with the non commutativity that exists between the terms appearing in the Hamiltonian: $\hat{T}$ and $\hat{V}$. To overcome this difficulty, several expansions of the term $e^{-\tau(\hat{T}+\hat{V})}$ have been proposed. The simplest choice consists in performing a first order expansion, employing Trotter's formula [107, 108]

$$e^{-\beta(\hat{K}+\hat{V})} = \lim_{M \to \infty} \left( e^{-\tau\hat{K}} e^{-\tau\hat{V}} \right)^M \tag{2.89}$$

The approximation $e^{-\tau(\hat{K}+\hat{V})} \approx e^{-\tau\hat{K}} e^{-\tau\hat{V}}$, with $\tau = \beta/M$, is usually called *primitive action* and it guarantees the convergence when the time step appearing in the exponentials is small. In the next section, we explain how to include higher order terms in the action in order to improve the convergence with the number of beads. In this approximation and working in coordinate space, the two operators on the right hand side of Eq. (2.89) read [124]

$$\langle \mathbf{R}_\alpha | e^{-\tau\hat{K}} | \mathbf{R}_{\alpha+1} \rangle = \left( \frac{1}{4\pi\lambda\tau} \right)^{dN/2} e^{-\frac{\sum_{i=1}^{N}(\mathbf{r}_{\alpha+1}-\mathbf{r}_\alpha)^2}{4\lambda\tau}} \tag{2.90}$$

$$\langle \mathbf{R}_\alpha | e^{-\tau\hat{V}} | \mathbf{R}_{\alpha+1} \rangle = e^{-\tau\sum_{i<j} V(\mathbf{r}_{ij,\alpha})} \delta(\mathbf{R}_\alpha - \mathbf{R}_{\alpha+1}) \tag{2.91}$$

where we have introduced $\lambda = \frac{\hbar^2}{2m}$. Notice that we use Greek letters for bead indexes and Latin ones for particle labels. Thus, the complete density matrix of Eq. (2.87) reads:

$$\rho(\mathbf{R}_1, \mathbf{R}_{M+1}; \beta) = \left( \frac{1}{4\pi\lambda\tau} \right)^{dNM/2} \int \prod_{\alpha=2}^{M} d\mathbf{R}_\alpha e^{-\frac{\sum_{i=1}^{N}(\mathbf{r}_{\alpha+1}-\mathbf{r}_\alpha)^2}{4\lambda\tau}} e^{-\sum_{i<j} V(\mathbf{r}_{ij,\alpha})} \delta(\mathbf{R}_\alpha - \mathbf{R}_{\alpha+1}) \tag{2.92}$$

The above dN(M-1) dimensional integral allows us to understand the classical isomorphism that we have advanced in the previous section. The kinetic part, coming from the propagator of Eq. (2.90), introduces an harmonic coupling between adjacent beads that have the same particle index. On the other hand, Eq. (2.91) represents the interaction via a two body potential of beads with the same imaginary time index. This is easily interpreted as a mapping into a classical system of polymers in which each particle of the system is represented by a polymer composed of M beads. The beads interact with their neighbors via a harmonic coupling, which allows to imagine them as connected by elastic springs [125].

It is worth remembering that the primitive approximation discussed so far is accurate only up to first order in $\tau$ (although it can be trivially extended to second order by employing the alternative expansion $e^{-\tau(\hat{K}+\hat{V})} \approx e^{-\frac{\tau}{2}\hat{V}} e^{-\tau\hat{K}} e^{-\frac{\tau}{2}\hat{V}}$). This can be a good approximation in some cases, although in general an improved action would be required, for example, in the study of superfluid phases, that usually appear at very low temperatures. At such low temperatures, a huge number of beads would be needed to achieve convergence, which may exponentially slow-down the method, and, as a matter of fact, make the calculations unfeasible.

### 2.4.3 The Chin Action

To obtain higher order approximations to the action in Eq. (2.89) it is necessary to use of the Baker–Campbell–Hausdorff formula. In this formula, higher order corrections to the primitive approximation are included by evaluating commutators involving different combinations of $\hat{K}$ and $\hat{V}$. Up to fourth order, this reads:

$$e^{-\tau(\hat{K}+\hat{V})} \approx e^{-\tau\hat{K}} e^{-\tau\hat{V}} e^{-\frac{\tau^2}{2}[\hat{K},\hat{V}]} e^{-\frac{\tau^3}{12}\left([\hat{K},[\hat{K},\hat{V}]]+[\hat{V},[\hat{V},\hat{K}]]\right)}. \tag{2.93}$$

The previous expression was first employed by Takahashi-Imada [126] and Li and Broughton [127], to deduce the following approximation:

$$e^{-\tau(\hat{K}+\hat{V})} \approx e^{-\tau\hat{K}} e^{-\tau\hat{V}} e^{-\frac{\tau^3}{24}[\hat{V},[\hat{V},\hat{K}]]}, \tag{2.94}$$

that involves the evaluation of only one commutator. The inclusion of the last term improves the accuracy of the method up to fourth order for the trace. It is easily shown that the double commutator, that involves derivatives of the potential reads

$$[\hat{V},[\hat{V},\hat{K}]] = 2\lambda |\nabla V|^2. \tag{2.95}$$

In order to have a physical intuition about this term, it is worth to define a force acting on a single particle $\mathbf{F}_i = \sum_{j\neq i} \nabla_i V(r_{ij})$, such that

$$[\hat{V},[\hat{V},\hat{K}]] = 2\lambda \sum_{i=1}^{N} |F_i|^2. \tag{2.96}$$

Similar to the computation of the potential term, the evaluation of the force involves only beads with the same imaginary time index. The computational cost at each imaginary time step due to the inclusion of this additional term, is compensated by the significant reduction in number of beads needed to obtain convergence. Further improvement over this scheme can be obtained by performing a symplectic expansion, that is, introducing some coefficients $t_i$, $v_i$ and $w_i$ such as:

$$e^{-\tau \hat{H}} = \prod_{j=1}^{M} e^{-\tau t_j \hat{K}} e^{-\tau v_j \hat{V}} e^{-\tau \omega_j [\hat{V},[\hat{V},\hat{K}]]} + \mathcal{O}(\tau^5) \qquad (2.97)$$

The idea is to fix the parameters $\{t_j, v_j, \omega_j\}$ to obtain a certain target accuracy. It is important to notice that, when using it in MC, the possible values of these parameters are restricted so that the probability distribution defined by Eq (2.97) is non-divergent and is positive definite, which imposes restrictions on the coefficients. Moreover, the Sheng-Suzuki theorem [128, 129] states that it is not possible to go beyond fourth order in $\tau$, if the coefficients $\{t_j, v_j, \omega_j\}$ are restricted to be positive .

Including the terms of Eq. (2.97) that have the double commutator (making $\{\omega_j\} \neq 0$), similarly to what is done in the Takahashi-Imada action [130–132], but it allows various equivalent decompositions. Indeed, Chin and Chen introduced a complete family of actions that include such kind of terms. Later on, this was also applied to PIMC calculations [133, 134], where it has demonstrated its improved efficiency. Let us define a new effective potential that includes some of the terms coming from the double commutator,

$$\hat{W}_{a_1} = \hat{V} + \frac{u_0}{v_1} a_1 \tau^2 [[\hat{V}, \hat{K}], \hat{V}], \qquad (2.98)$$

such that the Chin action (CA) becomes

$$e^{-\tau \hat{H}} = \prod_{j=1}^{M} e^{-\tau v_1 \hat{W}_{a_1}} e^{-\tau t_1 \hat{K}} e^{-\tau v_2 \hat{W}_{1-2a_1}} e^{-\tau t_1 \hat{K}} e^{-\tau v_1 \hat{W}_{a_1}} e^{-2\tau t_0 \tau \hat{K}}, \qquad (2.99)$$

from where it is clear that, in this implementation, each time step is split into three smaller imaginary time intervals. The thermal density matrix in Chin approximation is obtained from equations (2.96), (2.98) and (2.99):

$$\rho_{CA}(\mathbf{R}_1, \mathbf{R}_{M+1}; \beta) = \left(\frac{1}{4\pi\lambda\tau}\right)^{dNM/2} \int \prod_{\alpha=2}^{M} d\mathbf{R}_\alpha e^{-S_{CA}(\mathbf{R}_\alpha \mathbf{R}_{\alpha+1}, \tau)}, \qquad (2.100)$$

with the Chin action given by the following expression, in which we introduce the labels A and B, to denote the two additional intermediate coordinates introduced at each time step to perform the integration with this action [135]:

$$S_{CA}(\mathbf{R}_\alpha \mathbf{R}_{\alpha+1}, \tau) = \frac{1}{4\lambda\tau} \sum_{i=1}^{N} \left( \frac{1}{t_1}(\mathbf{r}_{\alpha,i} - \mathbf{r}_{\alpha A,i})^2 + \frac{1}{t_1}(\mathbf{r}_{\alpha A,i} - \mathbf{r}_{\alpha B,i})^2 + \frac{1}{2t_0}(\mathbf{r}_{\alpha B,i} - \mathbf{r}_{\alpha+1,i})^2 \right)$$
$$+ \tau \sum_{i<j}^{N} \left( \frac{v_1}{2} V(r_{\alpha,ij}) + v_2 V(r_{\alpha A,ij}) + v_1 V(r_{\alpha B,ij}) + \frac{v_1}{2} V(r_{\alpha+1,ij}) \right)$$
$$+ 2\tau^3 u_0 \lambda \sum_{i=1}^{N} \left( \frac{a_1}{2} |\mathbf{F}_{\alpha,i}|^2 + (1-2a_1)|\mathbf{F}_{\alpha A,i}|^2 + a_1 |\mathbf{F}_{\alpha B,i}|^2 + \frac{a_1}{2} |\mathbf{F}_{\alpha+1,i}|^2 \right),$$

$$(2.101)$$

It is worth noticing that, to achieve the desired fourth order expansion some conditions must be satisfied by to the parameters $t_0, t_1, v_1, v_2, u_0$ and $a_1$ that appear in the previous expression. Here we follow the choice of Ref. [130]:

$$t_1 = \frac{1}{1} - t_0$$
$$v_1 = \frac{1}{6(1-2t_0)^2}$$
$$v_2 = 1 - 2v_1$$
$$u_0 = \frac{1}{12 \left( 1 - \frac{1}{1-2t_0} + \frac{1}{6(1-2t_0)^2} \right)}$$

$$(2.102)$$

which leaves us with only two independent parameters, $a_1$ and $t_0$, satisfying the following restrictions in order to be able to build a positive definite probability density from it:

$$0 \le a_1 \le 1$$
$$0 \le t_0 \le \frac{1}{2} \left( 1 - \frac{1}{\sqrt{3}} \right).$$

It has been shown that, with a correct choice of these two parameters, this action can be made to work effectively up to sixth order for the energy [133]. In analogy to what is done in Ref. [135], we introduce some definitions that will allow us to write the expression in Eq. (2.101) in a more compact way:

$$T^t_{MN} = \sum_{\alpha=1}^{M} \sum_{i=1}^{N} \left( \frac{1}{t_1}(\mathbf{r}_{\alpha,i} - \mathbf{r}_{\alpha A,i})^2 + \frac{1}{t_1}(\mathbf{r}_{\alpha A,i} - \mathbf{r}_{\alpha B,i})^2 + \frac{1}{2t_0}(\mathbf{r}_{\alpha B,i} - \mathbf{r}_{\alpha+1,i})^2 \right)$$

$$(2.103)$$

$$V_{MN} = \sum_{\alpha=1}^{M} \sum_{i<j}^{N} \left( \frac{v_1}{2} V(r_{\alpha,ij}) + v_2 V(r_{\alpha A,ij}) + v_1 V(r_{\alpha B,ij}) + \frac{v_1}{2} V(r_{\alpha+1,ij}) \right) \quad (2.104)$$

$$W_{MN} = \sum_{\alpha=1}^{M} \sum_{i=1}^{N} \left( \frac{a_1}{2} |\mathbf{F}_{\alpha,i}|^2 + (1-2a_1)|\mathbf{F}_{\alpha A,i}|^2 + a_1 |\mathbf{F}_{\alpha B,i}|^2 + \frac{a_1}{2} |\mathbf{F}_{\alpha+1,i}|^2 \right), \quad (2.105)$$

such that the action in Eq. (2.101) reads,

$$S_0 = \frac{1}{4\lambda\tau} T_{MN}^t + \tau V_{MN} + 2\tau^3 u_0 \lambda W_{MN} \qquad (2.106)$$

This notation will be very useful to derive relevant estimators, such as those for the energy, as detailed in sec. 2.6.

### 2.4.4 The staging algorithm

When a stochastic algorithm devised to perform PIMC calculations is implemented, bead movements (updates) have to be proposed. This updates are accepted or rejected using the Metropolis algorithm described in section 2.1.3. In the case of the PIMC method, the probability distribution used for the Metropolis algorithm is defined by the action. When performing calculations at very low temperatures, the chains constituting the classical polymers are long, and moving one bead at a time is not efficient. On the other hand, if we propose to perform random collective movements of a large fraction of the total number of beads, the acceptance ratio may become very low. Such a low acceptance ratio can slow down the simulations in a critical way.

A smart solution to this problem comes from the *staging algorithm*. In this technique, movements involving beads in a segment of a polymer are sampled directly from the action of the free problem, that is, from the kinetic part of the complete action. In this way, the acceptance or rejection of the movements depends only on the potential part of the action.

With this idea in mind, we consider a segment of a polymer composed of $l$ beads: the thermal density matrix in this segment reads:

$$\rho(\mathbf{R}_j, \mathbf{R}_{j+l}; \beta) = \int d\mathbf{R}_{j+1}...d\mathbf{R}_{j+l-1} \prod_{k=j}^{j+l-1} \rho(\mathbf{R}_k, \mathbf{R}_{k+1}; \tau) \qquad (2.107)$$

In the following, we focus only on one of these polymers, and so, for the sake of simplicity, we write $r_j$ instead of $R_j$, having in mind that $j$ is the bead index inside the chosen ring-polymer. Our aim is to be able to write the product of density matrices as a product of terms for each independent bead:

$$\rho(\mathbf{r}_j, \mathbf{r}_{j+1}; \tau) \rho(\mathbf{r}_{j+1}, \mathbf{r}_{j+2}; \tau) ... \rho(\mathbf{r}_{j+(l-1)}, \mathbf{r}_{j+l}; \tau) =$$

$$\rho(\mathbf{r}_j, \mathbf{r}_{j+l}; l\tau) \times \left[ \frac{\rho(\mathbf{r}_j, \mathbf{r}_{j+1}; \tau) \rho(\mathbf{r}_{j+1}, \mathbf{r}_{j+l}; (l-1)\tau)}{\rho(\mathbf{r}_j, \mathbf{r}_{j+l}; l\tau)} \right] \times$$

$$\left[ \frac{\rho(\mathbf{r}_{j+1}, \mathbf{r}_{j+2}; \tau) \rho(\mathbf{r}_{j+2}, \mathbf{r}_{j+l}; (l-2)\tau)}{\rho(\mathbf{r}_{j+1}, \mathbf{r}_{j+l}; (l-1)\tau)} \right] \times ...$$

$$\left[ \frac{\rho(\mathbf{r}_{j+(l-2)}, \mathbf{r}_{j+(l-1)}; \tau) \rho(\mathbf{r}_{j+(l-1)}, \mathbf{r}_{j+l}; \tau)}{\rho(\mathbf{r}_{j+l-2}, \mathbf{r}_{j+l}; 2\tau)} \right] \tag{2.108}$$

In order to be able of sampling each of these terms in brackets independently, we would like to write them in the form:

$$\exp\left[ -\frac{m_k}{2\hbar^2\tau}(\mathbf{r}_{j+k+1} - \mathbf{r}_{j+k+1}^*)^2 \right], \tag{2.109}$$

which can be achieved by defining the staging coordinate and reduced mass for each one of the beads:

$$\mathbf{r}_{j+k+1}^* = \frac{\mathbf{r}_{j+l} + \mathbf{r}_{j+k}(l-(k+1))}{l-k} \tag{2.110}$$

$$m_k = m\left( \frac{l-k}{l-(k+1)} \right). \tag{2.111}$$

With the above definitions, Eq. (2.108) can be expressed as a product of Gaussians of the form of (2.109), meaning that the beads in between the ones with index $j$ and $j+l$ can be sampled directly from that Gaussian distributions. We freeze the two extremities of the chain segment, and update the coordinates of the intermediate beads in the following manner:

$$\mathbf{r}_{j+k+1}' = \mathbf{r}_{j+k+1} + \eta\sqrt{\frac{\hbar^2\tau}{m_k}} \tag{2.112}$$

with $\eta$ random number sampled from a normal Gaussian distribution.

A complete derivation for the staging algorithm for the Chin action, can be found in Appendix A of Phd. Thesis of G. Ferré [135].

### 2.4.5 The Worm algorithm

Simulating a quantum many-body systems implies being able to implement correctly the permutations between identical particles. The method that we have proposed up to now is incomplete in this sense, as it treats all the particles as if they were distinguishable. To implement the correct quantum statistics, the thermal density matrix presented above has to be rewritten. Its expression, for a system of N bosons

or fermions reads

$$\rho_1^{B/F}(\mathbf{r}_1, \mathbf{r}_2; \beta) = \frac{1}{N!} \sum_{\mathcal{P}} (\pm 1)^P \rho_1(\mathbf{r}_1, \mathcal{P}\mathbf{r}_2; \beta), \qquad (2.113)$$

with $\mathcal{P}$ a permutation of the particles labels, and $P$ the number of transpositions on each permutation that generates $\mathcal{P}$. The sum runs over all possible permutations of the $N$ particle labels. In the above expression the $+$ stands for bosons, and the $-$ for fermions. As we have already commented in sec. 2.3.5, the sum of positive and negative terms in the fermionic case, leads to the sign problem, which actually can make the signal-to-noise ratio to be unacceptable when solving the problem numerically. In DMC, this was cured by employing the *Fixed-Node* technique. Although it has not been as successful, some attempts have been done in the same direction in the PIMC framework by introducing the *Restricted-Path* [136] prescription; that constraints the sampling to those paths that preserve the sign. In this Thesis, we do not use the path integral method to study fermionic systems, so in the following we focus on the particular case of a system of bosons.

In the first implementation of PIMC, permutations between many particles were sampled as one update, which made the method really inefficient when sampling permutations between more than two particles [100]. A great improvement in this direction was the development, by Prokof'ev *et al*, with the development of the *Worm Algorith* [103, 137, 138, 124]. The main idea behind the worm algorithm is to extend the space of configurations. To the Z ensemble (Z-sector), represented by the usual ring-polymer configurations, the G-sector is added. The later includes configurations in which the polymer with particle index $i$ is open so that $r_{M+1}^i \neq r_1^i$. In order to be able to sample the two sectors and jump between one and another, two new polymer coordinate updates are introduced in the PIMC algorithm:

- *Open*: In this update we propose to open one of the ring-polymers (that was originally closed) at a certain bead position. The beads included in the segment of distance $l$, starting from one of the two new extremities are updated. Considering that we open the chain between the bead 1 and $M$, we update bead with index $j \in [M - j, M + 1]$,

$$\mathbf{r}_i = \{\mathbf{r}_1, \mathbf{r}_2, ..., \mathbf{r}_M, \mathbf{r}_{M+1} = \mathbf{r}_1\} \longrightarrow \mathbf{r}_i' = \{\mathbf{r}_1, \mathbf{r}_2, ..., \mathbf{r}_{M-l+1}, ... \mathbf{r}_M', \mathbf{r}_{M+1}' \neq \mathbf{r}_1'\}$$
$$(2.114)$$

what allows us to jump from the Z-sector to the G-sector.

- *Close*: In this case we propose a movement that closes a polymer which was already open, so that our sampling jumps from the G-sector to Z-sector. This movement consists on updating the beads included in the segment of length $l$ starting from one of the extremities of the open chain as

$$\mathbf{r}_i = \{\mathbf{r}_1, \mathbf{r}_2, ..., \mathbf{r}_M, \mathbf{r}_{M+1} \neq \mathbf{r}_1\} \longrightarrow \mathbf{r}'_i = \{\mathbf{r}_1, \mathbf{r}_2, ..., \mathbf{r}'_{M-l+1}, ...\mathbf{r}'_M, \mathbf{r}'_{M+1} = \mathbf{r}'_1\}$$
(2.115)

Although the previous movements allow to work in the Z and G sectors simultaneously, they are not enough to sample permutations. In order to do so, we have to include a third new update, called *swap*. When working in the G-sector, we propose an update in which the extremity of the open polymer is matched to a bead with a different particle index. To do so, the path is reconstructed according to the free particle thermal density matrix, similar to what is done when a staging movement is proposed (see sec. 2.4.4). If $i$ is the index of the open polymer and $j$ the one corresponding to the swap partner, the swap update reads:

$$\mathbf{r}^j = \{\mathbf{r}_1^j, \mathbf{r}_2^j..., \mathbf{r}_M^j, \mathbf{r}_{M+1}^j = \mathbf{r}_1\} \longrightarrow \mathbf{r}'^j = \{\mathbf{r}_1'^j = \mathbf{r}_{M+1}^i, \mathbf{r}_2'^j, ..., \mathbf{r}_{l-1}'^j, \mathbf{r}_l^j..., \mathbf{r}_M^j, \mathbf{r}_{M+1}^j = \mathbf{r}_1^j\}$$
(2.116)

The above movements allows us not only to change the permutation table, $p(i) \neq i$, but to sample permutations involving many particles just by proposing iteratively two-body permutations.

### 2.4.6 PIMC Algorithm

In this section we present a schematic representation on how the PIMC algorithm is implemented. Taking advantage of the classical isomorphism, at each iteration, our system is represented by a set of coordinates $\{\mathbf{R}_1, \mathbf{R}_2, ..., \mathbf{R}_M\}$ with $\mathbf{R}_j = \{\mathbf{r}_j^1, \mathbf{r}_j^2, ..., \mathbf{r}_j^N\}$ the set of coordinates of the $N$ particle system at the jth imaginary time step. This coordinates are updated by proposing the following movements, that are accepted or rejected according the Metropolis algorithm:

1. Center of mass movement: In this update all the beads corresponding to a certain particle index, or involved in a permutation, are displaced a distance $\Delta\mathbf{r}$, so that:

$$\mathbf{r}_i = \{\mathbf{r}_\alpha\} \longrightarrow \mathbf{r}'_i = \{\mathbf{r}'_\alpha = \mathbf{r}_\alpha + \Delta\mathbf{r}\}, (\alpha = 1, M)$$
(2.117)

   This movement is computationally expensive, as one has to recalculate the action for all the beads in the simulation. For this reason, center of mass updates are proposed only once in a while, after a certain number of iterations.

2. We propose an open or close movement, according to equations (2.114) and (2.115) depending if we are in the Z or the G-sector.

3. A staging movement is proposed, according to the prescription presented in section 2.4.4.

4. If we are sampling the G-sector, we proposed a number $N_{swap}$ of swap movements to generate permutations.

5. We compute the observables of interest. The computation of observables in PIMC, will be discussed in detail in section 2.4.7.

6. The above procedure is looped over from point the beginning until the desired statistical precision is obtained.

### 2.4.7 Computation of observables

The expectation value of any observable $\hat{O}$ in PIMC reads

$$\langle \hat{O} \rangle = \int \prod_{\alpha=1}^{M} d\mathbf{R}_\alpha O(\mathbf{R}_\alpha) \rho(\mathbf{R}_\alpha, \mathbf{R}_{\alpha+1}; \tau), \qquad (2.118)$$

which, in a Monte Carlo implementation, using $\rho$ as a probability distribution to sample from, implies that we can obtain an estimation of it through the expression

$$\langle \hat{O} \rangle \approx \langle \hat{O} \rangle_\rho = \frac{1}{ZM} \sum_{\alpha=1}^{M} O(\mathbf{R}_\alpha). \qquad (2.119)$$

In the above equation, the sum over beads only stands when one is computing observables in the Z-sector. In this case all the polymers are closed, and one can take advantage of the symmetry that exist between them to improve the efficiency in the evaluation of observables. However, when we computing off-diagonal observables in the G-sector, such as the one-body density matrix, only the beads that have the same index as the extremities of the worm can be used. This is explained in more detail in sec. 2.6, where different observables are discussed.

## 2.5 Path Integral Ground State

The Path integral formalism can be extended to zero temperature calculations, in the method know as *Path Integral Ground State* (PIGS) [100, 139–141]. It takes advantage of two powerful tools in many-body physics that we have already introduced: the variational principle and the imaginary time propagation. (see sections 2.2.1 and 2.1.4). On one hand, the variational principle states that the expectation value of $\hat{H}$, evaluated over a trial wave function $\Psi_T$, constitutes an upper bound to the real ground state of the system. On the other hand, we know that one can obtain the exact ground state of a system by performing imaginary time propagation over a variational ansatz, as long as it is not strictly orthogonal to the exact ground state wave function:

$$\phi_0(\mathbf{R}) = \lim_{\tau \to \infty} \Psi(\mathbf{R}, \tau) = \lim_{\tau \to \infty} \int d\mathbf{R} \, G(\mathbf{R}, \mathbf{R}'; \tau) \Psi_T(\mathbf{R}'). \tag{2.120}$$

Similarly to what we have already commented (for DMC and PIMC methods), solving the above equation is not always possible because of the lack of knowledge of the propagator in the first place. However, taking advantage of the convolution property, we can rewrite it as:

$$\phi_0(\mathbf{R}_M) = \lim_{M \to \infty} \int \prod_{\alpha=1}^{M-1} d\mathbf{R}_\alpha \, G(\mathbf{R}_{\alpha+1}, \mathbf{R}_\alpha; \delta\tau) \Psi_T(\mathbf{R}_1), \tag{2.121}$$

where $\delta\tau = \tau/M$. For solving the problem numerically, the number of time slices has to be fixed to a possibly large but certainly finite certainly finite number $M$, then the PIGS estimation from the ground state wave function reads

$$\Phi_{PIGS}(\mathbf{R}_M) = \int \prod_{\alpha=1}^{M-1} d\mathbf{R}_\alpha \, G(\mathbf{R}_\alpha, \mathbf{R}_{\alpha+1}; \delta\tau) \Psi_T(\mathbf{R}_1), \tag{2.122}$$

meaning that, strictly speaking, for a fixed number of integration steps (beads) one obtains a variational approximation to the real ground state of the system. However, the most remarkable property of the PIGS method is that it provides a systematic procedure to keep the bias of the ground state estimations under control: Just by increasing the number of beads, this difference can be made arbitrarily small. Indeed, for our purposes, it is enough to maintain it under the desired statistical uncertainty arising from the employment of MC integration methods. That is why, despite on its first formulation it was called *Variational PIMC* [100], it is nowadays considered to be an *exact method*.

### 2.5.1 Evaluation of observables

With the previous scheme in mind, one can compute local properties of the system as:

$$\begin{aligned}
\langle \phi_0 | \hat{O} | \phi_0 \rangle &= \frac{\int d\mathbf{R}_M O(\mathbf{R}_M) \Psi^*_{PIGS}(\mathbf{R}_M) \Psi_{PIGS}(\mathbf{R}_M)}{\int d\mathbf{R}_M \Psi^*_{PIGS}(\mathbf{R}_M) \Psi_{PIGS}(\mathbf{R}_M)} \\
&= \frac{\int d\mathbf{R}_0 ... d\mathbf{R}_{2M} \prod_{\alpha=0}^{2M-1} O(\mathbf{R}_M) \Psi^*_T(\mathbf{R}_{2M}) G(\mathbf{R}_{\alpha+1}, \mathbf{R}_\alpha; \delta\tau) \Psi_T(\mathbf{R}_0)}{\int d\mathbf{R}_0 ... d\mathbf{R}_{2M} \prod_{\alpha=0}^{2M-1} \Psi^*_T(\mathbf{R}_{2M}) G(\mathbf{R}_{\alpha+1}, \mathbf{R}_\alpha; \delta\tau) \Psi_T(\mathbf{R}_0)}
\end{aligned} \tag{2.123}$$

For Monte Carlo purposes, the probability distribution to be considered here is

$$P(\mathbf{R}_0, ..., \mathbf{R}_{2M}) = \frac{\prod_{\alpha=0}^{2M-1} \Psi^*_T(\mathbf{R}_{2M}) G(\mathbf{R}_{\alpha+1}, \mathbf{R}_\alpha; \delta\tau) \Psi_T(\mathbf{R}_0)}{\int d\mathbf{R}_0 ... d\mathbf{R}_{2M} \prod_{\alpha=0}^{2M-1} \Psi^*_T(\mathbf{R}_{2M}) G(\mathbf{R}_{\alpha+1}, \mathbf{R}_\alpha; \delta\tau) \Psi_T(\mathbf{R}_0)}. \tag{2.124}$$

An analog to the classical isomorphism that we have discussed for the PIMC method can also be introduced in this framework. The main difference in the present case is that there is no need to impose periodicity in imaginary time since $\mathbf{r}_{2M} \neq \mathbf{r}_0$, so that the classical polymers are now open. Another important remark here is that $\tau$ is just a parameter in PIGS, and has nothing to do with a physical temperature. The optimal value of $\tau$ is reach to obtain convergence when measuring observables.

In the two extremities of the chain, we impose a trial wave function $\Psi_T$ that is propagated in imaginary time to the center of the chain, where $\phi_0$ is sampled. As a consequence, the evaluation of observables is only possible on the center of the chain. This constitutes a disadvantage in efficiency of PIGS method in comparison to PIMC, where the symmetry that exists between all the beads in a close polymer improves the efficiency in the evaluation of properties.

As a final remark, we comment the special case in which the operator $\hat{O}$ to be measured commutes with the Hamiltonian. In this special case,

$$\langle \phi_0 | \hat{O} | \phi_0 \rangle = \lim_{\tau \to \infty} \langle \Psi_T | e^{-\tau \hat{H}} \hat{O} e^{-\tau \hat{H}} | \Psi_T \rangle = \lim_{\tau \to \infty} \langle \Psi_T | e^{-\tau \hat{H}} e^{-\tau \hat{H}} \hat{O} | \Psi_T \rangle \quad (2.125)$$

that tells us that we can simply compute $\hat{O} \Psi_T$ instead of $\hat{O} \Psi_{PIGS}$ for this particular case. We have written the last equality to emphasize the similarity with the evaluation of local estimators in VMC. However, it is important to remark that the imaginary time propagation present in PIGS makes this estimator statistically exact, once convergence is achieved.

Finally we can comment a few words about the statistics between indistinguishable particles. As long as the propagator that appears in Eq. (2.123) is symmetric under the exchange of identical particles, it is enough to impose the correct symmetry (or antisymmetry) in the wave function at the extremities of the chain in order to obtain the correct ground state, at least from the theoretical point of view. In this sense, the PIGS method has an advantage over PIMC: there is no need to explicitly sample permutations, and thus, there is no need to use the worm algorithm described in section 2.4.5 to find the ground state of the system. However, the *worm* is still useful when evaluating off-diagonal properties such as the One-body density matrix.

## 2.6 Quantum Monte Carlo Estimators

In this section we introduce the way in which observables for the many-body system represented by the Hamiltonian in Eq. (2.2) are computed in the different Monte Carlo methods. The energy of the system is the driving quantity that we evaluate and it represents also one of the quantities that can be evaluated in a unbiased way, not only with PIMC and PIGS, but also in DMC. For this reason we use it as a first example for all the methods. For other observables we stick to the general definition, writing explicit expressions only for the cases in which they clarify the text.

### 2.6.1 Energy per particle

Following the Schrödinger equation (2.1), and working in the bracket notation, the expectation value of the energy in the state $|\Psi\rangle$ is

$$E = \frac{\langle \Psi^* | \hat{H} | \Psi \rangle}{\langle \Psi^* | \Psi \rangle}. \tag{2.126}$$

**Variational Monte Carlo**

As we have already commented in sec. 2.2.2, an upper bound to the ground state energy can be obtained by sampling the local energy of Eq. (2.27), over the probability distribution given by the squared trial wave function

$$\langle E \rangle_{VMC} = \langle E_L \rangle_{\Psi_T^2} \tag{2.127}$$

with $\langle \rangle_{\Psi_T^2}$ the average evaluated over the probability distribution defined by $|\Psi_T(\mathbf{R})|^2$.

**Diffusion Monte Carlo**

In section 2.3.6, we commented briefly how do we compute observables in DMC. It is straightforward to see that the same expression of the local energy used in VMC works also in DMC. In this way and from equation (2.68), the expectation value over the asymptotic mixed probability distribution $f(\mathbf{R}, \tau \to \infty)$,

$$\langle E \rangle_{DMC} = \langle E_L \rangle_{f(\mathbf{R}, \tau \to \infty)}, \tag{2.128}$$

constitutes a pure estimator for the exact ground state energy. This estimation yields the exact energy for bosons (up to some statistical noise). On the contrary, when the Fixed-Node technique, described in sec. 2.3.5, is employed to study fermionic systems, it becomes variational. In this later case, the expression for the local energy implies evaluating derivatives of the orbitals included in the Slater determinant. An exhaustive discussion of how does this should be done, both using a basis of plane waves and with Backflow correlations, can be found in the Phd. Thesis of Víctor Grau [117].

**Path Integral Monte Carlo**

**The thermodynamic estimator:** In the path integral framework, the energy per particle is evaluated, making use of Eq. (2.100), as follows:

$$E = -\frac{1}{Z}\frac{\partial Z}{\partial \bar{\beta}} = -\frac{1}{MZ}\frac{\partial Z}{\partial \tau} = -\frac{1}{MZ}\left(-\frac{3dNM}{2\tau}Z - Z\frac{\partial S_{CA}}{\partial \tau}\right). \tag{2.129}$$

Making use of the notation introduced for the Chin action in Eq. 2.106, we obtain :

$$E = \left\langle \frac{3dN}{2\tau} - \frac{1}{M} \left( \frac{1}{4\tau^2 \lambda} T^t_{MN} - V_{MN} - 6\tau^2 u_0 \lambda W_{MN} \right) \right\rangle_Z . \qquad (2.130)$$

The brackets $\langle \rangle_Z$ on the previous expression, indicate that we average over the configurations $\{R\}$ in the Z-sector. Apart from the total energy, it is also possible to compute the kinetic contribution alone:

$$\begin{aligned}
K &= \frac{m}{\beta Z} \frac{\partial Z}{\partial m} = -\frac{\lambda}{\beta Z} \frac{\partial Z}{\partial \lambda} = -\frac{\lambda}{M\tau Z} \frac{\partial Z}{\partial \lambda} \\
&= -\frac{3dNM}{2\lambda} Z - Z \frac{\partial S_0}{\partial \lambda} \\
&= -\frac{3dNM}{2\lambda} Z - Z \left( -\frac{1}{4\lambda^2 \tau} T^t_{MN} + 2\tau^3 u_0 W_{MN} \right) .
\end{aligned} \qquad (2.131)$$

So kinetic energy reads:

$$K = \left\langle \frac{3dN}{2\tau} - \frac{1}{M} \left( \frac{1}{4\lambda\tau^2} T^t_{MN} - 2\tau^2 u_0 \lambda W_{MN} \right) \right\rangle \qquad (2.132)$$

The first term in the above equation resembles the energy of an ideal gas $E_{IG} \sim dN/2\beta$. The extra factor of 3 is due to the number of beads used with the Chin action at each time step. The above expressions are usually referred as the *thermodynamic estimators* for the total and kinetic energy. By subtracting equations (2.130) and (2.132), the potential energy is obtained. The terms in (2.132) can be large when $\tau$ is small, giving place to a large variance, due to the cancellation between the two terms.

**The Virial estimator:** As it was shown by Herman, Bruskin and Berne [142, 100], an improved estimator for the energy can be obtained by integrating by parts over the imaginary time variables

$$\begin{aligned}
E_V = \Bigg\langle &\frac{dN}{2\beta} + \frac{1}{12\lambda M^2 \tau^2} \sum_{\alpha=1}^{M} \sum_{i=1}^{N} \left( \mathbf{r}_{M+\alpha,i} - \mathbf{r}_{\alpha,i} \right) \left( \mathbf{r}_{M+\alpha-1,i} - \mathbf{r}_{M+\alpha,i} \right) \\
&+ \frac{1}{2\beta} \sum_{\alpha=1}^{M} \sum_{i=1}^{N} \left( \mathbf{r}_{\alpha,i} - \mathbf{r}^C_{\alpha,i} \right) \frac{\partial}{\partial \mathbf{r}_{\alpha,i}} \left( U(\mathbf{R}_\alpha) \right) \\
&+ \frac{1}{M} \sum_{\alpha=1}^{M} \frac{\partial U(\mathbf{R}_\alpha)}{\partial \tau} \Bigg\rangle .
\end{aligned} \qquad (2.133)$$

The fourth term in Eq. (2.133), is the same that appears for the potential part in the thermodynamic estimator. While the centroid coordinates $\mathbf{r}^C_{\alpha,i}$ introduced in the third

term are defined as:

$$\mathbf{r}_{\alpha,i}^C = \frac{1}{2M} \sum_{l=0}^{M-1} \left( \mathbf{r}_{\alpha+l,i} + \mathbf{r}_{\alpha-l,i} \right). \tag{2.134}$$

In the primitive approximation, $U(\mathbf{R}_\alpha)$ stands for the potential part. On the contrary, when using more elaborated actions, this term has to be generalized. In the case of the Chin action this term reads:

$$U(\mathbf{R}_\alpha) = \tau \sum_{i<j}^N \left( \frac{v_1}{2} V(\mathbf{r}_{\alpha,ij}) + v_2 V(\mathbf{r}_{\alpha A,ij}) + v_1 V(\mathbf{r}_{\alpha B,ij}) + \frac{v_1}{2} V(\mathbf{r}_{\alpha+1,ij}) \right) \tag{2.135}$$

$$+ 2\tau^3 u_0 \lambda \sum_{i=1}^N \left( \frac{a_1}{2} |\mathbf{F}_{\alpha,i}|^2 + (1-2a_1)|\mathbf{F}_{\alpha A,i}|^2 + a_1 |\mathbf{F}_{\alpha B,i}|^2 + \frac{a_1}{2} |\mathbf{F}_{\alpha+1,i}|^2 \right)$$

The Virial estimator for the Chin action was derived in appendix B of the Phd. Thesis by Ferré [135], where isotropic potentials are considered. The derivation for an anisotropic potential is straightforward following the indications presented there, and for most of the terms appearing in the expressions it is enough to write the derivatives in Cartesian coordinates. Here we summarize the expressions needed to compute the Virial estimator for an anisotropic potential:

$$E_V = \frac{dN}{2\beta} + \frac{1}{M} \left( \frac{1}{12\lambda M \tau^2} T_{MN}^{\text{off}} + \frac{1}{2} T_{MN}^V + 2\tau^2 u_0 \lambda Y_{MN} + V_{MN} + 6\tau^2 u_0 \lambda W_{MN} \right) \tag{2.136}$$

with $V_{MN}$ and $W_{MN}$ computed from equations (2.104) and (2.105) and the following definitions:

$$T_{MN}^{\text{off}} = \sum_{\alpha=1}^{M} \sum_{i=1}^{N} \left( \frac{1}{t_1} \left( \vec{r}_{M+\alpha,i} - \vec{r}_{\alpha,i} \right) \left( \vec{r}_{\alpha,i} - \vec{r}_{\alpha A,i} \right) \right.$$
$$+ \frac{1}{t_1} \left( \vec{r}_{M+\alpha A,i} - \vec{r}_{\alpha A,i} \right) \left( \vec{r}_{\alpha A,i} - \vec{r}_{\alpha B,i} \right)$$
$$\left. + \frac{1}{2t_0} \left( \vec{r}_{M+\alpha B,i} - \vec{r}_{\alpha B,i} \right) \left( \vec{r}_{\alpha B,i} - \vec{r}_{\alpha+1,i} \right) \right) \tag{2.137}$$

$$T_{MN}^{V} = \sum_{\alpha=1}^{M} \sum_{i=1}^{N} \left( \frac{v_1}{2} (\vec{r}_{\alpha,i} - \vec{r}_i^C) \vec{F}_{\alpha,i} + v_2 (\vec{r}_{\alpha A,i} - \vec{r}_i^C) \vec{F}_{\alpha A,i} \right.$$
$$\left. + v_1 (\vec{r}_{\alpha B,i} - \vec{r}_i^C) \vec{F}_{\alpha B,i} + \frac{v_1}{2} (\vec{r}_{\alpha+1,i} - \vec{r}_i^C) \vec{F}_{\alpha+1,i} \right) \tag{2.138}$$

$$Y_{MN} = \sum_{\alpha=1}^{M} \sum_{i=1}^{N} \sum_{\substack{j=1 \\ j \neq i}}^{N} \sum_{a=1}^{d} \sum_{b=1}^{d} \left( \frac{a_1}{2} (r_{\alpha,i} - r_{\alpha,i}^C)^a T(\alpha,i,j)_a^b (F_{\alpha,i} - F_{\alpha,j})_b \right.$$
$$+ (1 - 2a_1)(r_{\alpha A,i} - r_{\alpha,i}^C)^a T(\alpha A,i,j)_a^b (F_{\alpha A,i} - F_{\alpha A,j})_b$$
$$+ a_1 (r_{\alpha B,i} - r_{\alpha,i}^C)^a T(\alpha B,i,j)_a^b (F_{\alpha B,i} - F_{\alpha B,j})_b$$
$$\left. + \frac{a_1}{2} (r_{\alpha+1,i} - r_{\alpha,i}^C)^a T(\alpha+1,i,j)_a^b (F_{\alpha+1,i} - F_{\alpha+1,j})_b \right) \tag{2.139}$$

In fact, all the expressions are the same that those that were obtained in Appendix B of [135] except for the tensor $T(\alpha,i,j)_a^b$ that appears on the expression of $Y_{MN}$. In order to include the anisotropy of the potential, it cannot be reduced to radial derivatives. In Cartesian coordinates this term reads

$$T(\alpha,i,j)_a^b = \frac{\partial V(r_{\alpha,ij})}{\partial (r_i)_b \partial (r_{ij})^a} \tag{2.140}$$

**Path Integral Ground State**

In PIGS, observables can only be computed at the center of the chain, where the imaginary time propagation guarantees that one is sampling the ground state of the system. However, as it was shown in section 2.5.1, any observable that commutes with the Hamiltonian, can be also computed at the extremities of the chain. That allows us to compute $\hat{H}\Psi_T$ instead of $\hat{H}\Psi_{PIGS}$, and use the local energy estimator as in VMC method. Then the energy of the system satisfies the relation (*cf.* Eq. (2.125)):

$$\langle E \rangle_{PIGS} = \langle E_L \rangle_{\Psi_T^2, \tau \to \infty} \tag{2.141}$$

where the notation used on the left hand side indicates that we are sampling in a chain where the convergence on $\tau$ is guaranteed. That means that the imaginary time propagation of $\Psi_T$ to the center of the chain converges to the ground state of the system.

### 2.6.2 Pair Distribution Function

One observable that provides an intuitive understanding of the structure of a many-body system is the two-body radial distribution function. It is proportional to the probability of finding two particles at positions $\mathbf{r}_1$ and $\mathbf{r}_2$ simultaneously. In coordinate representation, it is given by the following expression:

$$g(\mathbf{r}_1, \mathbf{r}_2) = \frac{N(N-1)}{\rho^2} \frac{\int |\Psi(\mathbf{R})|^2 d\mathbf{r}_3...d\mathbf{r}_N}{\int |\Psi(\mathbf{R})|^2 d\mathbf{r}_1...d\mathbf{r}_N}. \qquad (2.142)$$

It is useful to particularize the above expression to the case of a homogeneous and isotropic system. In this case, $g(\mathbf{r}_1, \mathbf{r}_2)$ depends only on the relative distance $\mathbf{r}_{1,2} = \mathbf{r}_1 - \mathbf{r}_2$. In this context, where the density is a constant ($n = N/L^d$), the radial distribution function becomes

$$g(\mathbf{r}) = \frac{N(N-1)}{\rho^2 L^d} \frac{\int |\Psi(\mathbf{R})|^2 \delta(\mathbf{r}_{1,2} - \mathbf{r}) d\mathbf{R}}{\int |\Psi(\mathbf{R})|^2 d\mathbf{R}}. \qquad (2.143)$$

In order to improve the efficiency of the calculation, the previous expression can be written as:

$$g(\mathbf{r}) = \frac{2}{\rho N} \frac{\int |\Psi(\mathbf{R})|^2 \sum_{i<j} \delta(\mathbf{r}_{i,j} - \mathbf{r}) d\mathbf{R}}{\int |\Psi(\mathbf{R})|^2 d\mathbf{R}}. \qquad (2.144)$$

In chapter 4 we evaluate properties of a two component Fermi liquid, and particularly we evaluate the radial distribution function corresponding to atoms of the same and of different species. In this case, Eq.(2.142) generalizes to:

$$g(\mathbf{r}_1, \mathbf{r}_2) = \frac{N_\alpha(N_\beta - \delta_{\alpha\beta})}{\rho_\alpha \rho_\beta} \frac{\int |\Psi(\mathbf{R})|^2 d\mathbf{r}_3...d\mathbf{r}_N}{\int |\Psi(\mathbf{R})|^2 d\mathbf{r}_1...d\mathbf{r}_N}. \qquad (2.145)$$

where the Greek indexes label different species. Similarly to what we did for single component systems, for implementing it in a QMC algorithm, a more suitable expression can be written considering all possible pairs of particles:

$$g_{\alpha,\beta}(\mathbf{r}) = \frac{N_\alpha(N_\beta - \delta_{\alpha\beta})}{\rho_\alpha \rho_\beta L^d} \frac{\int |\Psi(\mathbf{R})|^2 \sum_{i<j} \delta(\mathbf{r}_i^\alpha - \mathbf{r}_j^\beta - \mathbf{r}) d\mathbf{R}}{\int |\Psi(\mathbf{R})|^2 d\mathbf{R}}. \qquad (2.146)$$

We usually use the notation $g_{\uparrow\uparrow}(\mathbf{r})$ and $g_{\uparrow\downarrow}(\mathbf{r})$ for intra-species and inter-species correlations respectively. In section 4.5, we study the limiting case of an impurity immersed in a bath of $N^\uparrow$ particle system of density $n = N^\uparrow/L^d$, where the above expression can also be used. In this particular case, it reads:

$$g_{\uparrow\downarrow}(\mathbf{r}) = g_{\uparrow I}(\mathbf{r}) = \frac{1}{\rho} \frac{\int |\Psi(\mathbf{R})|^2 \sum_{j=1}^{N^\uparrow} \delta(\mathbf{r}_{\downarrow j} - \mathbf{r}) d\mathbf{R}}{\int |\Psi(\mathbf{R})|^2 d\mathbf{R}} \qquad (2.147)$$

where the label $I$, indicates that only correlations involving the impurity are taken into account. In the case of homogeneous, translationally invariant systems, the evaluation of pair distribution functions is implemented in QMC methods by accumulating statistics of the relative distances between pairs of particles in an histogram.

### 2.6.3 Static Structure Factor

Although the radial distributions have been measured directly in some systems [143–145], generally it is simpler to estimate the static structure factor $S(\mathbf{k})$, that can be measured from scattering experiments. The static structure factor $S(\mathbf{k})$ is related to the Fourier transform of $g(\mathbf{r})$ as follows:

$$S(\mathbf{k}) = 1 + \rho \int d\mathbf{r} e^{i\mathbf{k}\mathbf{r}}(g(\mathbf{r}) - 1). \tag{2.148}$$

As all functions obtained with MC include statistical noise and the finite size of our simulation box impede the accurate determination of the above integral, it is usually be preferable to evaluate the static structure factor directly in the reciprocal lattice rather than performing the Fourier transform of $g(\mathbf{r})$. For this reason, we make use of the alternative estimator:

$$S(\mathbf{k}) = \frac{1}{N} \left\langle \sum_i e^{-i\mathbf{k}\mathbf{r}_i} \sum_j e^{i\mathbf{k}\mathbf{r}_j} \right\rangle. \tag{2.149}$$

As we carry out our simulations in a box of size $L_a$ with periodic boundary conditions on each of the $d$ spatial dimensions, the components of the vectors that we are allowed to use are discretized according to $\mathbf{k}^a = \left\{\frac{2\pi n^a}{L_a}\right\}_{a=1,d}$ with $n^a$ being integers.

### 2.6.4 One-Body Density Matrix

When studying BEC systems and their properties, a fundamental quantity to be taken into account is the *One-Body Density Matrix* (OBDM). The OBDM, is the inverse Fourier transform of the momentum distribution $n(\mathbf{k})$, that tells us the occupation of the state with momentum $\mathbf{k}$ in the system. In a BEC system, the state with $k = 0$ has a macroscopic occupation, which is reflected in a delta peak in the distribution at $(k = 0)$. On a system where interactions are present, higher momentum states $\mathbf{k} > 0$ are also populated. However, when performing simulations of finite systems with Periodic Boundary Conditions (PBC), the lowest momentum that can be accessed is $k_0 = \frac{2\pi}{L}$, and the study of the low $k$ behavior is seriously affected by finite size effects. To tackle this problem, one usually studies its Fourier transform, the OBDM, that can be evaluated in the coordinate representation according to the expression:

$$\rho_1(\mathbf{r}_1, \mathbf{r}'_1) = \frac{\int d\mathbf{r}_2 ... d\mathbf{r}_N \Psi^*(\mathbf{r}_1, \mathbf{r}_2, ... \mathbf{r}_N) \Psi(\mathbf{r}'_1, \mathbf{r}_2, ... \mathbf{r}_N)}{\int d\mathbf{r}_1 ... d\mathbf{r}_N |\Psi(\mathbf{r}_1, \mathbf{r}_2, ... \mathbf{r}_N)|^2}, \tag{2.150}$$

which is an off-diagonal quantity. Similarly to what was discussed for $g(\mathbf{r})$, for systems with translational invariant the OBDM depends only on the difference between $\mathbf{r}_1$ and $\mathbf{r}'_1$, $\rho_1(\mathbf{r}_1 - \mathbf{r}'_1)$. In anisotropic systems, on the contrary, the OBDM depends both on the magnitude and the direction of the relative vector $\mathbf{r}_1 - \mathbf{r}'_1$, making its computation more expensive compared to the isotropic case. In A. Macia's Phd. Thesis [92], where dipolar systems in 2D were studied, it was shown the convenience of computing the OBDM using an expansion in partial waves

$$\rho_1(\mathbf{r}) = \sum_{m=0}^{\infty} \rho_{1m}(r) \cos(2m\theta), \tag{2.151}$$

with $\rho_{1m}(r)$ the m-th mode contribution to $\rho_1(\mathbf{r})$. We use this expression for computing the OBDM in chapter 3. The main interest on the OBDM relies on its asymptotic behavior, that is related to the condensate fraction of the system

$$n_0 = \lim_{|\mathbf{r}_1 - \mathbf{r}'_1| \to \infty} \rho_1(\mathbf{r}_1, \mathbf{r}'_1), \tag{2.152}$$

which is the fraction of the system populating the zero momentum state.

Before explaining how do we calculate this observable with the different QMC methods, it is worth to remark that, similarly to what we do when computing $g(r)$, we obtain the OBDM building up frequency histograms of relative distances between $\mathbf{r}_1$ and $\mathbf{r}'_1$ .

**VMC estimator**

A variational estimation of the OBDM can be obtained by sampling the quantity:

$$\rho_1(\mathbf{r}_1, \mathbf{r}'_1) = \frac{\int d\mathbf{r}_2 ... d\mathbf{r}_N \frac{\Psi_T^*(\mathbf{R})}{\Psi_T^*(\mathbf{R}')} |\Psi_T(\mathbf{R}')|^2}{\int d\mathbf{R} |\Psi_T(\mathbf{R})|^2}. \tag{2.153}$$

with $\mathbf{R} = \{\mathbf{r}_1, \mathbf{r}_2, ... \mathbf{r}_N\}$ and $\mathbf{R}' = \{\mathbf{r}'_1, \mathbf{r}_2, ... \mathbf{r}_N\}$.

**DMC estimator**

In DMC, the OBDM can also be obtained by sampling $\frac{\Psi_T^*(\mathbf{r}_1, \mathbf{r}_2, ... \mathbf{r}_N)}{\Psi_T^*(\mathbf{r}'_1, \mathbf{r}_2, ... \mathbf{r}_N)}$ from the mixed probability distribution. The expression in Eq. (2.150) reads in the DMC framework:

$$\rho_1(\mathbf{r}_1, \mathbf{r}'_1) = \frac{\int d\mathbf{r}_2 ... d\mathbf{r}_N \frac{\Psi_T^*(\mathbf{R})}{\Psi_T^*(\mathbf{R}')} f(\mathbf{R}')}{\int d\mathbf{R} f(\mathbf{R})}. \tag{2.154}$$

The above expression is, in general, biased by the choice of $\Psi_T$. Moreover, the forward walking technique described in section 2.3.6.2 can not be easily applied to non-diagonal operators. On the contrary, using the estimators in equations (2.153) and (2.154) we can extrapolate our results as in Eq. (2.72), which gives us also an idea of how large the systematic error is.

**PIMC estimator**

In the Path integral framework, the OBDM is obtained from the thermal density matrix:

$$\rho_1(\mathbf{r}_1, \mathbf{r}_1') = \frac{V}{Z} \int d\mathbf{r}_2...d\mathbf{r}_N \rho(\mathbf{R}, \mathbf{R}'; \beta) \tag{2.155}$$

The computation of the OBDM in PIMC (and, in general, of any other off-diagonal operator), can be efficiently performed in the G-Sector. Making use of the Worm algorithm described in sec. 2.4.5, we can study the system in a configuration in which all the ring-polymers are closed except one (*worm*). In this way, we can compute $\rho_1$ as:

$$\rho_1(\mathbf{r}_1, \mathbf{r}_1') = \frac{V}{NZ} \left\langle \delta(\mathbf{r}_1^{worm} - \mathbf{r}_{M+1}^{worm} - \mathbf{r}) \right\rangle . \tag{2.156}$$

Unfortunately, at odds to what happens with diagonal properties, here one can not take advantage of the symmetry that exists between the beads in a closed polymer, and thus, the efficiency is affected. This problem is partially solved, once in the G-sector, by proposing various movements of the head and tail of the worm, and computing $\rho_1$ after each one of these movements. An important advantage of the Worm, compared to other schemes, is that it automatically gives the correct normalization for the OBDM at the origin, $\rho_1(0) = 1$.

### 2.6.4.1 PIGS estimator

Regarding T=0 calculations, it is possible to compute the $T = 0$ OBDM using the PIGS method. In analogy with the previous methods, one has to evaluate this property in a configuration where one of the chains is open at its center. It is important to remark that, in this particular case, the chains are not ring polymers anymore, but they are all open and connected on their extremities to a trial wave function. In order to compute off-diagonal quantities with this method, we cut one of the chains in the central region, and thus we evaluate

$$\rho_1(\mathbf{r}_1, \mathbf{r}_1') = \frac{V}{NZ} \left\langle \delta(\mathbf{r}_{c_1}^{worm} - \mathbf{r}_{c_2}^{worm} - \mathbf{r}) \right\rangle , \tag{2.157}$$

where the indexes $c_1$ and $c_2$ label the two beads of the worm corresponding to the central place of the chain where it is open. This works well when one propagates the trial wave function at the extremities for a long enough imaginary time, so that we can guarantee that we are sampling the actual ground state of the system.

### 2.6.5 Superfluid fraction

*Superfluidity* can be defined as the property of a fluid that flows with zero viscosity. This astonishing property constitutes a macroscopic manifestation of the underlying microscopic quantum nature of the system [146]. Usually, in realistic systems, in which interactions are important, only a fraction of the system will be in the superfluid state. This superfluid fraction is defined in terms of the fraction of the system that do not respond to the movements of the wall of the bucket containing it. Considering a recipient with cylindrical symmetry rotating around its axis, the superfluid fraction is obtained as

$$\frac{n_s}{n} = 1 - \frac{I_{eff}}{I_c},\tag{2.158}$$

with $I_c$, the classical momentum of inertia of the system and $I_{eff}$ the effective momentum of inertia observed in the rotating quantum system. The classical momentum of inertia definition reads

$$I_c = \sum_{i=1}^{N} m_i r_\perp^i{}^2,\tag{2.159}$$

with $r_\perp^i$ the distance of the ith particle from the rotating axis. On the other hand, $I_{eff}$ can be defined in terms of the work done at an infinitesimal rotation rate:

$$I_{eff} = \left(\frac{d^2F}{d\omega^2}\right)_{\omega=0} = \left(\frac{d\hat{L}_z}{d\omega}\right)_{\omega=0},\tag{2.160}$$

with $F$ the free energy and $\hat{L}_z$ the angular momentum around the rotating axis that we consider to be the Z-axis.

#### PIMC estimator

Treating a system in rotation is not simple when working with QMC methods. Pollock and Ceperley [147], by mapping the problem to a toroidal geometry (or equivalently, simulating the system with PBC), proposed and expression for the computation of the superfluid density. Their proposal is based on the computation of the so called *Winding number*, that can be easily implemented in PIMC. This quantity takes into account the diffusion of the *world lines* (polymers) at large imaginary times. They just concluded that,

$$\frac{n_s}{n} = \frac{mk_BT}{2\hbar^2 n}\sum_{a=1}^{d}\langle W_a^2\rangle,\tag{2.161}$$

where $\mathbf{W}_a$ (with $a = x, y, z$) is the *Winding number* along one of the directions of space, in units of the box length $L_a$

$$W_a = \frac{1}{L_a} \sum_{i=1}^{N} \sum_{j=1}^{M} (\mathbf{r}_{i,j+1} - \mathbf{r}_{i,j}). \tag{2.162}$$

with $N$ and $M$ the total number of particles and beads respectively. The implementation in PIMC of the estimator in Eq. (2.161) is fairly natural and has a relatively good efficiency when employing the Worm algorithm. However, it is not possible to use this estimator at $T = 0$, where the world lines are artificially shortened and matched to a trial wave function (*cf.* section 2.5 for details about the PIGS method).

**DMC estimator**

In Ref [148] an estimator of the superfluid fraction at $T = 0$ was introduced. Indeed, it constitutes an extension of the Winding number estimator to the limit of infinitely large polymer chains. Its mathematical expression reads:

$$n_s = \lim_{\tau \to \infty} \frac{1}{4N\tau} \left( \frac{D_s(\tau)}{D_0} \right), \tag{2.163}$$

where $D_s(\tau) = \left\langle (\mathbf{R}_{CM}(\tau) - \mathbf{R}_{CM}(0))^2 \right\rangle$ and $D_0 = \hbar^2/(2m)$. It relates the superfluid fraction of a system at T=0 with the diffusion of the center of mass of the system in imaginary time.

As a final remark, it is worth noticing that both the Winding number estimator (Eq. 2.161) and the diffusion one (Eq. 2.163) can be split on their spatial components such as $n_s = \frac{1}{d}(\sum_{a=1}^{d} n_s^a)$. This is of especial interest when dealing with anisotropic systems. Actually, we will take advantage of this property on chapter 3, when we describe the superfluid phase of the stripe phase in a 2D system of bosonic dipoles.

## 2.6.6   Systems with an added impurity

To close this section about the evaluation of the observables that have been computed in this Thesis, we summarize four quantities that are of interest in the study of a system in which one impurity is included, as will be the case for section 4.5, where we study the Fermi polaron. In this Thesis, we use the label $\uparrow$ for the particles of the majority specie constituting the bath , and $\downarrow$ or $I$ for the impurity. In particular, we will focus on the polaron energy, the polaron effective mass and the quasi-particle residue. Finally, we also include the definition of the excess of volume parameter.

#### 2.6.6.1 The polaron energy

One of the driving quantities when studying quantum systems with an added atomic impurity is the *polaron energy*. It is defined as the chemical potential related to adding the impurity in the medium, that is, the energy cost of adding an impurity to the pure system at fixed volume. This is easy to evaluate in the Monte Carlo framework as it is simply the difference of two energies:

$$\varepsilon_p = \Big[ E(N_\uparrow, 1) - E(N_\uparrow, 0) \Big]_V \ . \tag{2.164}$$

where we have introduce the energy of the system with an added impurity $E(N_\uparrow, 1)$, and the one of the pure system $E(N_\uparrow, 0)$ (both taken at fixed volume). There are two complications related to the evaluation of the above quantity. The first one is that, although the energy is easily evaluated in QMC, the polaron energy comes from the difference of two different (but similar) energies of order $N$. For this reason, the MC statistical noise makes it difficult to evaluate it with for a large number of particles. On the other hand, the polaron problem is usually thought as the ultra-dilute limit of a two component mixture. Due to that, the above quantity has an implicit dependence on the number of particles in the simulation, so finite size effects have to be treated carefully to give reliable results in the limit $1/N_\uparrow \to 0$.

#### 2.6.6.2 The Effective mass of an impurity

One might also be interested in the excitation spectrum of the impurity interacting with the bath in which it is immersed. At low momentum, and in certain cases, the excitation spectrum can be approximately described by a quasi-particle picture. To do this, we think of a free quasi-particle displacing in the medium with an *effective mass* $m^*$, so that the excitation spectrum reads

$$\epsilon_p(k) = \epsilon_p(k=0) + \frac{\hbar^2}{2m^*} k^2 + \mathcal{O}(k^4), \tag{2.165}$$

which is accurate up to second order in $k$ [149, 150]. In principle one could think that it is possible to compute the effective mass from the above expression. However, introducing an impurity in a state of momentum $\mathbf{k}$ in a DMC simulations, makes us deal with the sign problem (*cf.* section 2.3.5 for details). In this way, the energies $\epsilon_p(k)$ coming from a Fixed-Node calculation are upper bounds, leading to lower bounds of $m^*$.

On the contrary, a diffusion estimator, whose implementation in DMC is quite similar to the one that we have already discussed for the superfluid density, constitutes a better approach. The effective mass of the impurity can be obtained from the asymptotic long imaginary time diffusion of the impurity through the medium [151, 152]. This estimator reads

$$\frac{m}{m^*} = \lim_{\tau \to \infty} \frac{1}{4\tau} \frac{D_s^I(\tau)}{D_0},$$ (2.166)

with $D_0 = \frac{\hbar^2}{2m}$ being the free-particle diffusion constant and $D_s^I(\tau) = \langle (\mathbf{r}_I(\tau) - \mathbf{r}_I(0))^2 \rangle$ the squared imaginary-time displacement of the impurity.

### 2.6.6.3 Quasi-particle residue

Related to the above quantities, there is another property that is useful to describe the quasi-particle nature of the polaron: *the quasi-particle residue.* It is defined as the overlap between the wave function describing the system with an added interacting impurity and that of the system with a non-interacting impurity in the momentum state $k = 0$ [153],

$$Z = \left| \langle \Phi^{\mathrm{NI}} | \phi \rangle \right|^2,$$ (2.167)

with $|\phi\rangle$ the wave function describing the system with the interacting impurity and $|\Phi^{\mathrm{NI}}\rangle$ the bath with a non-interacting impurity. For the simpler case that one can consider, in which particles of the bath do not interact between each other, and assuming that they are fermions as it is the case studied in chapter 4, $|\Phi^{\mathrm{NI}}\rangle$ reduces to $|FS + 1\rangle$, which stands for a Fermi sea with a non-interacting impurity.

In a bosonic system the momentum distribution shows a peak at zero momentum when the system is in the BEC regime. The situation is different in fermionic systems, where the Fermi Statistics makes the momentum distribution to be populated at least up to the Fermi surface (located at the Fermi momentum $k = k_F$ even in the non-interacting case. Indeed, the momentum distribution shows a jump at the Fermi surface which equals 1 in the ideal case, and takes a value $Z < 1$ in the interacting case. Considering the impurity as the zero-density limit of a Fermi sea, we obtain the relation $Z = n_\downarrow(k = 0) - n_\downarrow(k = 0^+)$. At odds to what was discussed for the momentum distribution on section 2.6.4, the components at $k > 0$ persist only as a finite size effect, as they should scale with the inverse of the Volume [154, 155].

Similarly to what is done to extract the condensate fraction in bosonic systems, it is also interesting to evaluate the inverse Fourier transform of the impurity momentum distribution, which corresponds to evaluate the OBDM including only correlations between the impurity with the bath. While its integral over volume would yield $n_\downarrow(k = 0)$ for a finite system, its asymptotic value at $r \to \infty$ is a better estimate of $Z$, since the finite-size component is automatically removed. We thus evaluate the quasi-particle residue, in the DMC framework, from the following estimator:

$$Z = \lim_{|\mathbf{r}_\downarrow' - \mathbf{r}_\downarrow| \to \infty} \left\langle \frac{\Psi_T(\mathbf{R}_\uparrow, \mathbf{r}_\downarrow)}{\Psi_T(\mathbf{R}_\uparrow, \mathbf{r}_\downarrow')} \right\rangle.$$ (2.168)

The previous quantity suffers from the same evaluation difficulties as the OBDM on its evaluation. The DMC estimator of Eq. (2.168) is biased by the choice of $\Psi_T$ used for the importance sampling. Thus, an extrapolation of the DMC results, employing information coming from the VMC estimator (*cf.* Eq. (2.72)), has to be done, both to improve the quality of the results and to estimate the systematic bias.

### 2.6.6.4 The excess of volume parameter

The last observable that we study in this Thesis is the *excess of volume*. The excess of volume constitutes a measure of the effective volume occupied by the impurity in comparison to that of an average bath particle. The difference between these two quantities, can come out due to different physical mechanisms, for example: a difference in the masses, different inter-particle interaction or different quantum statistics between indistinguishable particles.

Considering a system with a very low concentration of impurities at fixed pressure $P$, the total density of the mixture can be related to that of the pure system conformed only of atoms of the majority specie:

$$n(P, x) = n(P, x = 0)(1 + \alpha x)^{-1}, \tag{2.169}$$

with $x$ the concentration of the impurities and $\alpha$ the excess volume parameter. It was shown by Saarela, Kurteen and collaborators [156, 157] that, in the limit $x \to 0$, it is possible to approximate $\alpha$ by evaluating the $k = 0$ value of the static structure factor $S_{\uparrow\downarrow}(k)$ between the impurity and the bath particles:

$$S_{\uparrow\downarrow}(0) = -(1 + \alpha) \ , \tag{2.170}$$

with $S_{\uparrow\downarrow}(k)$ the Fourier transform of the radial distribution function $g_{\uparrow\downarrow}(r)$ introduced in section 2.6.2,

$$S_{\uparrow\downarrow}(k) = n \int d\mathbf{r}\, e^{i\mathbf{k}\cdot\mathbf{r}} \left( g_{\uparrow\downarrow}(r) - 1 \right). \tag{2.171}$$

It is important to pay attention to the normalization factor in front of the previous expression, $n$. As we are interesting the ultra-dilute case $x \to 0$, this density coincides with the density of atoms of the majority specie. In our MC simulations we perform simulations with a finite number of particles, that is $x = 1/(N^\uparrow + 1)$, and we usually evaluate the static structure factor as in a mixture. Finally it is worth noticing that usually in MC simulations, the evaluation of the static structure $S_{\uparrow\downarrow}(k)$ in a two component mixture is done with a prefactor $\sqrt{n^\uparrow n^\downarrow}$ which is different from the factor $n$ appearing in Eq. (2.171).This means that, to extract the correct magnitude of $\alpha$ from the low k behavior in a calculation with PBC in which we actually evaluate the $S^{\uparrow\downarrow}(k)$ for a finite mixture, we have to take into account the additional factor $\sqrt{n^\uparrow n^\downarrow}/n$.

Moreover, the sign of $\alpha$ carries also qualitative valuable information to deduce whether the impurity induces an excess or deficit of volume: the excess of volume coefficient $\alpha$ would be $\alpha > 0$ or $\alpha < 0$ respectively.

# Chapter 3

# Superfluid properties of dipolar bosons in two dimensions

In this chapter we study the superfluid properties of a system of dipolar bosons that are fully polarized and in which atoms are restricted to move in the $XY$ plane. We also consider that the dipolar moments form a certain tilting angle $\alpha$ with the $Z$ axis. The phase diagram at zero temperature of this system was already studied [91, 47] in terms of the density $n$ and $\alpha$, revealing the existence of three different phases: gas, stripe and solid. Here we focus on the characterization of the superfluid properties across that phase diagram. Our calculations allow to address the question of whether the stripe phase of this system could be a candidate for a supersolid: a state of matter which was first predicted in 1969 [54], where two U(1) symmetries are broken simultaneously. One of these symmetries is related to the breaking of phase invariance, as it happens in superfluids, and the other one to the breaking of the continuous traslational symmetry. The simultaneous breaking of these two symmetries could lead one to think that the system exhibits two, apparently, contradictory properties: the simultaneous existence of spatial long-range order and supporting a super-flow. In a two-dimensional (2D) system, the superfluid properties have their own peculiarities, and the transition from the superfluid to normal phase follows the Berezinskii–Kosterlitz–Thouless (BKT) scenario [85, 86], whose main properties are summarized in this chapter.

The zero-temperature techniques described in Chapter 2 (Diffusion Monte Carlo (DMC) and Path Integral Ground State (PIGS)), allow us to evaluate the superfluid properties of the system in the ground state, revealing that both the gas and the stripe phase are superfluid and exhibit a finite condensate fraction. Our study is completed with the characterization of the thermal transition that exists between the superfluid phases and the normal ones. This transition was already studied for the isotropic case ($\alpha = 0$) by Filinov *et al.* [158], and here we extend it to the anisotropic phases that the system exhibit when $\alpha \neq 0$. To this end, we perform finite temperature calculations with the Path Integral Monte Carlo (PIMC) method. Our results show that the BKT scaling holds not only when anisotropy is present in the system but also when a continuous transnational symmetry is broken. The complete characterization of the BKT transition for the stripe phase suggests that the dipolar stripe phase is a candidate for the supersolid state of matter.

## 3.1 Introduction

The possible existence of a supersolid state of matter has been a long standing topic in physics since Andreev suggested its existence in 1969 [54]. From the theoretical point of view, a supersolid is associated to the breaking of two U(1) symmetries. The first one is related to the loss of continuous translational invariance as a consequence of the presence of a crystalline structure, and the second one corresponds to the appearance of a non-trivial global phase, as it corresponds to a Bose-condensed state [159].

The first attempts to find a supersolid phase were linked to Helium, due to its extreme quantum character and to its experimental versatility. Indeed, a lot of excitement emerged at the beginning of this century, when a experimental group claimed for its detection [160]. Years later, however, after a careful analysis of additional experimental data, this possibility had to be excluded [161]: the deviations from the conventional rotational moment of inertia that were originally reported, turned out to be the consequence of elastic effects. These difficulties for finding experimental evidence of a supersolid phase kept open the debate about how a supersolid should be really defined [162]. Some aside work has been devoted to find superfluid phases in solids with vacancies, as it was originally proposed by Andrew-Lifshitz, or in the lattice, where the spatial invariance is artificially broken. However the search of an intrinsic supersolid has been unfruitful for years.

More recently, with the development of ultracold gases experiments, the possibility of finding supersolid phases has reborn. Although conventional gaseous BEC systems are not able to break translational symmetry, and thus, they are not good candidates to present a supersolid phase, in recent years some systems with richer interactions have become available in the laboratory. In 2017, two experimental groups, almost simultaneously, claimed to have observed a phase with supersolid properties in systems in which spin-orbit coupling is present. Both experiments were performed in a reduced geometry: in the first one, the momentum dependence of the synthetic spin-orbit coupling induces a density modulation that gives rise to a stripe phase with phase coherence [163], whereas, in the second one, this was achieved by the coupling of atoms to the modes of the cavity containing them [164].

Also in the context of ultracold gases, dipolar systems offer new possibilities to investigate new supersolid phases of matter. Signatures of such phases have been experimentally reported for dipolar atoms confined in a trap with cylindrical symmetry [52, 51, 53], following the idea of a previous theoretical work [57]. In these experimental setups, dipolar atoms are polarized along a direction perpendicular to the trap axis, leading to the formation of dipolar clusters. For a certain combination of experimental parameters phase coherence between the different clusters is found. Some efforts have also been put on the study of the excitation spectrum of such a system that have been experimentally characterized [58, 60, 61]. These studies come to complement similar results that have also been also obtained in spin-orbit systems [164].

In the decade of 1990, and in the context of condensed matter, stripe phases started to catch interest. They appear due to the non-homogeneous structures that are present in some materials [165, 166], and it has been found that their presence offers a mechanism for obtaining high temperature superconductors [167]. Regarding dipolar systems, although their stripe phase have not been experimentally achieved yet (for a quasi-2D stripe phase realization see *cf.* [168]), much work has been done in the study of its properties [91, 46, 47, 169–173]. In this chapter we study the superfluid properties of the dipolar stripe phase, both at zero and finite temperature, showing that the long-range spatial structure that characterize them is compatible with the presence of a finite superfluid fraction. Equivalent results have been found in systems in which anisotropy and/or long-range interactions are present. The closest to our study is the study of the phase diagram of 2D dipolar bosons in the lattice, where similarly to the continuous case studied in this Thesis, a supersolid stripe phase is found [173]. Finally, it is also worth mentioning that a superfluid stripe phase has been studied in the Hubbard model with an isotropic long-range interaction. In this case, the rotational symmetry is broken spontaneously by the interplay between the long-range character of the inter-particle interaction considered with the lattice, that makes the atoms to occupy certain lattice positions in order to minimize the energy [174].

Regarding two-dimensional systems, the superfluid transition is different from the usual three-dimensional (3D) scenario, as the scaling of the superfluid properties of a 2D system has its own peculiarities. The differences can be easily understood in terms of the properties of long-range correlations [175]: while in 3D off-diagonal long-range order is allowed at low temperatures, in 2D this is only possible in the limit of zero temperature. However, two-dimensional systems support quasi-long-range order, reflected on an algebraic decay of their off-diagonal correlations. In this sense, the transition from the superfluid to the normal phase has to be thought as the transition from a phase with quasi-long-range order to a normal one (with exponential decay of the off-diagonal correlations). This phenomenology was first studied by Berezinskii in 1971 [85] and soon later by Kosterlitz and Thouless [86]. Differently to what happens in 3D systems, in two dimensions the superfluid fraction of the system performs a jump and vanishes at the critical temperature $T_{BKT}$. This jump follows a universal law that was first studied by Nelson and Kosterlitz [176], and its mathematical expression reads

$$\frac{n_s(T_c, L)}{n} = \frac{2mk_B}{\pi\hbar^2}\frac{T_c}{n}. \tag{3.1}$$

In section 3.4 we take advantage of this universal jump to determine the critical temperature of the superfluid transition, in analogy to what has been done in other systems. The first studies in specific systems came in the context of condensed matter physics, where the BKT transition was studied in Helium films [177–179] and Coulomb layers [180]. More recently, as it has happened with many other condensed matter problems, ultracold gases in pancake geometries [87, 90, 88] have proven themselves as valid platforms to study this phenomenon. The validity of the BKT scenario has been demonstrated even in systems where disorder is present [181, 182]. Here, we show that neither the presence of anisotropy nor the breaking of translational invariance invalidates the BKT universal relations.

## 3.2 The system

We study a strictly 2D system of bosonic dipoles. We consider that all the dipoles are polarized along the same direction of the space, such that they form an angle $\alpha$ (tilting angle) with the Z axis. Without loss of generality we also consider that the field polarizing the dipoles is contained in the $XZ$ plane. Such a system is described by the following $N$-particle Hamiltonian:

$$H = -\frac{\hbar^2}{2m}\sum_{j=1}^{N}\nabla_j^2 + \frac{C_{dd}}{4\pi}\sum_{i<j}^{N}\left[\frac{1 - 3\lambda^2\cos^2\theta_{ij}}{r_{ij}^3}\right], \tag{3.2}$$

with $N$ the total number of particles, $\lambda = \sin\alpha$, and $(r_{ij}, \theta_{ij})$ the polar coordinates associated to the position vector of particle $j$ with respect to particle $i$. The constant $C_{dd}$ is proportional to the square of the (electric or magnetic) dipole moment of the components, assumed all of them to be identical. We usually use dipolar units, obtained from the characteristic dipolar length $r_0 = mC_{dd}/(4\pi\hbar^2)$, and the dipolar scale of energy $\epsilon_0 = \frac{\hbar^2}{mr_0^2}$, so that we can write the Hamiltonian of Eq. (3.2) in a dimensionless form:

$$H = -\frac{1}{2}\sum_{j=1}^{N}\nabla_j^2 + \sum_{i<j}^{N}\left[\frac{1 - 3\lambda^2\cos^2\theta_{ij}}{r_{ij}^3}\right]. \tag{3.3}$$

The ground state phase diagram of this system has been studied in previous works as a function of density $n$ and tilting angle $\alpha$ [91, 47]. In Fig. 3.1 we show the region of the phase diagram that we study in this chapter, that is, the region in which the stripe phase appears, and its vicinity.

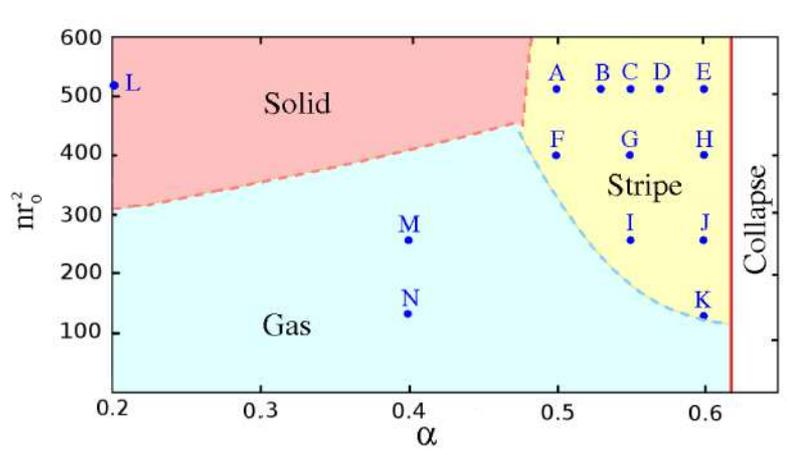

Fig. 3.1 . Phase diagram of the 2D dipolar Bose gas at zero temperature. Letters indicate the set of points corresponding to fixed density and polarization angles explored in this Thesis.

## 3.3 Zero Temperature Study

In this section we report the zero temperature calculations that we have performed to study the ground state of the system represented by the Hamiltonian of Eq. (3.2). The methods that we have employed for this purpose are DMC and PIGS – for details about this methods, see sections 2.3 and 2.5 respectively.

### 3.3.1 Details of the calculation

Although both DMC and PIGS methods give exact results for the ground state of the system, their efficiency is improved when a trial wave function $\Psi_T$ is used for importance sampling. For the calculations presented in this chapter we have defined $\Psi_T$ to be of the form:

$$\Psi_T(\mathbf{R}) = \Psi_{1B}(\mathbf{R})\Psi_{2B}(\mathbf{R}), \tag{3.4}$$

with $\Psi_{1B}$ and $\Psi_{2B}$ containing one and two-body correlation terms, respectively. For the study of the gas phase we only consider two-body correlations, which is the same as setting $\Psi_{1B} = 1$. This choice preserves the continuous translational invariance symmetry that is characteristic of a gas. The two-body term is taken to be of the Jastrow form [111]

$$\Psi_{2B}(\mathbf{R}) = \prod_{i<j}^{N} f(\mathbf{r}_{ij}), \tag{3.5}$$

where $f(\mathbf{r}_{ij})$ is the Jastrow factor that introduces two-body correlations. In contrast to what happens in many physical condensed matter systems, such as Helium, in our case the Jastrow factor has to incorporate the anisotropy that the dipolar interaction introduces in the system, and as a consequence, it depends not only on the magnitude of $\mathbf{r}_{ij}$ but also on its direction. We construct the Jastrow factor $f(\mathbf{r}_{ij})$ as the zero energy solution of the two-body dipolar problem, matched with a phononic solution at a certain distance $r_M$ (used as a variational parameter [91, 183]). It reads

$$f(\mathbf{r}) = \begin{cases} A\phi_{2B}(\mathbf{r}) & if \quad r \leq R_M \\ \\ Be^{-C\left(1/r + 1/(L-r)\right)} & if \quad r > R_M \end{cases} \qquad (3.6)$$

where $\phi_{2B}$ corresponds to the two-body zero energy solution that includes the anisotropy of the system in the case of $\alpha \neq 0$. As we are doing simulation in a box of length $L = \sqrt{N/n}$, correlations are cut for distances higher than $r > L/2$, and so, we also impose the following conditions on the two-body Jastrow correlation functions $f(\mathbf{r})$: $f(|\mathbf{r}| > L/2) = 1$, $f'(|\mathbf{r}| > L/2) = 0$. These conditions, together with the conditions of continuity and derivability of the wave function at $r = R_M$, fix the values of the constants $A$, $B$, and $C$.

When the structural properties of a system are studied with DMC, the estimation of some observables, such as the static structure factor, is not exact but can be biased by the trial wave function. In this case, obtaining exact results is still possible (see section 2.3.6.2 for details about the pure estimator technique), but it can be computationally expensive if the quality of the trial wave function that is used is low. For this reason, when studying the stripe phase, we include a one-body term as written in Eq. (3.4), that explicitly takes into account the density modulation of the system. Here, it is important to remark that, even if this term is not included, the stripe phase still appears as the ground state at the proper region of the phase diagram. The mathematical expression of the one-body term that we employ reads

$$\Psi_{1B}(\mathbf{r}) = \prod_{i=1}^{N} \exp\left[\eta_{str}\cos\left(\frac{2\pi N_{str}y_i}{L_y}\right)\right] , \qquad (3.7)$$

with $y_i$ the $y$-coordinate for the ith particle, $L_y$ the box side length along that direction, and $N_{str}$ the number of stripes contained in the simulation box. As a check for the validity of this trial wave function it is worth to remark that $\eta_{str}$ is a variational parameter that is consistently found to be zero in the gas phase. Similarly to what happens when simulating solids, the number of stripes in the box has to be commensurated with the box length in order to correctly reproduce the thermodynamic limit. This implies that the number of stripes $N_{str}$ and the number of particles $N$ in the simulation box are related to each other. With this restriction in mind, we define the another variational parameter that we have to optimize to improve the quality of $\Psi_T$: $\Delta y_{str} = L_y/N_{str}$, that forces us work with a simulation box with an aspect ration $L_y/L_x \neq 1$.

The values of the variational parameters, $\eta_{str}$ and $\Delta y_{str}$, related to the one-body term of Eq. (3.7), are optimized using VMC (for details of this method, see section 2.2), and their optimal values are summarized in Table 3.1. The implicit relation of the aspect ratio of the box $L_y/L_x$, the number of stripes and $N_{str}$ and the number of particles, make the optimization of $\Delta y_{str}$ an intricate problem. Furthermore, as calculations are performed with a cut-off on the potential at a $r = L/2$ ($L = \mathsf{min}(L_x, L_y)$), strong deformations of the box would include additional undesired finite size effects on the calculation, that would be reflected on a non-trivial dependence of the energy per particle as a function of $\Delta y_{str}$.

To give an idea on how big these finite size effects are, on the left panel of Fig. 3.2 we show the optimization of $\Delta y_{str}$ using two methods: in the first one (purple points) we try to work with a square box ($L_y/L_x \approx 1$) what allows to minimize the finite size effects coming from introducing a cutoff in the potential at $r = L/2$. In this case, the number of stripes inside the box and the number of particles are not constant. In the second method, the number of stripes inside the box is fixed and the box is deformed as $\Delta y_{str}$ changes. Finally the process is repeated with a larger number of particles to check its evolution with finite size effects are under control. The use of these two methods allows us to estimate the values of $\Delta y_{str}$ and their uncertainties, as shown in Table 3.1. On the right panel of Fig. 3.2 we show the optimization of the parameter $\eta_{str}$ once the optimal $\Delta y_{str}$ is fixed. The optimal values of $\eta_{str}$ are also shown in Table 3.1. The inclusion of the one-body term in the trial wave function together with the use of the pure estimators technique leads to a static structure factor in agreement with the PIGS prediction, which is exact [92].

| $nr_0^2$ | $\Delta y_{str}(r_0)$ | $\eta_{str}$ | $\Delta y_{str}(r_0)$ | $\eta_{str}$ | $\Delta y_{str}(r_0)$ | $\eta_{str}$ |
|---|---|---|---|---|---|---|
| 512 | 0.049(5) | 0.10(1) | 0.050 (4) | 0.24 (2) | 0.052(5) | 0.90(8) |
| 400 | 0.056 (5) | 0.03(1) | 0.057 (4) | 0.32 (3) | 0.060(2) | 0.80(5) |
| 256 | — | — | 0.069 (5) | 0.011(1) | 0.080(2) | 0.45 (3) |
| 128 | — | — | — | — | 0.115 (4) | 0.03 (1) |

Table 3.1 Optimal variational parameters for the stripe trial wave function employed in the DMC calculations.

For the PIGS simulations, and since exact results are guaranteed by a proper propagation in imaginary time, we have adopted a much simpler choice for the trial wave function. In this case we employ the zero-energy solution of the isotropic two-body problem, matched with a phononic tail, as is explicit written in Eq. (3.6).

### 3.3.2   Results at zero temperature

In order to characterize the superfluidity of the ground state of the dipolar gas and stripe phases, we compute the superfluid fraction at the points that are labeled in the phase diagram of Fig. 3.1. These points are selected so that they cover the region of the phase diagram in which the system appears in its stripe form, and its vicinity. To this end, we perform DMC simulations using the diffusion estimator for the superfluid fraction that was introduced in Eq. (2.163). With the aim of performing a deeper analysis, we have split it into the contributions coming from the different directions of space, as shown in the following equation

$$\frac{n_s}{n} = \frac{1}{2n}(n_s^X + n_s^Y). \qquad (3.8)$$

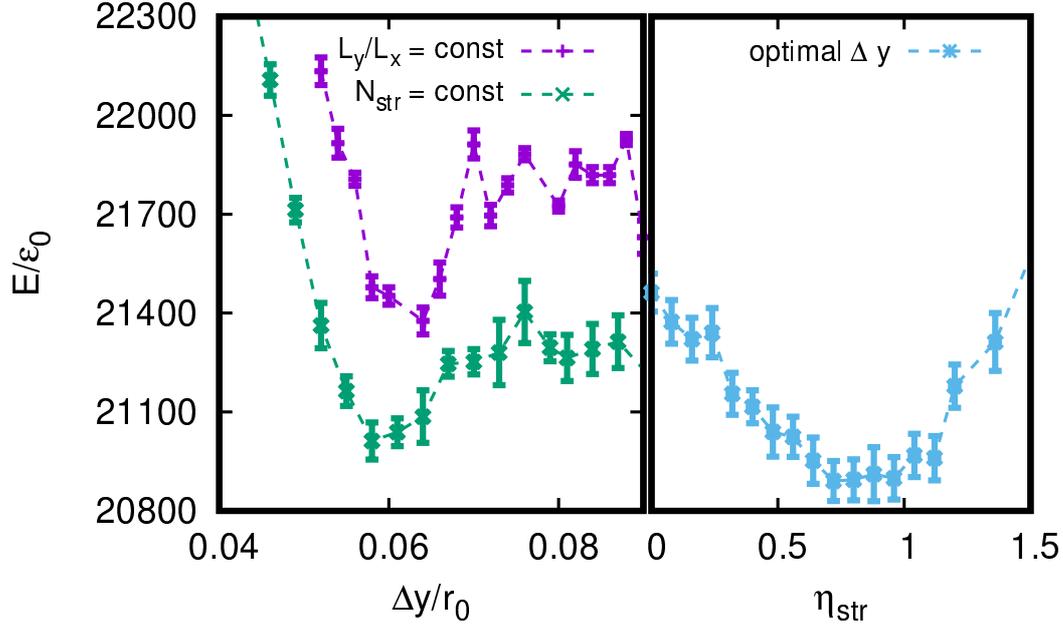

Fig. 3.2 Variational optimization of the parameters in the one-body term of the stripe trial wave function. Left panel: optimization of the optimal distance between stripes $\Delta y_{str}$ with two methods: purple points represent calculations in which the distance between stripes is varied maintaining fixed the aspect ratio of the box length $L_y/L_x \approx 1$ for $N \approx 155$. Green points correspond to same optimization but maintaining fixed the number of stripes in the box and varying its aspect ratio $L_y/L_x \neq 1$ (N = 135). Right panel: optimization of the dimensionless parameter $\eta_{str}$ for the optimal value of $\Delta y_{str}$. Energies and distances in dipolar units.

In figure 3.3 we report the DMC results for the total superfluid fraction, and its two spatial contributions $n_s^X/n$ and $n_s^Y/n$, for the stripe phase. On the left panel, we fix the density to $nr_0^2 = 512$ and we vary the tilting angle (points A-E in the phase diagram of Fig.3.1), showing that the stripe phase has always a finite superfluid signal $n_s/n$ whose smaller values are around $\frac{n_s}{n} \sim 0.5$ (Blue points), corresponding to the highest values of $\alpha$. We also report its separate contributions. As we increase the tilting angle and we approach to the collapse line, the superfluid density across the $Y$ direction decreases significantly, but it always has a non-zero value (green symbols). Furthermore, the values of $n_s^X/n$ are compatible with a superfluid signal of 1 (red symbols). On the right panel of the same figure, we plot the superfluid densities for different points of the phase diagram where the tilting angle is fixed to $\alpha = 0.6$ (points E, H, J, K in Fig. 3.1), close to the collapse line. Similarly to what is seen in the previous case, as we go deeper into the stripe regime, the superfluid signal across the $Y$ direction is highly suppressed, though, it remains finite at the highest density considered here ($n_s^Y/n \sim 2\%$). It is worth to remark that, near the gas-stripe transition line (points F, I, and K in the phase diagram of Fig. 3.1), the total superfluid fraction approaches to its maximum value, in contrast with what happens near the solid-stripe transition line, where it presents a lower, although still large, value. In Table 3.2, we summarize our results for the superfluid fractions at zero temperature.

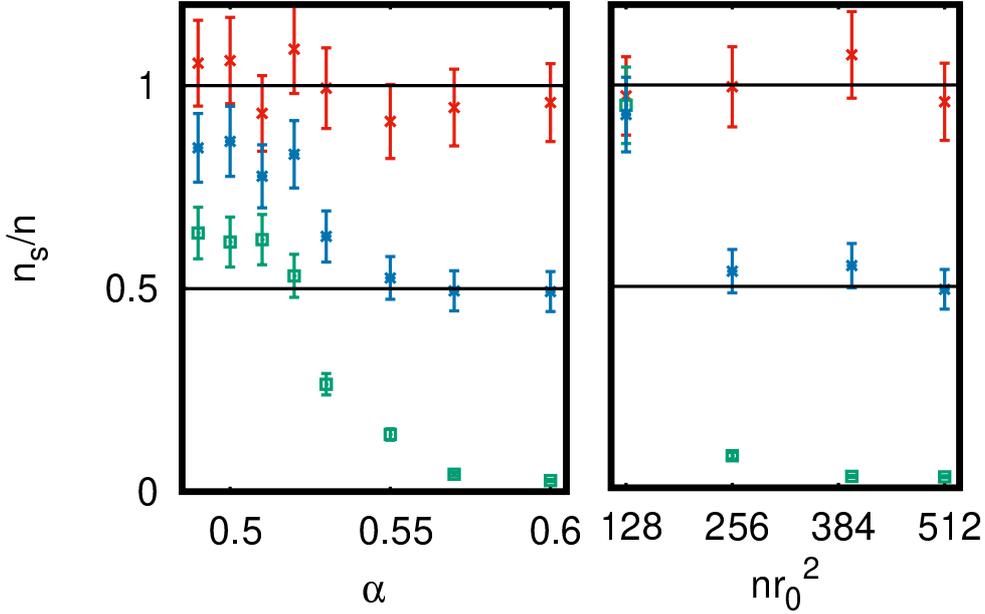

Fig. 3.3 . Superfluid fractions along the $X$ direction $n_s^x/n$ (red crosses), along the $Y$ direction $n_s^y/n$ (green squares) and total $n_s/n$ (blue stars). The left panel shows the dependence of these quantities on the polarization angle at the fixed density $nr_0^2 = 512$. The right panel corresponds to $\alpha = 0.6$ and different densities. In all the cases the system remains in the stripe phase.

A direct measure of the off-diagonal long-range order is provided by the one-body density matrix. In DMC the OBDM is not a pure estimator, and the application of the pure estimators technique described in section 2.3.6.2 is not straightforward. However, its exact estimation can be obtained in PIGS, with the help of the *worm* algorithm (*cf.* section 2.6.4). When performing calculations at T=0, and thus, employing PIGS instead of PIMC, the worm is created by opening one of the polymers at its center, where the propagation in imaginary time guarantees the sampling of the real ground state. The long-range asymptotic value of the OBDM provides an estimation of the condensate fraction of the system

$$\lim_{r \to \infty} n_1(r) = n_0/n. \tag{3.9}$$

| | $nr_0^2$ | $\alpha$ | $\frac{n_0}{n}$ | $\frac{n_s}{n}$ | $\frac{n_s^X}{n}$ | $\frac{n_s^Y}{n}$ |
|---|---|---|---|---|---|---|
| A | 512 | 0,50 | 0.00030(4) | 0.86(8) | 1.06(8) | 0.61(8) |
| B | 512 | 0.53 | 0.00055(6) | 0.62(6) | 0.99(8) | 0.26(3) |
| C | 512 | 0.55 | 0.0029(3) | 0.53(5) | 0.92(8) | 0.14(2) |
| D | 512 | 0.57 | 0.0031(3) | 0.49(5) | 0.95(8) | 0.043(4) |
| E | 512 | 0.60 | 0.0047(5) | 0.49(5) | 0.95(8) | 0.027(3) |
| F | 400 | 0.50 | 0.0038(3) | 1.05(8) | 1.07(8) | 1.04(8) |
| G | 400 | 0.55 | 0.0042(4) | 0.63(6) | 1.001(7) | 0.26(3) |
| H | 400 | 0.60 | 0.0052(4) | 0.55(5) | 1.07(8) | 0.028(3) |
| I | 256 | 0.55 | 0.015(1) | 1.05(8) | 1.03(8) | 1.08(8) |
| J | 256 | 0.60 | 0.011(1) | 0.54(5) | 1.00(8) | 0.080(6) |
| K | 128 | 0.60 | 0.071(4) | 0.95(7) | 0.97(7) | 0.93(7) |
| L | 512 | 0.20 | 0 | 0 | 0 | 0 |
| M | 256 | 0.40 | 0.019(2) | 1 | 1 | 1 |

Table 3.2 Superfluid densities and condensate fraction for the points shown in Fig. 3.1. Figures in parenthesis are the error bars.

It is worth to remember that 2D systems can only support off-diagonal long-range order at $T = 0$ and thus, only in this limit there could be a condensate state. In Fig. 3.4, we show two examples of OBDM evaluated with the zero temperature estimator of Eq. (2.157). In particular, we evaluate it at density $nr_0^2 = 512$ and for two different polarizations: $\alpha = 0.2$ and $\alpha = 0.55$ where the system is in the solid and the stripe phase respectively (points L and C in the phase diagram of Fig 3.1, respectively). The OBDM for these two phases is plotted both across the $X$ and $Y$ directions, and their behavior hints that their asymptotic value is independent of the direction. Although we can not compute it for distances longer than $L/2$, the previous statement is clear keeping in mind that $n_1(\mathbf{r})$ is the Fourier transform of the momentum distribution $n(\mathbf{k})$. In the solid phase, an exponential decay of the OBDM comes out, reflecting that no off-diagonal long-range order is present in this phase, not even at zero temperature. On the contrary, when this same quantity is computed in the stripe phase it clearly shows a small but finite condensate fraction. Results for the condensate fraction, evaluated over the points of the phase diagram in Fig. 3.1 are summarized in Table 3.2.

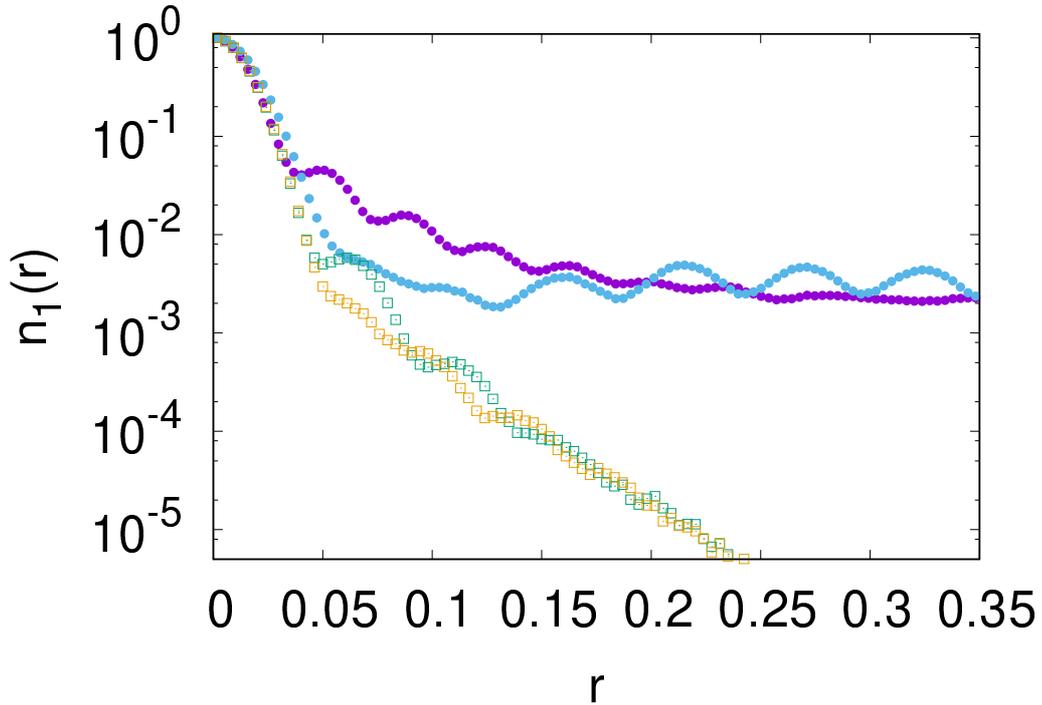

Fig. 3.4 . One-body Density Matrix of the 2D dipolar Bose system at the density $nr_0^2 = 512$ in the stripe phase for $\alpha = 0.55$ (filled circles), and in the solid phase for $\alpha = 0.20$ (empty squares). Purple circles and green squares: cuts along the X direction; blue circles and orange squares: cuts along the Y direction. Distance $r$ is measured in units of $r_0$. Error bars are smaller than 10% of each measure and have not been included for the sake of clarity.

Similarly to what happens with the superfluid fraction, $n_0/n$ is always finite in the stripe phase, with its larger values in points close to the gas to stripe transition line.

## 3.4   The BKT Transition

The zero temperature study presented above is complemented in this section with the study of the same properties at finite temperature. For that purpose, we use the PIMC method introduced in section 2.4.

The inclusion of temperature in our analysis allows to study the thermal transition that occurs between any superfluid phase and its correspondent normal phase as temperature is increased. In figure 3.5, we schematically represent the transitions that are observed: For the gas phase (bottom panel), a transition from a superfluid gas to a normal gas occurs at the critical temperature $T_{BKT}$, whereas for the stripe phase, the transition from a superfluid to a normal stripe phase is followed, if temperature continues increasing, by the melting of the stripes to a gas phase at the fusion temperature $T_F$. Due to the spatial order that characterizes the superfluid stripe phase, its superfluid signal is weaker against thermal fluctuations when compared to the gas phase, which is reflected in a lower superfluid transition temperature.

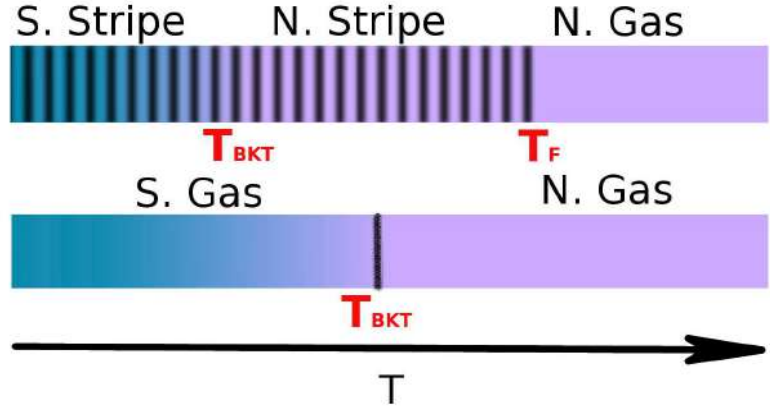

Fig. 3.5 Schematic representation of the thermal phase transition that occurs in the gas and stripe phases of the bosonic dipolar system. N and S labels stand for normal and superfluid phases respectively.

However, as we have already commented in the introduction of this chapter, the superfluid to normal fluid transition has some peculiarities in two-dimensional systems, that make them different from their equivalent three-dimensional ones. While in 3D the superfluid fraction decreases continuously to zero at the critical temperature, in two dimensions this happens abruptly at the exact transition temperature. This abrupt jump in the superfluid density is known as *universal jump*, and follows the universal law that was advanced in the introduction [176]:

$$\frac{n_s(T_c, L)}{n} = \frac{2mk_B}{\pi\hbar^2}\frac{T_c}{n}.$$

(3.10)

The above is closely related to the nature of the superfluid transition in 2D systems, that concurrently is closely related to the fact that in 2D off-diagonal long-range order is only possible at zero temperature. This makes that, at odds to what happens in 3D, at finite T only a transition between a phase with off-diagonal quasi-long-range order to a normal one can occur. This is known as the Berezinskii–Kosterlitz–Thouless (BKT) transition, named after the first authors that studied its universal properties [85, 86]. In particular, and around the critical temperature, the correlation length $\xi$ (in units of $1/\sqrt{n}$) presents a singularity of the form

$$\xi(T) \sim e^{a/t^{1/2}},$$

(3.11)

where $t = (T/T_c - 1)$ and $a$ being a non-universal parameter depending on the density and other properties of the system [184]. In our case, as we perform calculations with a finite number of particles inside a simulation box of length $L$, all the correlations are artificially cut at the edge of the box, so we can assume $\xi(T) \sim L$, as is usually done in finite-size scaling. This means that, when trying to determine the critical temperature $T_c$, one obtains only an estimation that is affected by finite size effects, $T(L)$, instead of the value corresponding to the infinite system $T_c(L = \infty) = T_{BKT}$. Due to this, and in order to extract $T_{BKT}$, a scaling of the critical temperature with the length of the box is needed. Such a scaling is deduced from (3.11), and reads

$$T_c(L) = T_c(L = \infty) + \frac{b}{\ln^2(L\sqrt{n})} \, , \tag{3.12}$$

with $b$ a non-universal constant. The above relation, together with the universal jump of Eq. (3.10), allows us to determine the critical temperature $T_c(L = \infty) = T_{BKT}$ both in the gas and stripe phases.

### 3.4.1 Results at finite Temperature

**The Gas Phase**

To understand the method that we employ to obtain the critical temperature of the BKT transition, it is useful to have a look at figure 3.6: there, we show how the critical temperature in the thermodynamic limit can be extracted by taking advantage of the universal relations (3.10) and (3.12). As an example, we show results corresponding to density $nr_0^2 = 25$ and different tilting angles ($\alpha = 0.0, 0.2, 0.4, 0, 6$), all of them corresponding to the gas phase (*cf.* phase diagram of Fig. 3.1). In the left panel of that figure we show the superfluid densities as a function of temperature computed for different system sizes and for the particular case $\alpha = 0.6$. In the same plot, the expression for the Universal Jump of Eq. (3.10) is included (red line). The computation of superfluid densities has been done using the Winding number estimator that was introduced in section 2.6.5 (see Eq. (2.161)). For any physical conditions, the critical temperature of a 2D system has to follow the Universal Jump prediction, the critical temperature for a given system size $T_C(L)$ are obtained as the cuts of the universal jump predictions with the curves of the superfluid density as a function of temperature for that fixed system size. Once this temperatures are obtained, they can be used to extract its thermodynamic limit value with the help of Eq. (3.12), as its shown in the right panel of the same figure. These results show that the BKT scaling is still valid when the interactions that are present in the system are anisotropic. It is also worth to remark that, for the case $\alpha = 0$, we recover the results obtained by Filinov *et al* in Ref. [158]. Notice that we express temperatures in units of the dimensionless density $T/nr_0^2$ (inspired by Eq. (3.10), which still has units of temperature (In the reduce units that we use temperature has units of energy, $\epsilon_0$ as we take $k_b = 1$).

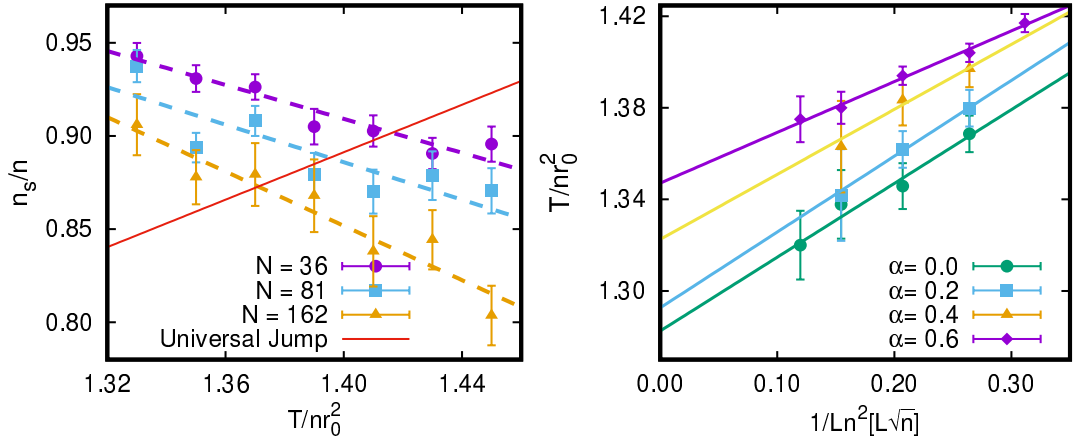

Fig. 3.6 Left panel: superfluid fraction as a function of temperature for different system sizes at density $nr_0^2 = 25$ and tilt angle $\alpha = 0.6$. Points are MC results, dashed lines are linear fits to PIMC data and the solid line is the universal jump of Eq. (3.10). The crossing points between the lines and the universal jump give the critical temperatures $T_C(L)$. Right panel: scaling of the critical Temperature $T_c(L)$ with the system size, as given by Eq. (3.12), at the same density and for different polarization angles. Points are PIMC data and solid lines are linear fits.

When the isotropic case ($\alpha = 0$) was studied in Ref [158], it was found a non-monotonic behavior of $T/nr_0^2$ as a function of the dimensionless density: In units of $nr_0^2$, the critical temperature increases at low densities and, above a characteristic value ($nr_0^2 \sim 1.4$), the behavior is the opposite. The authors of that work attribute this change in the density dependence to the appearance of the roton in the quasi-particle spectrum. This roton, has been observed to appear starting from density $nr_0^2 \simeq 1$ [185, 158, 91]. Motivated by this, we have studied how the tilting angle ($\alpha > 0$) influences the behavior of the critical temperature in these two regimes. The same analysis that we have shown in Fig. 3.6 at density $nr_0^2 = 25$ has been repeated for a much lower value of the density: $nr_0^2 = 0.01$ and for the same tilting angles $\alpha = 0.0, 0.2, 0.4, 0, 6$. Our PIMC results for the critical temperature are reported in Table 3.3, where it can be seen that the behavior of $T_c/nr_0^2$ with the tilting angle is the opposite for densities 0.01 and 25. Indeed, increasing $\alpha$ reduces (increases) the critical temperature at low (high) density. Therefore, the effect of increasing the tilting angle is the same to that found for the isotropic system when density (interaction strength) is decreased. This can be understood in terms of how does the s-wave scattering length change as the tilting angle is increased. Up to second order in $\lambda$, we can write [186, 183]:

$$a_s(\lambda) \simeq r_0 e^{2\gamma} \left(1 - \frac{3\lambda^2}{2}\right) , \qquad (3.13)$$

with $\gamma$ the Euler's Gamma constant. In this sense, the growth of $\alpha$ translates into an effective reduction of the interaction (or $nr_0^2$ when using dipolar units) which provides a qualitative understanding of why increasing $\alpha$ for given density, provides the same behavior as reducing the density maintaining the tilting angle constant.

**The Stripe Phase**

The same analysis that we have performed for the gas phase can be applied to extract the superfluid critical temperature of the stripe phase $T_{BKT}$. Finding $T_{BKT}$ for this phase is of relevance because it would confirm the existence of a supersolid state at finite temperature and not only in the ideal case of $T = 0$, as was already shown in section 3.3. Besides, it is also of relevance to show that the BKT scaling applies to phases in which continuous translational symmetry is broken, as its nature only depends on the dimensionality of the system.

On the two left panels of figure 3.7, we show the same analysis that we have already shown for the superfluid fraction as a function of the temperature in the gas phase, but for the points K ($nr_0^2 = 128$, $\alpha = 0.6$, top) and J ($nr_0^2 = 256$, $\alpha = 0.6$, bottom), both of them corresponding to the stripe phase. Similarly to what we did in the analysis of the gas phase data, in the right panels we show the scaling of the critical temperature, not only for the stripe phase, but also for the gas phase at the same densities and lower tilting angle ($\alpha = 0.4$ corresponding to points M and N in the phase diagram of Fig. 3.6). Although, due to its computational cost, it is difficult to perform simulations below temperatures $T/nr_0^2 = 0.4\epsilon_0$, their extrapolation to $T = 0$ seem to be in agreement with the zero temperature calculations of the previous section $\left[\frac{n_s}{n}\right]_{\alpha=0.6}^{nr_0^2=128} = 0.95(7)$ and $\left[\frac{n_s}{n}\right]_{\alpha=0.6}^{nr_0^2=256} = 0.54(5)$ (*cf.* Table 3.2). Here, it is important to remark that the superfluid signal along the $Y$ direction is weaker than that along $X$ direction, and that for the transition temperature its value is below $\left[\frac{n_s^Y}{n}\right]_{T_{BKT}^-} < 5\%$.

Results for the critical temperature of Fig. 3.7, are summarized in Table 3.3. By increasing the density, the critical temperature within the stripe phase decreases in a similar form to what was previously obtained for the gas in the regime $nr_0^2 > 1$. However, at fixed density, if one crosses the transition line from gas to stripe phase, the superfluid fraction is highly suppressed and, as the critical temperature is related to it by Eq. (3.10), it also decreases. In other words, due to the breaking of a continuous translational symmetry, the superfluidity in the stripes is thermally more fragile than what it is in the gas phase.

Similarly to what was done in the study at zero temperature, a deeper knowledge of the superfluid phases can be obtained by evaluating the one-body density matrix. As our system is 2D, and we perform calculations at finite temperature, the long-range behavior of the OBDM does not saturate to any fixed value at large distances that one could associate with a finite condensate fraction, as it was stated in Eq. (3.9). The reason is that at finite temperature long-range correlations decay either algebraically (superfluid) or exponentially (normal fluid). However, the BKT theory yields a prediction for the value of the exponent $\eta$ of the algebraic decay, characteristic of the 2D superfluid phase. In this case, the OBDM long-distance asymptotic behavior reads

$$n_1(\mathbf{r}) \sim r^{-\eta} \quad ; \quad r \to \infty, \tag{3.14}$$

with

$$\eta = \frac{mk_bT}{2\pi\hbar n_s}. \tag{3.15}$$

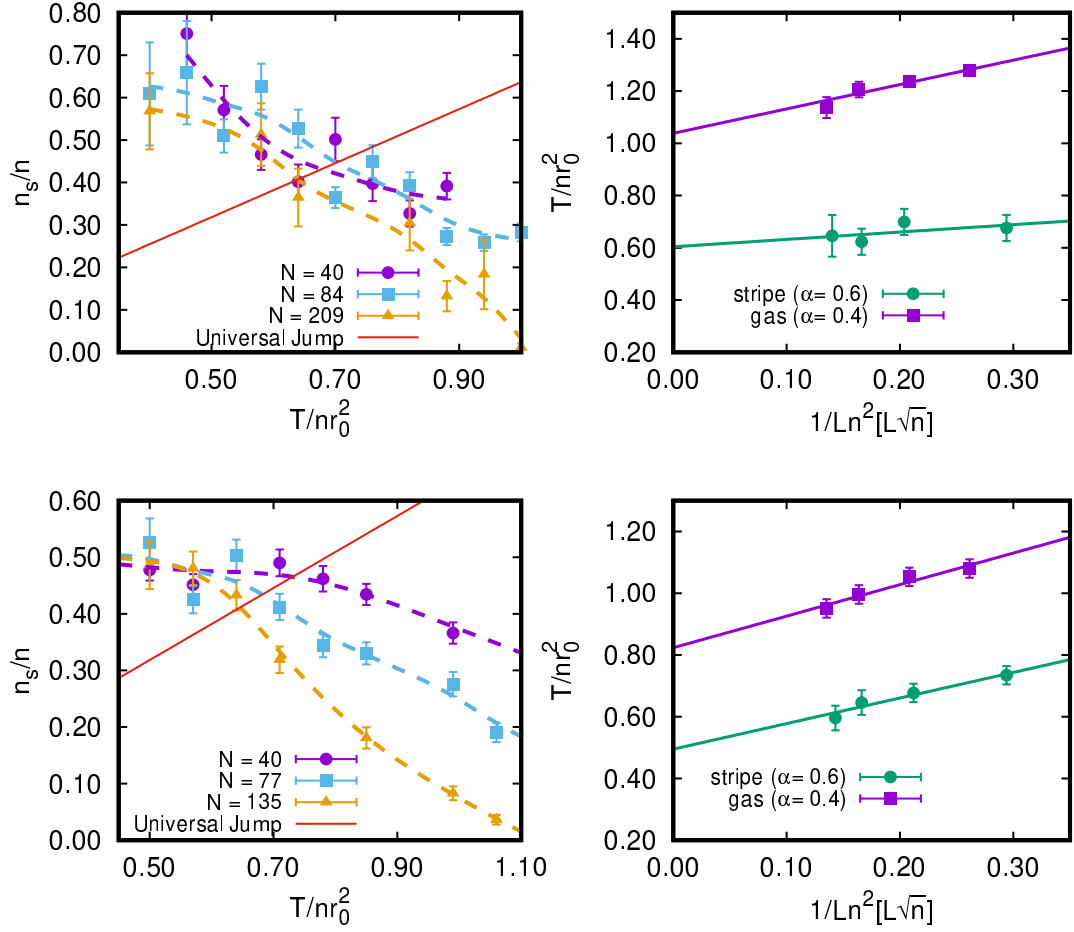

Fig. 3.7 Left panels: superfluid fraction as a function of temperature for different system sizes at densities $nr_0^2 = 128$ and $256$ (top and bottom respectively and tilting angle $\alpha = 0.6$, corresponding to the stripe phase. Points are PIMC results, dashed lines are guides to the eye, and the solid line is the universal jump (3.10). Right panel: scaling of the critical temperature $T_c(L)$ with the system size, as given by Eq. (3.12), at the same densities and two tilting angles: $\alpha = 0.4$ (gas, point M and N in Fig. 3.1 ) and $0.6$ (stripe, points K and J in Fig. 3.1 ). Points are PIMC data and solid lines are linear fits.

| Gas Phase | | | | | | | |
|---|---|---|---|---|---|---|---|
| $nr_0^2$ | $\alpha$ | $T_c/nr_0^2$ [$\epsilon_0$] | $n_s/n(T_c)$ | $nr_0^2$ | $\alpha$ | $T_c/nr_0^2$ [$\epsilon_0$] | $n_s/n(T_c)$ |
| 0.01 | 0.0 | 1.316(6) | 0.838(4) | 25 | 0.0 | 1.282(8) | 0.816(6) |
| 0.01 | 0.2 | 1.317(3) | 0.838(6) | 25 | 0.2 | 1.292(5) | 0.823(4) |
| 0.01 | 0.4 | 1.29(11) | 0.821(6) | 25 | 0.4 | 1.322(1) | 0.842(3) |
| 0.01 | 0.6 | 1.263(13) | 0.804(8) | 25 | 0.6 | 1.347(3) | 0.858(2) |
| 128 | 0.4 | 1.04(4) | 0.66(3) | 256 | 0.4 | 0.82(3) | 0.52(2) |

| Stripe Phase | | | | | | | |
|---|---|---|---|---|---|---|---|
| $nr_0^2$ | $\alpha$ | $T_c/nr_0^2$ [$\epsilon_0$] | $n_s/n(T_c)$ | $nr_0^2$ | $\alpha$ | $T_c/nr_0^2$ [$\epsilon_0$] | $n_s/n(T_c)$ |
| 128 | 0.6 | 0.56(7) | 0.38(4) | 256 | 0.6 | 0.49(4) | 0.31(3) |

Table 3.3 BKT critical temperatures, in dipolar units, for different values of the density $nr_0^2$ and tilting angle $\alpha$, and in both the gas and stripe phases. The superfluid fraction at the critical temperature is evaluated through Eq. (3.10). Numbers in parenthesis are the estimated errors.

| $T/nr_0^2$ | $T/T_{BKT}$ | $n_s/n$ | $\eta$ |
|---|---|---|---|
| 0.46 | 0.76 | 0.50(8) | 0.147 |
| 0.64 | 1.06 | 0.33(4) | 0.31(3) |
| 2.34 | 3.90 | 0 | - |

Table 3.4 Superfluid fraction and critical exponent evaluated at a density $nr_0^2 = 128$ and tilting angle $\alpha = 0.6$ with 209 particles.

Using the expression for the universal Jump (see Eq. (3.10)) it can be seen that the maximum value of this exponent ($\eta = 1/4$) takes place for the critical temperature. In Table 3.4 we summarize its values for different temperatures based on our calculations for the system at density $nr_0^2 = 128$ and tilting angle $\alpha = 0.6$ with 209 particles. We also show in that table a case in which the system is in the normal stripe phase ($T/nr_0^2$=2.34). In Fig. 3.8, we show the OBDM evaluated for the same temperatures as in Table 3.4, where the different algebraic and exponential decay behaviors can be seen above and below the BKT critical temperature. The thermal OBDM matrices have been computed in PIMC with the worm algorithm, taking advantage of the estimator of Eq. (2.156). For the two lower temperatures in the plot, the long-range behavior of the OBDM is well captured with an algebraic fit of the form shown in Eq. (3.14), with the exponent $\eta$ given by the BKT theory. On the other hand, an exponential fit is required to accurately describe the long-range behavior of the OBDM at the highest temperature. It is important to remark that in figure 3.8 we are only plotting the angular average (on angle $\theta$) of the OBDM, which is enough to study the long-range behavior of the OBDM.

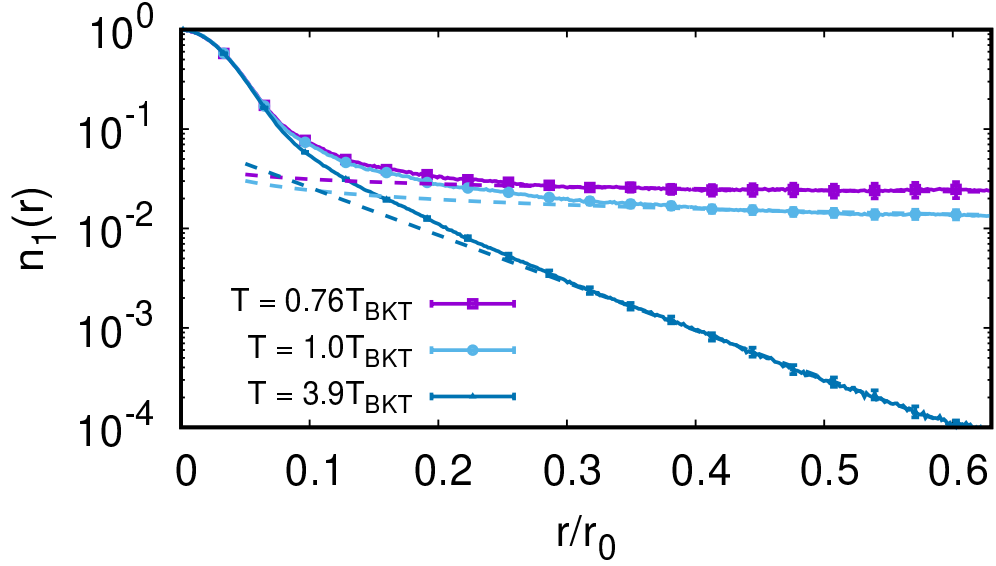

Fig. 3.8 One-body density matrix of the stripe phase ($nr_0^2 = 128$ and $\alpha = 0.6$) at different temperatures, above and below the transition temperature $T_{BKT}$. The straight lines correspond to fits to the asymptotic behavior when $r \to \infty$.

**Stripe Melting**

Similarly to what happens with a solid, as temperature increases, one expects that the spatial order of the stripe phase would disappear and, thus, a transition from the normal stripe phase (non-superfluid) to a gas should occur. Up to now, we have shown how temperature destroys the off-diagonal long-range order in the supersolid stripe phase; in what follows, by studying its structural properties, we show that if temperature keeps increasing, the diagonal long-range order is also destroyed. One good quantity to study this change from stripe to gas, is the static structure factor evaluated along the $Y$ direction (that is perpendicular to the stripe direction). To this aim, we employ the definition for the static structure factor that was introduced in Eq. (2.149), that in the PIMC framework reads

$$S_y(\mathbf{k}) = \frac{1}{NZ} \langle \hat{\rho}_{-\mathbf{k}_y} \hat{\rho}_{\mathbf{k}_y} \rangle \ . \tag{3.16}$$

By evaluating the strength of the main peak that appears on the static structure factor, we can study how thermal fluctuations destroy the spatial order along the $Y$ direction. This is shown in figure 3.9(for $nr_0^2 = 128$ and $\alpha = 0.6$) where it can be seen that the height of the peaks in the superfluid and normal stripe phases are compatible ($T = 0.93\,T_{BKT}$ and $T = 2\,T_{BKT}$), while it is clearly suppressed as temperature increases further , to almost vanish for $T \simeq 10\,T_{BKT}$. This is a clear symptom of the melting to a gas phase.

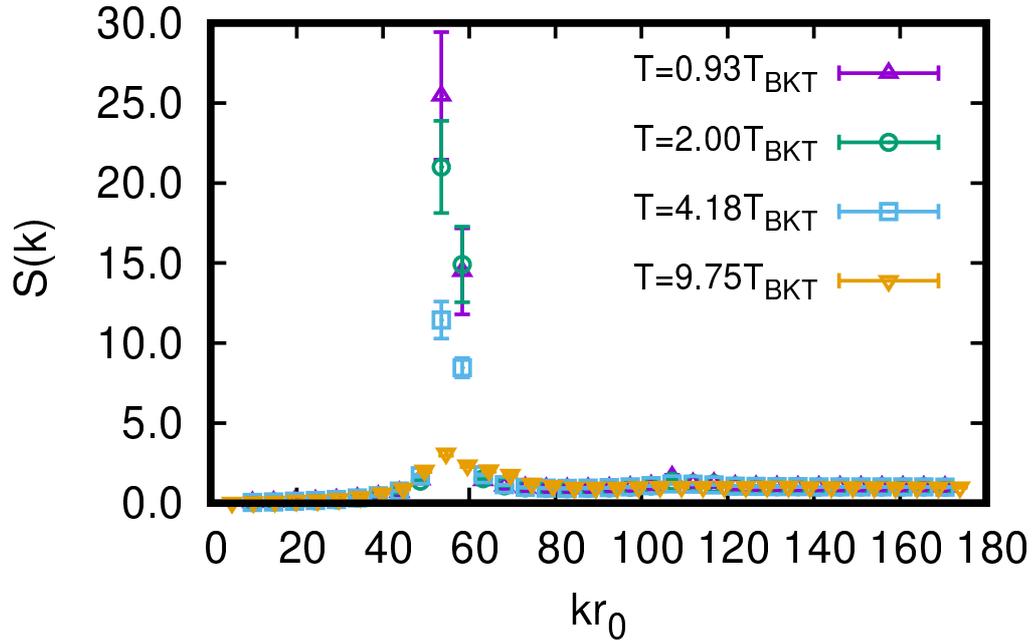

Fig. 3.9 Evolution with the temperature of the static structure factor $S_y(k)$ in the stripe phase with temperature for density $nr_0^2 = 128$ and tilting angle $\alpha = 0.6$.

The evolution of the stripe structure can also be qualitatively analyzed by looking at the spatial distribution of particles in the PIMC simulation. Figure 3.10 shows snapshots of PIMC simulations at the same temperatures of Fig. 3.9. As it is known, in the PIMC formalism, each particle is represented by a polymer, which helps to visualize its quantum delocalization. At temperatures below $T_{BKT}$, the snapshots reveal that there are paths connecting different stripes; when these crossing paths are of the length of the simulation box there is a nonzero winding number in that direction. When this is the case the superfluid fraction is finite. In the second frame of Fig. 3.10, these transverse paths have nearly disappeared and their length is shorter than the box side. Also in the $X$ direction the interconnections are not so abundant, and particles seem to be more localized. In the third frame, we still observe some reminiscence of the characteristic stripe order, although the presence of dislocations between the different stripes is clear. The appearance of these dislocations in the stripe phase at relative high temperature has been deeply studied in Refs. [170, 187]. Finally, the last frame corresponds to a temperature where the melting to a gas phase leads to the disappearance of the spatial order that is characteristic of the stripes.

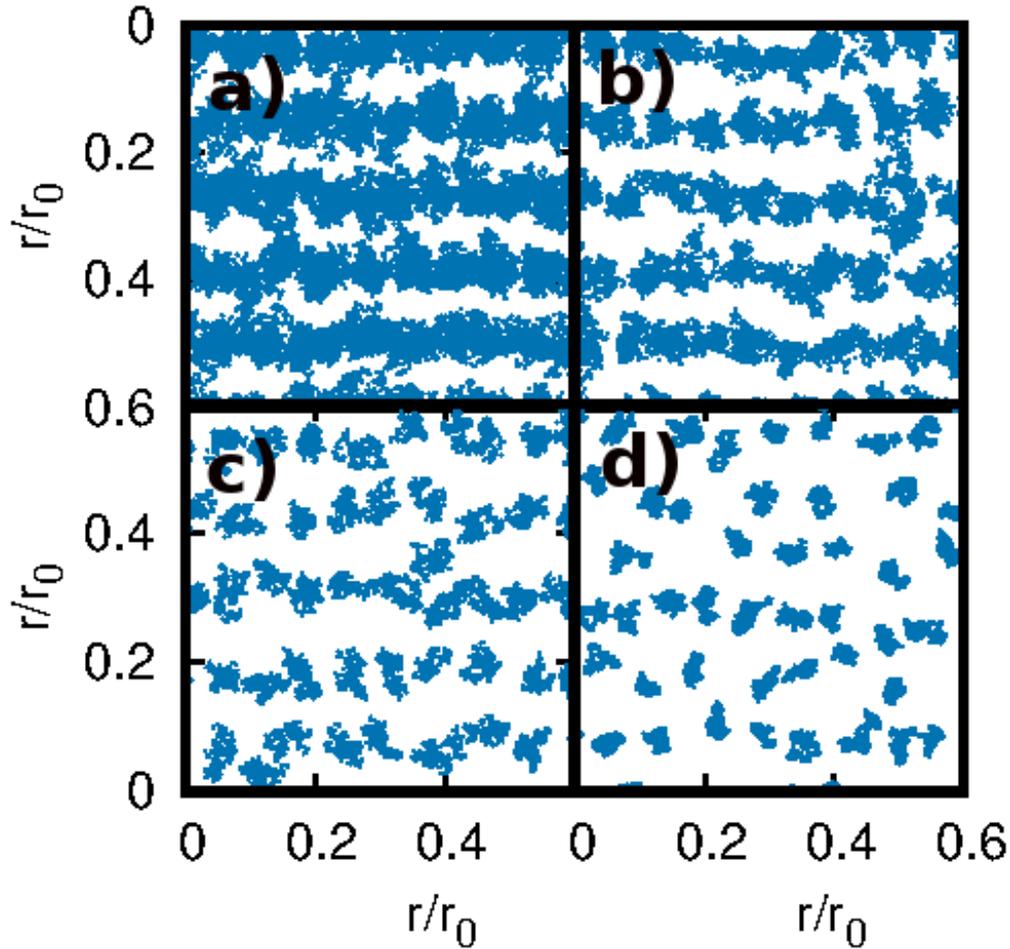

Fig. 3.10 Snapshots of the PIMC simulations of the stripe phase ($nr_0^2 = 128$, $\alpha = 0.6$) for increasing temperatures. The temperature increases from a) to d) panels. The values of $T$ are the same than in Fig. 3.9.

## 3.5 Quasi 1D behavior?

A relevant concern is whether the stripes could be described as an ensemble of 1D systems. If that was the case, particles inside each of the stripes would be so confined that interchanges among them would not be possible. Although both our results at zero and finite temperature indicate that this is not the case, here we want to show it explicitly. To this end we compare our QMC results for the stripe phase with the predictions of the Luttinger liquid (LL) theory, both at zero and finite temperature.

### 3.5.1 Zero Temperature

We start this 1D analysis at zero temperature. In the Luttinger Liquid theory, the system can be completely described by the sound velocity $c$. This parameter can be extracted from different observables of the system and, if the Luttinger liquid theory reproduces the physics of the system, any evaluation of $c$ should give the same value, no matter from what observable it is extracted. Here we focus on two 1D observables [188]. The first one is the static structure factor, that for low momentum, shows a linear dependence with $k$ of the form

$$S(\mathbf{k}) = \frac{\hbar^2}{2mc}k, \qquad k \to 0; \tag{3.17}$$

The other observable in which we are going to focus is the OBDM, whose asymptotic long distance behavior in the 1D LL model

$$n_1(\mathbf{r}) = Ar^{1/\eta_{LL}}, \qquad r \to \infty. \tag{3.18}$$

where $\eta_{LL} = \frac{\hbar^2}{m} \frac{2\pi n_l}{c}$ and $n_l$ the mean linear density along the stripes. The above expressions give $n_1 \to 0$ as $r \to \infty$, as it should be for a one-dimensional system where off-diagonal long-range order is not allowed, not even at zero temperature. Although as we have shown in section 3.3 the stripe phase presents a finite condensate fraction at zero temperature and hence, its asymptotic long-distance behavior can not be reproduce by a fit of the form of Eq. (3.18), in Fig. 3.11 we focus on the behavior of the OBDM in the intermediate distances regime. In that figure we show the OBDM computed along the $Y$ (green stars) and $X$ (purple open squares) directions at $\alpha = 0.6$ and $nr_0^2 = 512$. The solid lines are fits of the form $Ax^{-1/\eta_{LL}}$ for fixed $\eta_{LL}$ obtained from the slope of the static structure factor near the origin. These results show that the decay of the OBDM (in the intermediate distances regime) along the stripe direction is well captured by the LL approach. This fact hints a reminiscence of the Luttinger liquid that a pure 1D dipolar system would exhibit (for a study of the properties of 1D dipolar system see Ref. [189]). On the other hand, in the same figure we show that the same theory, applied to the OBDM along the $Y$ direction, fails to reproduce the PIGS results. The inset in Fig. 3.11 shows a snapshot of the system after thermalization in PIGS, for the same conditions $nr_0^2 = 512$ and $\alpha = 0.6$, where a pair of examples of particle exchange between different stripes are visible and have been highlighted. It is worth recalling that since simulations in PIGS are done with open chains (with variational wave functions at the end points), it is hardly possible to see long exchange lines crossing the whole simulation box when using this method.

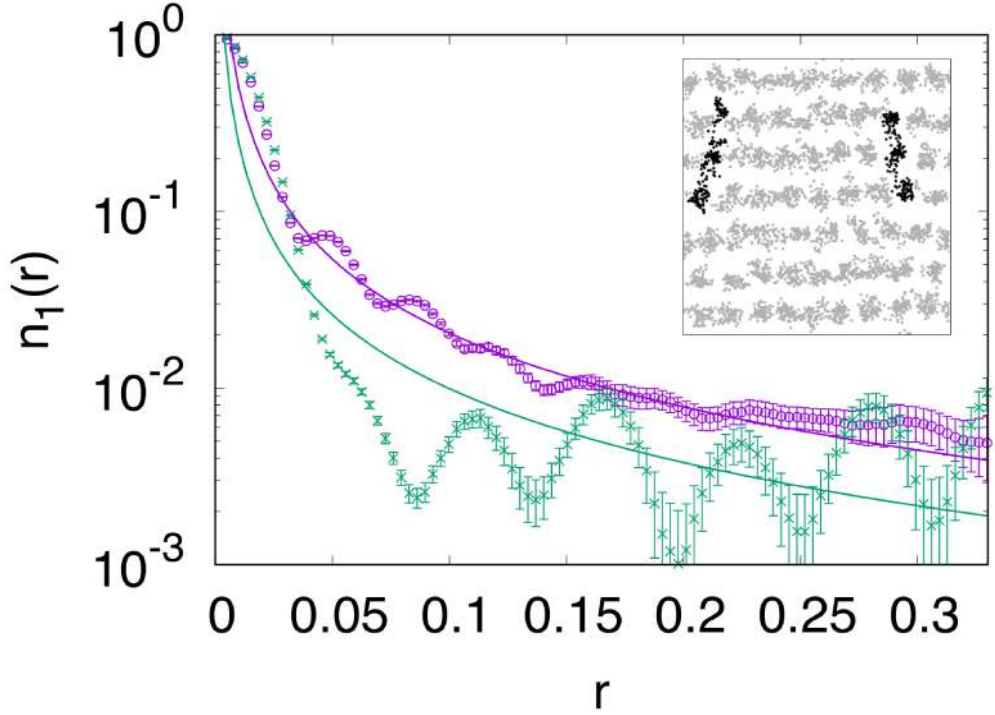

Fig. 3.11 . One-body density matrix along the $Y$ (green stars) and $X$ (purple open squares) directions at $\alpha = 0.6$ and $nr_0^2 = 512$. The solid lines are fits of the form $Ax^{-1/\eta}$ for fixed $\eta_{LL}$ obtained from the slope of the static structure factor near the origin. The inset shows a snapshot of the PIGS simulation, where some of the particle exchanges are highlighted in black.

### 3.5.2 Finite Temperature

Although a one dimensional system do not show superfluidity in the thermodynamic limit, a superfluid fraction can still appear as a finite size effect in system of length $L$. In this subsection we show that our results for the stripe phase of the dipolar system are not compatible with such a behavior.

In the framework of the Luttinger liquid theory, the superfluid fraction is predicted to scale with the system size $L$ and temperature $T$ as [80]

$$\frac{n_s}{n} = \frac{\gamma}{4} \frac{\left|\Theta_3''(0, e^{-\gamma/2})\right|}{\Theta_3(0, e^{-\gamma/2})}, \tag{3.19}$$

where $\Theta_3(z, q)$ is the Theta function, $\Theta_3''(z, q) = d^2\Theta_3(z, q)/dz^2$, and $\gamma = \frac{mk_bTL}{\hbar^2 n_l}$.

In figure 3.12, we show our PIMC results for the superfluid fraction evaluated for different system sizes $L$ and temperatures $T$ keeping the density fixed. They correspond to the points K ($nr_0^2 = 128$, $\alpha = 0.6$, left) and J ($nr_0^2 = 256$, $\alpha = 0.6$, right) in the phase diagram of figure 3.1, which is the same data set shown in Fig. 3.7 and used to extract the BKT transition temperature. It is obvious from this graph that the PIMC points do not collapse to a single law, as it would be the case of a 1D system (or a collection of them), and then we can assert that our data is not compatible with a one-dimensional theory. Furthermore, our PIMC points are shifted to larger values of $\gamma$ when compared to the LL prediction of Eq (3.19) (black line), showing that the superfluid signal in the stripe phase is more robust, against the product of system size and temperature $LT$, than it is in the limit of a 1D system described by the Luttinger Liquid theory.

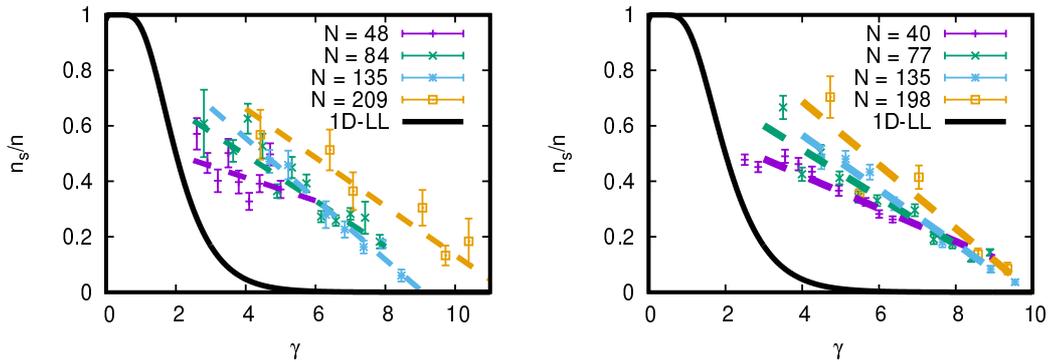

Fig. 3.12 Superfluid fraction of the stripe phase for different number of particles and as a function of the scaling parameter of Luttinger theory, essentially $LT$. The Luttinger Liquid prediction of Eq (3.19) is represented by the black line. Left (right) panel data correspond to points K ($nr_0^2 = 128$, $\alpha = 0.6$, left) and J ($nr_0^2 = 256$, $\alpha = 0.6$, right) in the phase diagram of Fig. 3.1. Dashed lines are lines to the eye.

## 3.6   Summary

Summarizing, in this chapter we have studied the two-dimensional fully-polarized system of bosonic dipoles. In particular we have performed a systematic study of the superfluid properties in the different phases that are present in the phase diagram of the system. In a first approximation to the problem, the employment of exact zero temperature QMC techniques allows us to determine that both the gas and stripe phases of the this system have finite superfluid and condensate fractions at $T = 0$. This study is complemented by the characterization of the superfluid to normal phase transition, that in 2D is of the BKT type. Furthermore, we have shown that the same BKT scaling applies for the gas and the stripe phases, no matter of the anisotropy of the system or the spatial long-range order present in the stripe phase. The scaling of the superfluid fraction with the temperature and system size, allows us to determine the critical temperature at which the BKT transition occurs. In the case of the stripe phase, a second phase transition occurs between the classical stripe and the gas phase, similarly to what happens with usual solid like phases. To give an estimation of the temperature at which this transition occurs $T_F$, we have focused on the study of the structural properties of the system trough the evaluation of the static structure factor. We find that the spatial order in the stripe stands up to a temperature $T_F$ that is roughly ten times larger than that at which the superfluid signal is loss ($T_{BKT}$). In the last part of the chapter we have compared our MC results for the stripe phase to the Luttinger liquid model, that reproduces the physics of quantum 1D systems at low temperatures. This comparison excludes the possibility of describing the stripe phase as an ensemble of 1D isolated systems.

# Chapter 4

# Two-component Fermi dipoles in two dimensions

In this chapter we study a two-component fermionic system of dipoles in two dimensions at zero temperature. We focus on the case in which all the dipoles are polarized along the perpendicular direction to the plane containing their movement, what makes the interaction between them isotropic. In section 4.3, we perform a study of the equation of state (EOS) of the unpolarized phase at low density. To this purpose, we employ the DMC method with the Fixed-Node (FN) approximation, that allows to perform calculations of fermionic systems avoiding the *sign problem* (*cf.* section 2.3.5 for details about the method). We compare our results in the weakly interacting regime to those of a Hard-Disk model [190] in order to determine the regime of universality. This study is completed in section 4.4 with the calculation of the Equation of State (EOS) of the system at higher densities, and discuss about the possible existence of a ferromagnetic phase as the ground state before the crystallization point. If that was the case, it would constitute an example of the *itinerant ferromagnetism* phenomena, that has been the object of discussions for quite some time in the context of ultracold fermionic gases with short-range interactions [29–36] until the recent claim for its experimental observation [28]. Itinerant ferromagnetism is so computationally demanding that it stresses state-of-the-art FN implementations. In particular, we show that the usual two-body backflow correlations are not enough to accurately describe this system. In the last part of this chapter (section 4.5), we study the repulsive Fermi polaron; defined as the physics of a single impurity immersed in a Fermionic bath with repulsive impurity-bath interaction. Similarly to the previous case, all the atoms in the system are polarized such that the interaction is isotropic, and we use the comparison with a short-range hard-disk model to study the universality of the problem in terms of the gas parameter $na_s^2$. We also report observables that allow to discuss the validity of the quasi-particle picture: the quasi-particle residue and the effective mass of the polaron.

## 4.1 Introduction

Since the first realization of a degenerate Fermi gas in the context of ultracold atoms [19], a lot of effort has been put in order to control these systems and make them an useful platform to study phenomena that are characteristic of fermionic systems. Here, we analyze some of these phenomena, without the purpose of performing an exhaustive review of Fermi degenerate gases, but with the aim of showing the relevance that their study has nowadays in the field of ultracold atoms.

The issue of superfluidity, that we have discussed in the previous chapter for bosonic systems, is more subtle when fermionic species are involved. Indeed, superfluidity in Fermi gases requires the existence of a condensate of pairs such as it was first described in the Bardeen-Cooper-Schrieffer (BCS) theory. This has been widely studied in experiments [25], not only in the spin-balanced case but also in the general case in which spin imbalance is present, what enriches the physics of these systems. Particularly the imbalance between the different spin components gives place to a situation in which not all the system is in the superfluid phase, but only a fraction of it. In this situation, new phenomenology has been studied as phase separation between the superfluid and the normal phase [191]. Indeed, the study of the pairing mechanism in these systems has caught also some attention both from the experimental [191, 192] and the theoretical point of view (*cf.* [193] for a Monte Carlo study and [194] for a mean-field (MF) approach to the phase diagram of this system).

Another interesting phenomena, related to the previous one, that has been demonstrated by studying Fermi gases close to a Feshbach resonance is the BEC-BCS crossover. This crossover occurs between a superfluid state of low-momenta Cooper pairs and the BEC condensation of molecular pairs as the interaction parameter is varied from the weakly to the strongly interacting regime, what permits to study the pairing mechanism in very different regimes (*cf.* [21–25] for experiments and [26, 27] for Monte Carlo studies)

Although it is the main subject of sections 4.4 and 4.5, it is worth to mention in this introduction two problems that have deserved a lot of attention in the previous years. The first one, the possible existence of a stable, fully polarized phase, is a longstanding topic in the field. Although its first prediction was related to the electron gas in three dimensions [195–201], a lot of theoretical work has been done to study it in ultracold gases. Finally in 2017, a claim for its observation was reported in a trapped system [28] (See section 4.4, for a discussion of this problem in two-dimensional dipolar systems). The second one is the Fermi polaron. In fact, the polaron problem was first put forward by Landau and Pekar [202] to study the properties of an electron embedded in a lattice and its interaction with its phonon modes. Later on, this problem has been generalized to other situations, as the case in which the medium is a fermionic bath. This particular case is discussed in section 4.5 for the two-dimensional dipolar system.

Studying quantum systems in reduced geometries is also of interest because quantum correlations are enhanced in them. In two dimensions the equation of state for the attractive and repulsive branches have been characterized [190] employing QMC for a hard-disk system. In this two-dimensional geometry all the above commented problems have been studied: the BEC-BCS crossover [203], itinerant ferromagnetism [204, 205] and two-dimensional Fermi polarons [206, 207].

Once experiments with fermionic dipolar atoms [43, 208–210] and molecules [211] have been achieved [43, 208–210], access to new phenomenology has came out. In particular it has been shown that the anisotropy, characteristic of dipolar interaction can induce the deformation of the Fermi surface in a gas of Er atoms [45]. The case of low concentration fermionic impurities into a bosonic bath has been studied both in dipolar condensates [212] and dipolar droplets [77], but to the best of our knowledge, the case of the dipolar Fermi polaron (an impurity interacting with a dipolar potential with a Fermi bath), remains unstudied.

Actually, 2D dipolar systems have been the object of many studies due to the tunability of anisotropic effects, which can be varied by polarizing the atoms or molecules along different directions. In what concerns to fermionic systems, the fully polarized state in the particular case in which all the dipoles are polarized in the perpendicular direction to the plane containing them was studied in Ref. [93] employing DMC. In this case, the gas to solid phase transition was characterized.

Other works in which bi-layer systems are analyzed constitute an extension of exact 2D studies. The inclusion of the second layer makes the anisotropic character of the dipolar interaction to be present even in the case in which the dipoles are polarized in the direction perpendicular to the layers, which enriches the phase diagram. In particular the impurity problem has been characterized in this geometry, showing a crossover, as a function of the inter-layer distance, from a free quasi-particle regime to a different one where localization effects appear [213]. The BEC-BCS crossover has been also studied in this geometry where the pairing mechanism is enhanced by the attractive part of the dipolar interaction [214]. In both cases, the QMC results show deviations from perturbation and mean-field schemes. Dipolar fermions have also been studied in a multi-layer geometry, using a mean-field approximation, where both the crystallization point and the formation of stripes are discussed [215].

In the following, we present the QMC studies that we have performed regarding two-dimensional dipolar systems of fermionic species. Our calculations complement previous studies about dipolar system, in particular, we discuss the universality of the equation of state of the balanced Fermi mixture, the possibility of finding an itinerant ferromagnetic phase, and the dipolar Fermi polaron.

## 4.2 The system

We consider a system of dipolar atoms confined in a two-dimensional geometry. In particular we consider the case in which all the components have the same mass and the same dipolar moment. The system is polarized along the normal direction to the plane containing the atoms, what makes the interaction isotropic. The Hamiltonian of the $N$-particle system reads

$$H = -\frac{\hbar^2}{2m} \sum_{i=1}^{N} \nabla_i^2 + \frac{C_{dd}}{4\pi} \sum_{i<j} \frac{1}{r_{ij}^3}, \tag{4.1}$$

which is in fact the same in equation (3.2) for $\alpha = 0$. We use the dipolar units that were introduced in the previous chapter ($r_0 = mC_{dd}/(4\pi\hbar^2)$ and $\epsilon_0 = \frac{\hbar^2}{mr_0^2}$, see section 3.2), so that we can write the Hamiltonian in dimensionless form. The other remarkable difference between this system and the one of the previous chapter is that now we work with two component fermions, with each component labeled as $\{\uparrow, \downarrow\}$ in analogy with a spin-1/2 Fermi system. The population imbalance between the two fermionic species is encoded into the polarization

$$P = \frac{N_\uparrow - N_\downarrow}{N_\uparrow + N_\downarrow}, \tag{4.2}$$

and is kept fixed during the simulation. In the above expression $N_\sigma$ with $\sigma \in \{\uparrow, \downarrow\}$, represents the number of particles of each specie present in the system. In particular in section 4.3 we fix the polarization to zero to study the low density equation of state, while in section 4.4 we compute the two extreme cases of $P = 0$ and $P = 1$. Finally, in section 4.5, in order to characterize the dipolar Fermi polaron, we study the limit $P \to 1$.

## 4.3 Unpolarized system at low density

In this section, we evaluate the equation of state of the 2D dipolar Fermi mixture. The single component bosonic dipolar system was already studied in a previous work [84], where it was shown that the mean-field approach fails to reproduce accurately the energy of the system, even at densities that are much lower than those where the limit of universality is expected to hold. In particular, for densities as low as $10^{-100}$, deviations in the energy of about 1% are found. This is a well known fact of two-dimensional systems and several beyond mean-field corrections have been proposed and tested for the bosonic case [216, 217, 84]. Here, we also discuss discrepancies with the mean-field approach for the two-component Fermi system, although the precision needed to discuss beyond mean-field effects accurately exceeds the scope of this work. The fully polarized equation of state for the dipolar fermionic system has also been studied by means of Monte Carlo, from the low density regime to the crystallization phase transition [93].

### 4.3.1 Details about the method

For the study of the fermionic dipolar system at zero temperature, we use the DMC method. As it was explained in chapter 2, performing calculations with fermions is harder than with bosons because one has to deal with the well known sign problem. To tackle this problem we use the Fixed-Node (FN) approximation. In order to implement this method one has to choose a trial wave function with a known nodal surface, and therefore the method becomes variational (see section 2.3.5 for details about the FN technique).

The wave function that we use for importance sampling when studying a fermionic system is the product of a symmetric $\Psi_S$ and an antisymmetric $\Psi_A$ terms

$$\Psi_T(\mathbf{R}) = \Psi_A(\mathbf{R})\Psi_S(\mathbf{R}). \tag{4.3}$$

In particular for the calculations of the system in the low density regime we use a Jastrow-Slater wave function. In this case the symmetric part is a product of Jastrow factors

$$\Psi_S(\mathbf{R}) = \prod_{i<j}^{N} f(\mathbf{r}_{ij}). \tag{4.4}$$

Similarly to what was done in the previous chapter to study the bosonic system, the Jastrow factor is constructed from the zero energy solution of the two-body dipolar problem, and is it matched at a certain distance $r_M$ to a phononic solution [91, 183]

$$f_J(r) = \begin{cases} AK_0\left(2\sqrt{\frac{r_0}{r}}\right) & r < R_M, \\ B\exp\left[-C\left(\frac{1}{L-r} + \frac{1}{r}\right)\right] & r > R_M, \end{cases} \tag{4.5}$$

where $K_0$ is the modified Bessel function and $R_M$ is taken as a variational parameter. The constants $A$, $B$ and $C$ are fixed by imposing the conditions of continuity and derivability at $r = R_M$, together with the conditions $f(|\mathbf{r}| > L/2) = 1$ and $f'(|\mathbf{r}| > L/2) = 0$.

For the antisymmetric part, we consider the product of two Slater determinants, one for each species ($D_\uparrow$ and $D_\downarrow$):

$$\Psi_A(R) = D_\uparrow\left(\mathbf{x}_1, \ldots, \mathbf{x}_{N_\uparrow}\right) \times D_\downarrow\left(\mathbf{x}_{N_\uparrow+1}, \ldots, \mathbf{x}_N\right). \tag{4.6}$$

Finally, as we are interested in studying the properties of the 2D dipolar Fermi system in the thermodynamic limit, we perform our DMC simulations inside a box with Periodic Boundary Conditions (PBC). The length of the box $L$ is fixed by the density of the system $n$ and the total number of particles $N$ equals $nL^2$. This allows also to define the partial densities for each component

$$n_\sigma = N_\sigma/L^2 \tag{4.7}$$

with $\sigma \in \{\uparrow, \downarrow\}$, so that $n = n_\uparrow + n_\downarrow$.

**finite-size effects treatment**

Although with our calculations we want to reproduce the physics of the infinite system, our simulations are performed with a fixed number of particles $N$ in a box of length $L$. This issue introduces undesired *finite-size effects* in the calculation. There are two finite-size effect corrections that we include in the energy with the aim of reproducing correctly the thermodynamic limit.

1. **Fermi kinetic correction**: It is a well known fact that the energy per particle of the Ideal Fermi Gas (IFG) has a non-monotonic behavior with the number of particles. This behavior appears also when treating with for interacting systems, as the ones that are studied in this Thesis. For this reason, in our calculations we subtract the following quantity to our DMC energies

$$\Delta E_{IFG}^N = E_{IFG}^N - E_{IFG} \tag{4.8}$$

where $E_{IFG}^N$ is the energy of an IFG composed of a finite number of particles $N$, and $E_{IFG}$ is the energy per particle of the infinite IFG. This correction has been used to study different fermionic systems (see, for instance, Ref. [33]).

2. **Evaluation of the potential tail**: The dipolar potential has a slow decaying tail, and thus an estimation of its tails outside the simulation box is needed. In our finite-size simulations we usually cut the range of the potential at a distance $R_{cut} = L/2$ with $L$ the length of the box, and assume (as and approximation) that there are no correlations for distances larger than $R_{cut}$. Then we introduce a potential correction for the energy as follows

$$\Delta E_{tail} = \frac{1}{2} \int_{L/2}^\infty \frac{1}{r^3} n(r) g(r) 2\pi r dr = \frac{2\pi n^{3/2}}{\sqrt{N}} \tag{4.9}$$

where for the last equality we have assumed that the system is homogeneous ($n = cte.$) and that the distance $L/2$ is large enough compared to the inter-particle correlation distance so that we can consider $g(r \geq L/2) = 1$.

### 4.3.2 Comparison with approximate theories

In order to benchmark our QMC calculations in the weakly interacting regime ($nr_0^2 \ll 1$), it is worth to compare them with other approximated theories. The simplest model that we can compare with is the Ideal Fermi Gas (IFG), that corresponds to setting $C_{dd} = 0$ in the Hamiltonian of equation (4.1). The ideal fermi gas energy per particle $E_{\text{IFG}}$ is given by

$$\frac{E_{\text{IFG}}}{\hbar^2/m} = \frac{1+P}{2}\pi n_\uparrow + \frac{1-P}{2}\pi n_\downarrow = \frac{\pi n}{2}\left(1+P^2\right), \qquad (4.10)$$

in the thermodynamic limit, with $P$ the polarization of the system, as defined in Eq. (4.2), and the partial densities, as defined in Eq. (4.7). The IFG energy is only a good approximation in the limiting case in which dipolar interaction is negligible compared to the Fermi kinetic contribution to the energy. In order to improve over this result, the contribution coming from the dipolar interaction can be approximated in the low density regime. With this aim it is helpful to use the Hartree-Fock (HF) approximation that, for the $P = 1$ case, gives an approximate expression for the energy per particle of the dipolar system [218]. In units of the IFG energy of Eq. (4.10), the HF scheme energy reads

$$\frac{E_{\text{HF}}}{E_{\text{IFG}}} = \frac{256}{45\sqrt{\pi}}\sqrt{nr_0^2}, \qquad (4.11)$$

obtaining a higher order approximation is possible through many-body perturbation theory [218]. Unfortunately, the same HF approximation cannot be applied to obtain an approximate result for the two-component mixture ($P = 0$). The reason is that the Fourier transform of $1/r^3$ in two dimensions is ill-defined, although some regularization schemes can be used, as it was done in Ref. [219]. The derivation of the result in Eq. (4.11) is possible due to the strong cancellation that occurs between the direct and exchange contributions to the Hartree-Fock energy. However, this cancellation is absent when two different fermionic species are considered. For this reason, the Hartree-Fock approximation can only be obtained for dipolar gases in the limit $P = 1$.

When the HF theory cannot be applied, we use a simpler mean-field approach that is also valid in the limit $nr_0^2 \ll 1$, where the properties of the gas do not depend on the microscopic details of the inter particle interaction but only on the two-dimensional scattering length $a_s$. It consists in replacing the dipolar repulsion between particles of different species with a zero-range interaction. For the $1/r^3$ repulsion, the two-dimensional scattering length in dipolar units reads [186, 183]

$$a_s = r_0 e^{2\gamma}, \qquad (4.12)$$

with $\gamma \simeq 0.5772$ the Euler constant. Notice that the above expression can be recovered from Eq. (3.13) by setting $\lambda = 0$. In this way, the approximate expression that we use to evaluate the energy per particle of an unpolarized mixture ($P = 0$) is

$$E = E_{\text{IFG}} + E_{\text{MF}}, \tag{4.13}$$

as it is usually done to give an approximate description of fermions with zero-range interactions. In two dimensions, the mean-field interaction energy reads [220]

$$\frac{E_{\text{MF}}}{\hbar^2/m} = \frac{\pi n}{\left| \log \left( c_0 n a_s^2 \right) \right|}, \tag{4.14}$$

which depends on a free parameter $c_0$, that is related to the peculiarities of the scattering theory in 2D [221]. At odds to what happens in three and one-dimensional systems, the dependence of $E_{\text{MF}}$ on the gas parameter $n a_s^2$ is weak, as it enters in the coupling constant through the logarithm. Following Ref. [190], we set $c_0 = (\pi/2) \exp(2\gamma) \simeq 4.9829$, which corresponds to setting an energy scale equal to twice the Fermi energy of the unpolarized case. For the beyond-mean-field contribution, several expressions have been proposed and tested in the bosonic case [216, 217, 84].

It is important to remark that, in Eq. (4.13), $E_{\text{MF}}$ only accounts for interactions between particles of different species. Notice that, due to the Pauli exclusion principle, the Hartree-Fock contributions $E_{\text{HF}}$ for same-species repulsion is proportional to $n^{3/2}$ (see Eq. (4.11)), and in the limit $n r_0^2 \to 0$ it yields a subleading correction with respect to same species energy encoded in $E_{\text{MF}}$. Furthermore, the inclusion of $E_{\text{HF}}$ in Eq. (4.13) would constitute an uncontrolled approximation, since it is unknown how $E_{\text{HF}}$ would combine with the beyond-mean-field correction for the opposite-spin interaction energy. The derivation of different beyond mean-field contributions has been the object of several studies (see, for example Ref. [84]), however, their evaluation is beyond both the scope and the precision of our MC study.

As a final remark, the relation between the two-dimensional scattering length and the dipolar length of Eq.(4.12) is interesting if one wants to compare the results for the dipolar system to other models. In particular we use it to compare with a system of Hard-disks (HD)[1]. Details about this model can be find in appendix B.

### 4.3.3 Results

#### 4.3.3.1 Low density equation of state

In order to determine the EOS in the low density regime we evaluate the energies per particle of the balanced Fermi mixture, that is the $P = 0$ state, using the DMC method[2]. In Fig. 4.1 we show these energies for the dipolar system (Blue points) in units of the IFG energy as a function of the gas parameter. For all the range of densities that we have evaluated, the energy is above the IFG prediction, as it corresponds to a purely repulsive dipolar gas. In the inset of the same figure we plot the same energies but in units of Eq. (4.13). As it can be seen up to values of the gas parameter $na_s^2 \simeq 10^{-3}$, the mean-field prediction is accurate (solid red line), with relative differences that are of the order of 1%. On the contrary when we simply compare with the IFG energy (dashed black line), the relative difference grows up to 10% for the lower density considered. Clear deviations from the mean-field prediction start to appear for values of the gas parameters highers than $10^{-3}$.

---

[1] All the calculations for the Hard-disks system shown in this chapter, that we use to compare with the dipolar system, have been performed Gianluca Bertaina. They have been originally published in Refs. [190] and [56].

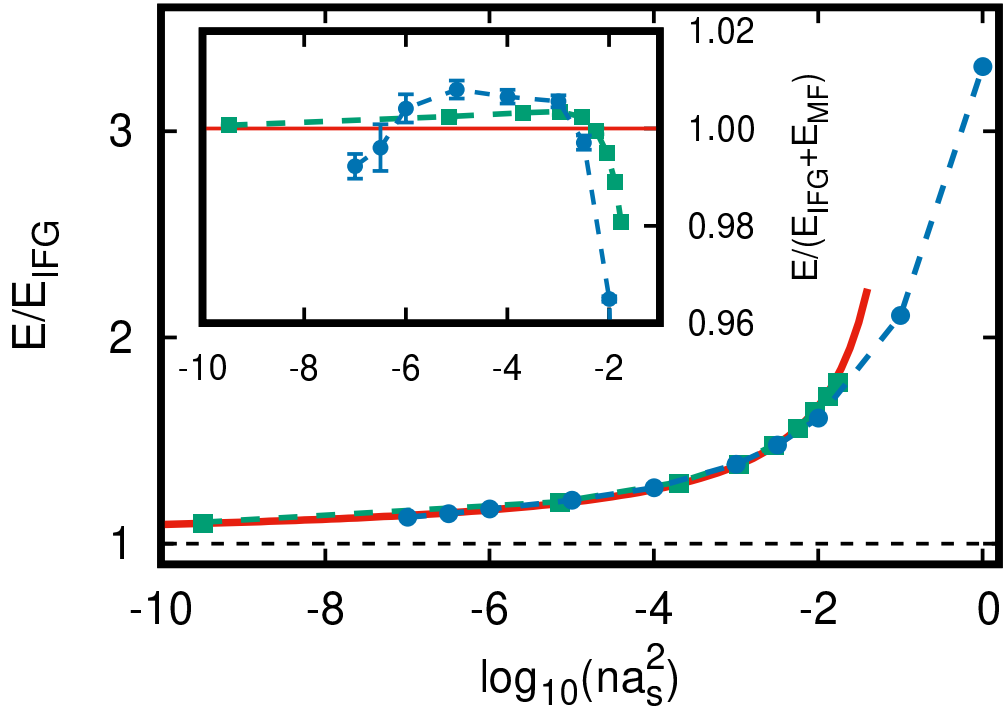

Fig. 4.1 Equation of state for the unpolarized system ($P = 0$) in units of the ideal-Fermi-gas energy $E_{\text{IFG}}$. DMC data for the dipolar gas (blue circles – blue dotted lines are a guide to the eye) are compared to the IFG and MF curves (black dashed line and red solid line) and to the DMC energy for hard disks (green squares, – green dotted lines are a guide to the eye from Ref. [190], with disk diameter $a_s = e^{2\gamma} r_0$). Inset: Same as in the main panel but in units of the MF equation of state of Eq. (4.13).

In the same figure, we show the results for the Hard-Disks model of Ref. [190]: The results show that both models predict the same energy for values of $nr_0^2$ up to $10^{-3}$, which puts a limit to the highest density at which universality holds. The same

---

[2]Most of the calculations regarding the dipolar system that are presented in the chapter have been done in collaboration with Tommaso Comparin, and have been originally published in Refs. [94] and [56]

comparison between the dipolar and the HD model was already performed for the equivalent bosonic system where it was found that the universal regime is reached for densities $nr_0^2 \leq 10^{-7}$ [216, 84], that is a regime in which the energy depends only on the gas parameter $na_s^2$ (Notice that, from Eq. (4.12), the conversion between the density in units of the two-dimensional scattering $a_s$ and in units of the dipolar length $r_0$ is roughly $a_s^2 \simeq 10r_0^2$). In our case, for the fermionic system we find that both the dipolar and the HD equation of state are quite similar, pointing to the existence of a regime of universality at low density. However, as it is shown in the inset of Fig. 4.1, residual differences between the two models can be identified. For the lower density considered the HD model seems to be well reproduced by the mean-field prediction, while the deviations of the dipolar model from the mean-field curve is systematically larger. Despite the dipolar potential in 2D can be formally contracted to a short-ranged model, it has a $1/r^3$ decay at long distances: This makes the dipolar system to reach universality at lower densities (see also the discussion about pair distribution functions in the following). Error bars include both statistical uncertainties and systematic errors, and they are smaller than the symbol sizes in the cases in which they are not shown.

Finally, it is worthy to point out that the Fixed-Node energies reported here have been computed employing plane waves as single particle orbitals in the Slater determinant of Eq. (4.6), which constitutes a good choice for the nodal surface at low densities. In section 4.4, we discuss the effects of improving the nodal surface by including Backflow correlations, and show that their effect is negligible in the regime $nr_0^2 \ll 1$. However, they become important when comparing the small energy differences between two phases with different polarization.

We have focused on the EOS of the non-polarized phase, as it is the ground state of the system in the low density regime. The fully polarized phase of the dipolar system was studied in Ref. [93] and the DMC equation of state was reported, showing that the Hartree-Fock prediction of Eq. (4.11) is a valid approximation to the ground-state energy at densities up to $nr_0^2 < 10^{-2}$. In the same work, it is also predicted the appearance of a solid phase at density $nr_0^2 \approx 50$.

#### 4.3.3.2 Radial distribution functions

To give a more complete description of the balanced mixture of dipolar fermions, we compute the same and different species radial distribution functions. For their computation we use the expressions in Eq. (2.147) for $g_{\uparrow\uparrow}(r)$ and $g_{\uparrow\downarrow}(r)$. It is also worth to compare our QMC results for the dipolar model in the low density regime to

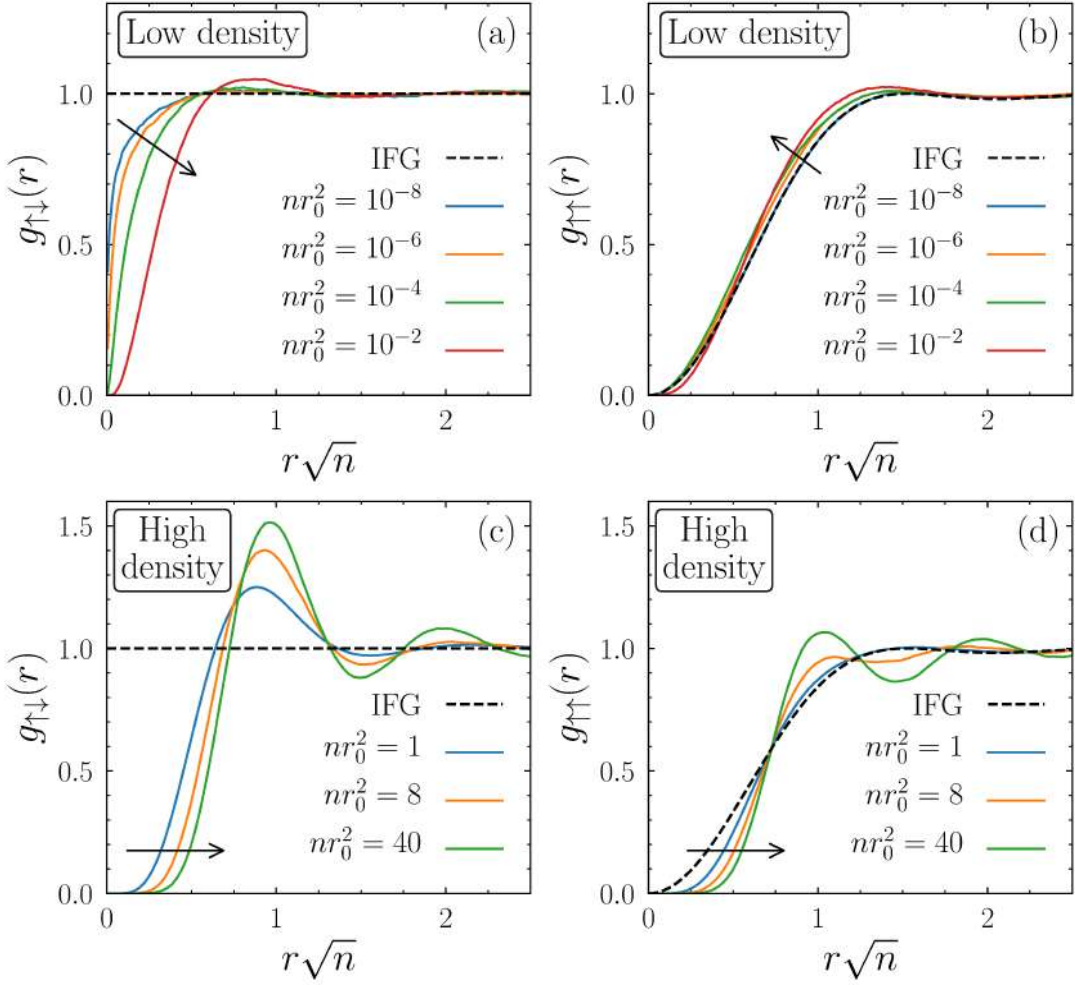

Fig. 4.2 Pair distribution functions for the $P = 0$ state, computed through DMC (solid lines), for different-species [panels (a) and (c)] and same-species pairs [panels (b) and (d)]. The analytic IFG distribution functions are also shown (dashed black lines). Arrows point towards increasing $nr_0^2$. The particle numbers are $N = 74$ (for $nr_0^2 < 1$), $N = 98$ (for $nr_0^2 = 1$) and $N = 122$ (for $nr_0^2 > 1$). The trial wave functions are $\Psi_{JS}$ for $nr_0^2 < 1$ and $\Psi_{BF}$ for $nr_0^2 \geq 1$. Statistical error bars are of the order of the line width.

the prediction of the IFG, whose analytical expressions read [222]

$$g_{\uparrow\uparrow}(r) = 1 - \left(\frac{2J_1(\tilde{r})}{\tilde{r}}\right)^2 \tag{4.15}$$

$$g_{\uparrow\downarrow}(r) = 1 \tag{4.16}$$

with $\tilde{r} = \sqrt{4\pi n_\uparrow} r$ and $J_1(r)$ is the Bessel function of first kind.

In the plots of Fig. 4.2 we show our results for $g_{\uparrow\uparrow}(r)$ and $g_{\uparrow\downarrow}(r)$. In these plots distances are re-scaled to the mean-inter-particle distance as $r\sqrt{n}$ to facilitate the comparison between different densities, with arrows indicating increasing density. In panels a) and b) we show results for low densities ($nr_0^2 \ll 1$). For comparison, we include in dashed lines the radial distributions of the Ideal Fermi Gas. As the density is decreased, the radial distribution function involving correlations between same species approaches to the one of the IFG: in particular for density $nr_0^2 = 10^{-8}$ the two curves are barely indistinguishable. Regarding the $g_{\uparrow\downarrow}(r)$, it approaches to the IFG prediction for all the range $nr_0^2 \lesssim 10^{-4}$, apart from a strong suppression at short distances due to the dipolar repulsion. In this regime the curve of $g_{\uparrow\downarrow}(r)$ ($g_{\uparrow\uparrow}(r)$) smoothly shifts its short distance wall to higher (lower) distances in units of $\sqrt{n}r$ as the density is increased, reflecting the relevance of the dipolar repulsion against the Fermi statistics between identical particles. However, as we enter in the non-universal regime (density $nr_0^2 = 10^{-2}$, red lines in the plot) the same species radial distribution function intersects the lower density curves and the different species one starts to develop Friedel oscillations: we interpret it as a crossover between weakly to the intermediate coupling regime, where the microscopic details of the interaction start to be important.

In panels c) and d) of Fig. 4.2, we show the different and same species radial distributions for densities that are in the strongly interacting regime. In the plot corresponding to $g_{\uparrow\downarrow}$, the dipolar repulsion induces strong oscillations, corresponding to the formation of shells of particles of different species around a given one. On the other hand, the same species radial distribution function shows that same species particles are kept farther apart: for $nr_0^2 = 1$ no peak is visible, and, more surprisingly, for density $nr_0^2 = 8$, Fermi repulsion suppress the first peak making it lower than the second one. For density $nr_0^2 = 40$, the first peak height dominates, although both peaks are still compatible.

Finally it is important to remark that the radial distributions in panels c) and d) of Fig. 4.2 are not computed with the simple Jastrow-Slater wave function $\Psi_{JS}$. In this case we have used a wave function that explicitly includes backflow corrections $\Psi_{BF}$ as it was explained in section 2.3.5. Through the comparison of the results obtained using $\Psi_{JS}$ and $\Psi_{BF}$ we find that the maximum relative difference, for the higher density studied here $nr_0^2 = 40$, appears around the first peak and can be as large as 6%. For densities $nr_0^2 < 1$ the inclusion of backflow corrections does not modify the functions $g(r)$. The importance of including backflow correlations is discussed in the next section, where it becomes crucial to approach the real ground state of the system.

## 4.4 Itinerant Ferromagnetism

In the low density regime, the ground state of the system is expected to be unpolarized (*cf.* Eq. (4.10)). However, as interactions become more important, we may ask ourselves about the possibility of having a polarized fluid as the ground state of the system. The first prediction for such a phenomena, that is usually referred in the bibliography as *itinerant ferromagnetism*, appeared in the study of the three-dimensional homogeneous electron gas [223]. Apart from the numerous studies about this topic in the electron gas, both in two and three dimensions [195–201], it has been subsequently studied in other quantum many body systems such as $^3$He [224–226], and more recently in the context of ultracold fermionic gases with short-range interactions [29–36].

This last system is of some importance because of two reasons: the first one is that it constitutes the closest example of the textbook Stoner Model of magnetism [227], and the second one is that in 2017, the first claim for the observation of an itinerant ferromagnetic state was reported in a experiment performed with ultracold $^6$Li atoms [28]. This achievement came after some years of research in which difficulties related to the instability towards the formation of molecules had to be overcame [29–31].

In this section, we present a study about the possibility of having a phase that exhibits itinerant ferromagnetism in a two-dimensional dipolar Fermi system. At odds to the case reported in the experiment of Ref. [28], it constitutes an example in which the long-range character of the dipolar potential cannot be neglected.

The study of itinerant ferromagnetism has historically represented a challenge and a test-bed for many-body theories. Progress on this subject was especially connected to technical advances in the field of DMC simulations, like the use of backflow correlations and of twist-averaged boundary conditions—see, for instance, Ref. [199]. In this work, we employ the DMC method for the case of a two-dimensional dipolar gas, and we show that the level of accuracy obtained with the commonly used Jastrow-Slater and backflow-corrected trial wave functions is not sufficient to determine whether the ground state becomes polarized at larger pressure. To go beyond this limitation, we benchmark our calculations comparing with the recently developed iterative-backflow trial wave functions [119, 120], finding finally no signature of a polarized ground state

### 4.4.1 Details about the method

The issue of itinerant ferromagnetism has stood as a hard-problem for many-body theories because of its subtle dependence on the quantum correlations. On the other hand, and precisely for this reason, it has been used to test many body theories, and the case of the Quantum Monte Carlo has not been different. Due to the observed small energy differences between the state with $P = 1$ (ferromagnetic phase) and the one with $P = 0$ (paramagnetic phase), the correct evaluation of the possible paramagnetic to ferromagnetic transition cannot be performed with the simple Jastrow-Slater wave function that we have used to determine the EOS in the low density regime. For this reason, we employ backflow corrected wave functions as introduced in section 2.3.5. Its simplest implementation consists on replacing the position coordinates inside the plane waves entering in the Slater determinant by the backflow coordinates, defined as

$$\mathbf{r}_i \rightarrow \mathbf{q}_i \equiv \mathbf{r}_i + \sum_{j \neq i} \left( \mathbf{r}_i - \mathbf{r}_j \right) f_{\mathrm{BF}}(r_{ij}). \tag{4.17}$$

The construction of the backflow wave function $\Psi_{\mathrm{BF}}$, requires a parametrization for the function $f_{\mathrm{BF}}$. In this work, we use the same Gaussian parametrization for as in Ref. [116]

$$f_{\mathrm{BF}}(r) = \lambda_{\mathrm{BF}} \exp\{-[(r - r_{\mathrm{BF}})/\sigma_{\mathrm{BF}}]^2\} \tag{4.18}$$

where $\lambda_{\mathrm{BF}}$, $r_{\mathrm{BF}}$ and $\sigma_{\mathrm{BF}}$ are parameters that have to be optimized using the variational principle. The simultaneous optimization of these parameters constitutes an intricate problem and sometimes gives place to different combinations of $\{\lambda_{\mathrm{BF}}, r_{\mathrm{BF}}, \sigma_{\mathrm{BF}}\}$ that provide the same FN energy. In Table 4.1 we list the parameters that we have used for the evaluation of the energy in the paramagnetic and ferromagnetic phases. Parameters $r_{\mathrm{BF}}$ and $\sigma_{\mathrm{BF}}$ give the position and width of the Gaussian function $f_{\mathrm{BF}}$, that is usually centered as a distance close to that where the first peak of in the $g_{\uparrow,\downarrow}$ appears. On the other hand, $\lambda_{\mathrm{BF}}$ gives an indication of how important backflow correlations are, and its value is always larger in the $P = 0$ state than in the $P = 1$ one.

| | $P = 0$ | | | $P = 1$ | | |
|---|---|---|---|---|---|---|
| $nr_0^2$ | $\lambda_{\mathrm{BF}}$ | $r_{\mathrm{BF}}$ | $\sigma_{BF}$ | $\lambda_{\mathrm{BF}}$ | $r_{\mathrm{BF}}$ | $\sigma_{BF}$ |
| 8 | 0.30 | 0.098 | 0.180 | 0.13 | 0.125 | 0.160 |
| 16 | 0.30 | 0.090 | 0.105 | 0.13 | 0.085 | 0.125 |
| 24 | 0.25 | 0.090 | 0.090 | 0.09 | 0.090 | 0.095 |
| 32 | 0.25 | 0.085 | 0.075 | 0.09 | 0.090 | 0.100 |
| 64 | 0.35 | 0.050 | 0.060 | 0.13 | 0.045 | 0.100 |

Table 4.1 Values of the optimized backflow parameters employed for the calculations that are presented in this section.

Finally, to benchmark our calculation we compare our DMC results with an independent implementation of the DMC algorithm by Markus Holzmann [118–120]. In this implementation different nodal surfaces, including two and three-body backflow correlations, and the iterative backflow procedure are employed. Some details about these calculations are summarized in appendix C. The use of different backflow nodal surfaces in VMC calculations allows to perform a zero variance extrapolation taking advantage of the fact the exact nodal surface would have zero variance [106] (see discussion before Eq. (2.23)). Assuming that $\Psi_T$ has a large overlap with the exact ground state wave function $\phi_0$, some bounds can be established that relate the variational estimation of the energy $E_T$ with the true ground state energy $E_0$ and its variance $\sigma_T^2$ [228, 229, 105]. In the limit of small variance, the following linear extrapolation can be written

$$E_T = E_0 + NA\sigma_T^2 \qquad \text{for } \sigma_T^2 \rightarrow 0 \tag{4.19}$$

with $A$ a fitting constant. This allows to perform an evaluation of the ground state of the system through the computation of $E_T$ and $\sigma_T^2$ with VMC. The validity of this approach has been tested both in ${}^4$He and ${}^3$He, where it is found that the best DMC available ground state energy estimation is compatible with the one obtained with the variance extrapolation method using VMC [119].

### 4.4.2 Results

To determine whether the ground state of the dipolar system is the paramagnetic or the ferromagnetic phase, we directly compare the energies per particle of the unpolarized $E_{P=0}$ and polarized $E_{P=1}$ states

$$\Delta E \equiv E_{P=0} - E_{P=1}. \tag{4.20}$$

The quantity $\Delta E$ is computed twice for each of the phases: one with the usual SJ nodal surface and the other one employing a BF wave function. We find that the correction in the energy per particle due to the inclusion of backflow correction is always greater in the $P = 0$ state. This is expected because the inclusion of backflow constitutes a correction mainly in s-wave channel which is highly suppressed in the $P = 1$ state due to Pauli exclusion [224, 230].

In both approximations (SJ and BF), we find that $\Delta E$ crosses from negative values at low density to positive values at larger densities. This signals a region at low density in which the ground state is the unpolarized phase, and a region in which the $P = 0$ state is unstable towards the appearance of a polarized phase at larger densities. As we are only comparing the two extreme cases of $P = 0$ and $P = 1$, the crossing point determines and upper bound to the possible appearance of a partially polarized phase, since in principle the ground state could have a polarization in the range $0 < P < 1$.

In Fig. 4.3, the quantity $\Delta E$ is plotted for the calculations performed both with JS and BF nodal surfaces. As it can be seen, the prediction from the transition density shifts towards higher densities when a better nodal surface is employed from $(nr_0^2)_{IF} = 20(2)$ with JS to $(nr_0^2)_{IF} = 26(4)$ with the backflow. In Table 4.2, the DMC energies for both phases are listed. For each of the phases we have included a column, labeled as "corr", where the correction in the energy per particle due to the inclusion of backflow is reported. As it can be seen, this correction is larger when the density increases, and it is always bigger in the $P = 0$ phase than in the polarized one, which is a well known fact from studies of other systems such as the electron gas, and ${}^3$He [199, 224, 230]. Calculations for the unpolarized (polarized) phase have been done using 122 (121) particles, which guarantees that finite-size effects are below 0.1%. This is shown in Fig. 4.4, where it can be seen that the finite-size scaling of the energy per particle is shown for a density that is clearly inside the paramagnetic phase domain ($nr_0^2 = 16$) and another one in the window where a possible ferromagnetic phase may exist, according to the results shown in Fig. 4.3.

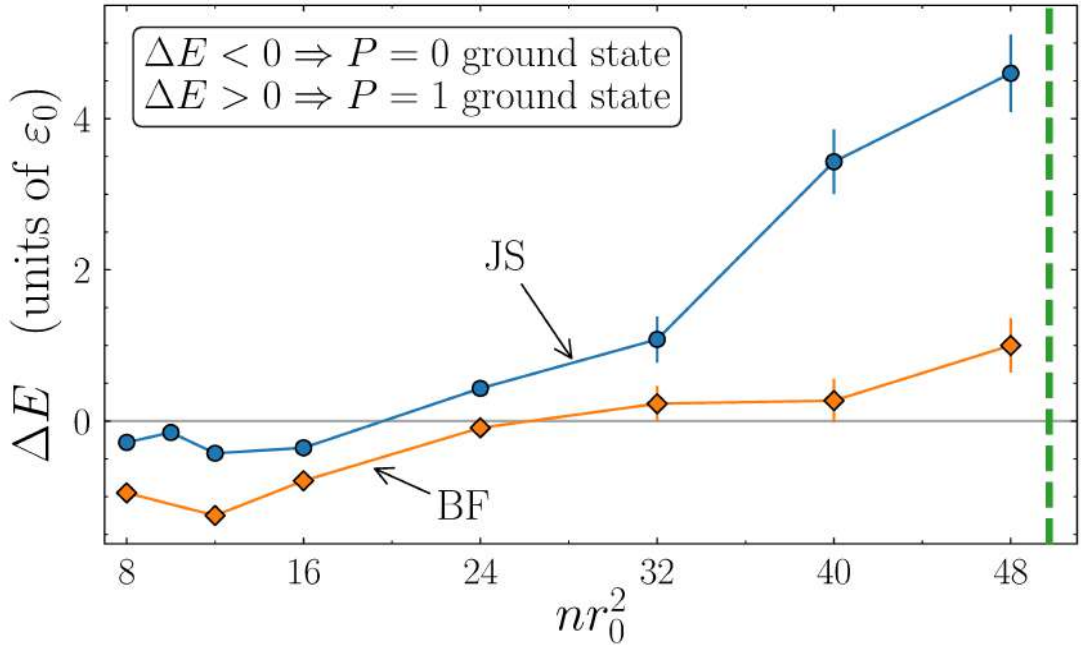

Fig. 4.3 Difference in the energy per particle between $P = 0$ and $P = 1$ states [*cf.* Eq. (4.20)], computed through DMC with the Jastrow-Slater (JS) or Backflow (BF) trial wave function (see Table 4.2). Statistical error bars are shown, and lines are a guide to the eye. The vertical dashed line marks the freezing density for the $P = 1$ state [93].

| | $P = 1$ | | | $P = 0$ | | |
|---|---|---|---|---|---|---|
| $nr_0^2$ | $E_{JS}$ | $E_{BF}$ | Corr. | $E_{JS}$ | $E_{BF}$ | Corr. |
| 8 | 168.66(1) | 168.63(1) | 0.03(1) | 168.38(2) | 167.68(4) | 0.70(4) |
| 12 | 295.98(2) | 295.92(2) | 0.06(3) | 295.55(6) | 294.67(5) | 0.88(8) |
| 16 | 442.22(2) | 442.04(2) | 0.18(3) | 441.87(7) | 441.25(6) | 0.6(1) |
| 24 | 781.21(4) | 780.8(1) | 0.4(1) | 781.6(1) | 780.71(7) | 0.9(1) |
| 32 | 1172.45(5) | 1171.9(1) | 0.6(1) | 1173.5(3) | 1172.1(2) | 1.4(4) |
| 40 | 1608.0(1) | 1607.7(1) | 0.3(2) | 1611.4(4) | 1607.9(2) | 3.5(5) |
| 48 | 2083.1(1) | 2083.0(2) | 0.1(2) | 2087.7(5) | 2084.0(3) | 3.7(6) |
| 64 | 3137.7(1) | 3137.4(2) | 0.3(2) | 3145.4(8) | 3139.8(3) | 5.6(9) |

Table 4.2 DMC energy per particle at different densities, in units of $\varepsilon_0$. Data for $P = 1$ ($P = 0$) are obtained with $N = 121$ ($N = 122$) particles. Columns marked as "Corr." indicate the energy correction $E_{JS} - E_{BF}$. Statistical errors are reported in parentheses.

As the main correction to the transition density to an itinerant ferromagnetic phase comes from the correction to the $P = 0$ phase, we may ask ourselves whether the polarized state would continue being the ground state of the system if we use a better

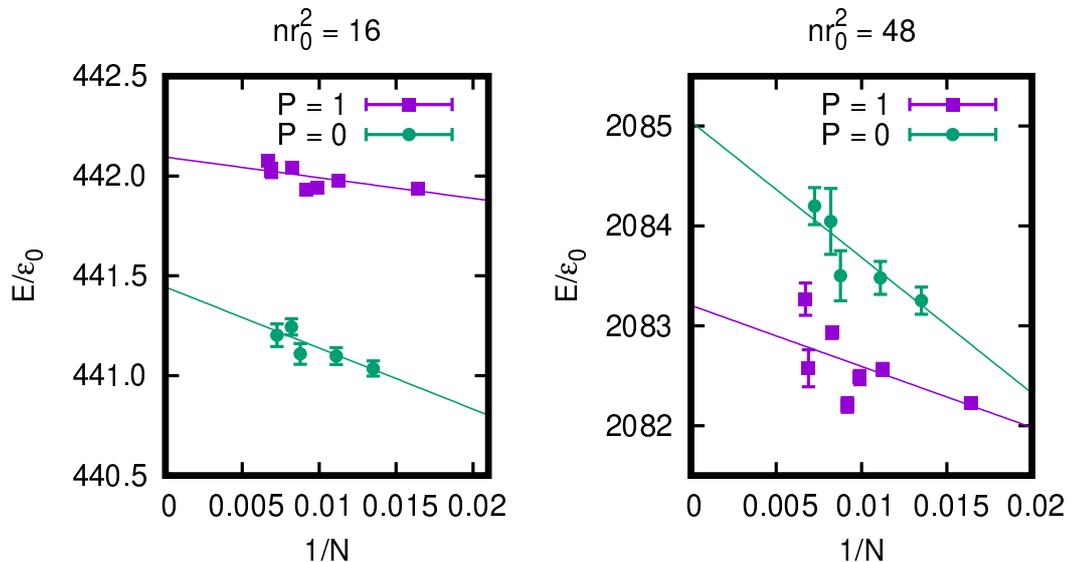

Fig. 4.4 finite-size scaling of the energy per particle in the paramagnetic ($P = 0$) and ferromagnetic ($P = 1$) phases of the two-dimensional dipolar systems of fermions. Left (right) panel corresponds to density $nr_0^2 = 16$ ($nr_0^2 = 48$).

nodal surface. As we have already explained, to answer this question one needs to employ different nodal surfaces and perform an extrapolation to zero variance using VMC. Details about the wave functions that we use to this purpose that include not only the usual two-body backflow correlations but also an iterative backflow procedure can be found in the appendix C. Results obtained with that method are plotted in figure 4.5, where $E_T$ energies evaluated with VMC employing different wave functions are plotted. In the same plot, we plot results for the two states $P = 0$ and $P = 1$ for a fixed density ($nr_0^2 = 40$ (left) and $nr_0^2 = 48$ (right)). As it can be seen in the plot, the larger correction when including backflow affects the unpolarized state. In the same plot, two linear extrapolations to zero variance are included (according to Eq. (4.19)) that allow to give an estimation of the value exact energy per particle in the two phases. These results seem to suggest that no itinerant ferromagnetic phase appear in the dipolar system before the crystallization point.

In Table C.1 of appendix C, the VMC energies of Fig. 4.5 are listed together with the DMC Fixed-Node energies evaluated with the same wave functions. These DMC results seems to be in qualitative agreement with the conclusions of the zero-variance extrapolation. For the higher density computed, the best choices for the nodal surface give results that are almost compatible in the polarized phase. On the other hand, the corrections are more clear in the unpolarized phase where the FN energy is systematically reduced. Furthermore, the best DMC energy for both of densities $nr_0^2 = 40$ and $nr_0^2 = 48$ are below the ones for the polarized phase for the same nodal structure quality. Finally, although the variance related to a certain trial wave function is not well defined in DMC, as we are sampling through the mixed probability distribution $f(\mathbf{R}) = \Psi_T(\mathbf{R})\phi_0(\mathbf{R})$, an extrapolation to zero variance of the DMC results (employing the VMC variance) leads to the same conclusions.

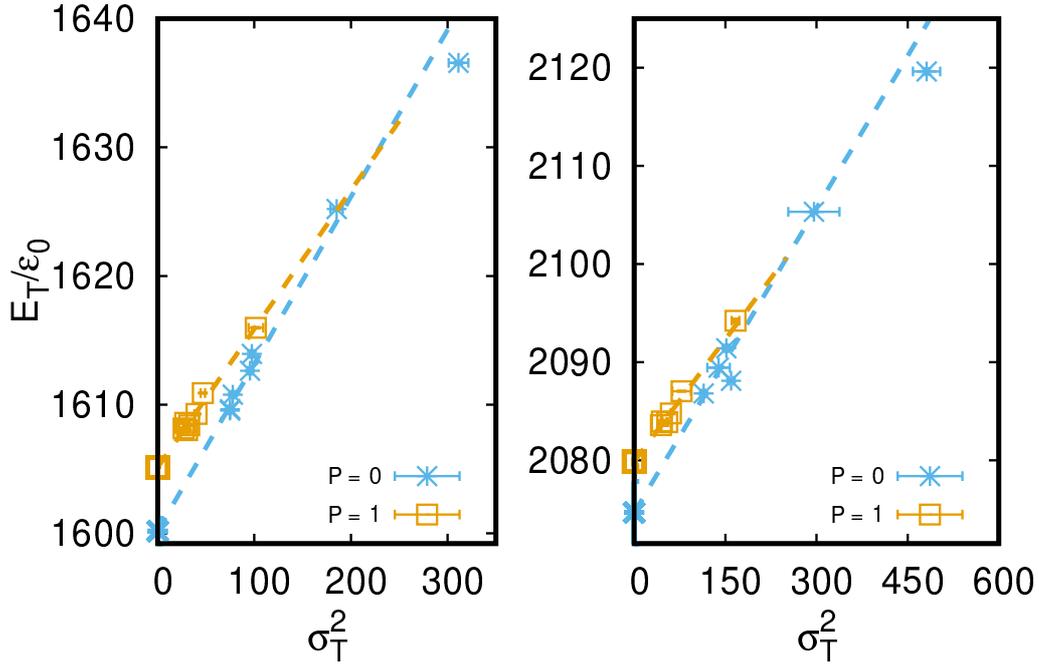

Fig. 4.5 Scaling of the energy per particle against variance evaluated with VMC at two different densities $nr_0^2 = 40$ (left) and $nr_0^2 = 48$ (right). Blue points correspond to the $P = 0$ state and yellow ones to $P = 1$, and different points of the same color correspond to calculations performed with different trial wave functions, as is explained on appendix C. Dashed lines correspond to a linear fit according to Eq. (4.19). Points at $\sigma_T^2 = 0$ correspond to the value of the linear extrapolation.

## 4.5   The dipolar Fermi polaron

The polaron problem was put forward by Landau and Pekar [231, 202] to study the interaction of an electron with a crystal lattice. In the strongly coupled regime, it was shown that the distortion of the lattice caused by the presence of the electron may induce a local potential that traps the electron. A few years later, Fröhlich developed a Hamiltonian formulation [232] to describe the coupling between the electron and the phonon modes. Using this model, a first variational ground-state solution for the intermediate coupling regime was derived by Feynman [233]. Some decades later, the picture was completed with exact results for the Fröhlich model Hamiltonian obtained using the diagrammatic Quantum Monte Carlo (QMC) method [234, 235]. The polaron (impurity) problem has also been studied in other fields of physics, such as condensed matter (*cf.* an impurity of $^3$He in bulk $^4$He [236, 151]) and nuclear matter [237].

As has been previously commented in this Thesis, the achievement of the Bose-Einstein condensate state (BEC) in the past decades has provided a new platform to tackle several problems and the case of the polaron is not different. The name Bose polaron was initially coined to indicate an impurity coupled to a BEC, and two-component mixtures of ultracold gases featuring a very small concentration of one of the components were proposed as candidate systems where to investigate the quasi-particle nature of the impurities [238, 239]. In recent years, these configurations have been realized in mixtures of both different hyper-fine levels of the same atomic species [240], and of different atoms [241, 242]. In these experiments, the polaron problem was investigated close to a Feshbach resonance, which allows to tune the interaction strength between the impurity and the bath. Two branches have been characterized at very low temperatures in systems where the effective interaction between the impurity and the bath is repulsive: the attractive polaron branch, corresponding to the ground state of the impurity in the medium, and the repulsive polaron branch, which consists in an excited state of the impurity [243, 152].

Furthermore, in the context of ultracold gases, Fermi degenerate systems offer new possibilities where the polaron picture can arise. Experimental measurements have been reported for a spin-down impurity "dressed" in a bath of a spin-up Fermi gas (*cf.* in $^6$Li [73] ) and for atomic mixtures such as $^{40}$K impurities into $^6$Li, where attractive and repulsive polaron branches have also been observed [244]. While the relation between the bosonic case and the Fröhlich formulation is straightforward, the fermionic equivalent problem (Fermi polaron) is more challenging and opens the door to a richer scenario. Different theoretical works [245–247] have studied the polaron as a first insight into some physical phenomena that are characteristic of the strongly interacting regime: the pairing mechanism that gives rise to the BEC-BCS crossover [71–73], the possible itinerant ferromagnetism in two-component systems [33, 29, 28, 94] or the Kondo effect in systems containing magnetic impurities [74].

The realization of quantum degenerate systems composed of atoms with large magnetic moment has motivated additional interest in the polaron problem. The dominant dipolar interaction between these atoms is of longer range and anisotropic. This was first achieved with Cr atoms [40, 41] and more recently also with Dy [42, 43] and Er [44, 45] that have a larger magnetic moment than Cr. Regarding the polaron problem, the report of experimentally accessible ultracold mixtures of Er and Dy [75] and the study of low concentration impurities of $^{163}$Dy in a $^{164}$Dy droplet [77] has motivated the study of the dipolar polaron in three [212] and in quasi-two dimensional configurations [248]. The dipolar polaron has also been studied in a bi-layer geometry, where localization effects are predicted near the crystallization point [213].

In two dimensions, quantum correlations are enhanced compared to the three-dimensional case. While the one particle-one hole picture has demonstrated its utility to study the Fermi polaron problem in 3D systems [245], it fails when trying to accurately reproduce the physics of the equivalent system in 2D [153] as it has been shown with the Diagrammatic Monte Carlo technique. Up to now, some efforts have been put in the study of the repulsive Fermi polaron, studied as the repulsive branch of a system with short-range interactions (*cf.* Refs. [249, 250] and [251, 252, 153] for experiment and theory, respectively). Here, we study the equivalent system but with dipolar interactions, which in principle would be accessible in current experiments.

### 4.5.1 The system

We study the repulsive Fermi polaron as the limit of high population imbalance ($P \to 1$) in a two-component system, whose species labeled as ↑ and ↓ in analogy with spin-1/2 particles. The system, consisting of $N = N_\uparrow + 1$ particles, contains a single atomic impurity immersed in a bath composed of $N_\uparrow$ atoms. We study this model in a system with dipolar interaction and we compare our results to those of a Hard-disk model (for details about the later, see appendix B), which allows to determine the regime of universality for different properties in this problem. In the dipolar model, one assumes dipolar interaction between all the particles in the system, and similarly to the other calculations presented in this chapter, that all the dipolar moments are polarized along the direction perpendicular to the plane of motion, so that the interaction between particles is isotropic.

Hereby, with the aim of reproducing the physics of a uniform infinite system, we simulate all the particles in a square box with periodic boundary conditions, and with the box side $L$ fixed by the density $n$ of the bath ($L = \sqrt{N_\uparrow/n}$ ). The $N$-particle Hamiltonian reads

$$\hat{H} = -\frac{\hbar^2}{2m}\nabla_\downarrow^2 - \frac{\hbar^2}{2m}\sum_{i=1}^{N_\uparrow}\nabla_i^2 + \sum_{i<j}^{N_\uparrow}V^{\text{bath}}(r_{ij}) + \sum_{j=1}^{N_\uparrow}V^{\text{int}}(r_{\downarrow j}), \qquad (4.21)$$

where $r_{ij} \equiv |\mathbf{r}_i - \mathbf{r}_j|$ is the distance between two bath particles and $r_{\downarrow j} \equiv |\mathbf{r}_\downarrow - \mathbf{r}_j|$ is the distance between a bath particle at $\mathbf{r}_j$ and the impurity position $\mathbf{r}_\downarrow$. Throughout this section, labels $i$ and $j$ refer to bath particles. $V^{\text{bath}}(r)$ is the two-body potential between the bath particles, and $V^{\text{int}}(r)$ is the interaction potential between the impurity and the bath.

### 4.5.2 Details about the method

The calculations presented in this section are performed using DMC with the FN prescription. The wave function employed for importance sampling is similar to that one described in section 4.3.1 for the study of the fermionic mixture (*cf.* Eqs. (4.3)-(4.6)). Thus, it is constructed as the product of an antisymmetric and a symmetric part $\Psi_T(\mathbf{R}) = \Psi_A(\mathbf{R})\Psi_S(\mathbf{R})$. The antisymmetric part is represented by a single Slater determinant of plane waves of dimension $N^\uparrow$, that reflects the Fermi statistics that exists between the bath particles. The choice of plane waves as single particle orbitals in the Slater determinant is justified and is also accurate for describing the system in the low density regime in which we focus here $nr_0^2 \ll 1$. On the other hand, the symmetric part is of the Jastrow form, with the Jastrow factors constructed from the zero energy solution of the two-body problem.

As described in section 4.3.1, the Jastrow factors between the ith and jth particles $f_{ij}(\mathbf{r}_{ij})$ are constructed from the zero energy solution of the two-body problem, matched at a certain distance $R_M$ (which we use as a variational parameter) with a phononic behavior (*cf.* Eq. (4.4)). In general, the bath/bath and bath/impurity correlations are significantly different, and therefore, not all the Jastrow factors are equal. This is implemented in our calculations by considering two different variational parameters $R_M^{\uparrow\uparrow}$ and $R_M^{\uparrow\downarrow}$ instead of only one. In this way, the symmetric part of the wave function is written as

$$\Psi_J(\mathbf{R}) = \prod_{j=1}^{N_\uparrow} f_{\uparrow\downarrow}(r_{j\downarrow}) \prod_{i<j}^{N_\uparrow} f_{\uparrow\uparrow}(r_{ij}) \ . \tag{4.22}$$

In both the same $f_{\uparrow\uparrow}$ and different $f_{\uparrow\downarrow}$ species Jastrow factors we impose the conditions $f_{\uparrow\uparrow}(L/2) = f_{\uparrow\downarrow}(L/2) = 1$, $f'_{\uparrow\uparrow}(L/2) = f'_{\uparrow\downarrow}(L/2) = 0$. In Fig. 4.6, we compare the variational energy for two densities and in two situations: in the first case we consider only one variational parameter (equivalent to set $R_M^{\uparrow\uparrow} = R_M^{\uparrow\downarrow} = R_M$, purple points), and in the second one $R_M^{\uparrow\uparrow}$ and $R_M^{\uparrow\downarrow}$ are optimized independently (green points). The dashed blue line in the plot corresponds to a calculation in which the variational parameter is set to its optimal value for the pure ferromagnetic phase $R_M^{\uparrow\uparrow} = R_M^{\uparrow\downarrow} = R_M^{FM}$. As it can be seen, a small effect on the variational energy can be found, although it disappears when DMC calculations are performed.

The importance for using a trial wave function of the form of Eq. (4.22), instead of one with only one variational parameter, comes out when we are interested in observables for which there is not straightforward to obtain an unbiased DMC estimator. This is the case of the quasi-particle residue $Z$ that was introduced in sec. 2.6.6.3. A discussion about this observable for the dipolar repulsive polaron is carried out in section 4.5.6. Here we just focus on the possible bias of this observable due to the employment of $\Psi_T$ for importance sampling. The quasi-particle residue is obtained as the asymptotic behavior of a OBDM obtained including only correlations between the impurity and the bath. In figure 4.7 we show results for the computation of such OBDM for the two densities $nr_0^2 = 10^{-2}$ (top) and $nr_0^2 = 1$ (bottom). In the left panels we show VMC results, while on the right ones correspond to DMC extrapolated quantities according to the expressions of Eqs. (2.72) (Squares) and Eq (2.73) (circles). In the left panel, purple squares correspond to calculations performed with only one variational parameter $R_M$ (A-model Wave function), while blue dots correspond to independently optimized $R_M^{\uparrow\uparrow}$ and $R_M^{\uparrow\downarrow}$ parameters (B-model Wave function). On the right panel green and purple points correspond to simulations performed with the A-model and red and blue ones to the B-model. As it can be seen, for both densities the different between variational $Z_{VMC}$ and extrapolated $Z_{extr}$ asymptotic values are reduced when the best variational wave function is used. In what follows we use the difference $\Delta Z = Z_{VMC} - Z_{extr}$ to estimate the systematic bias in the quasi-particle residue.

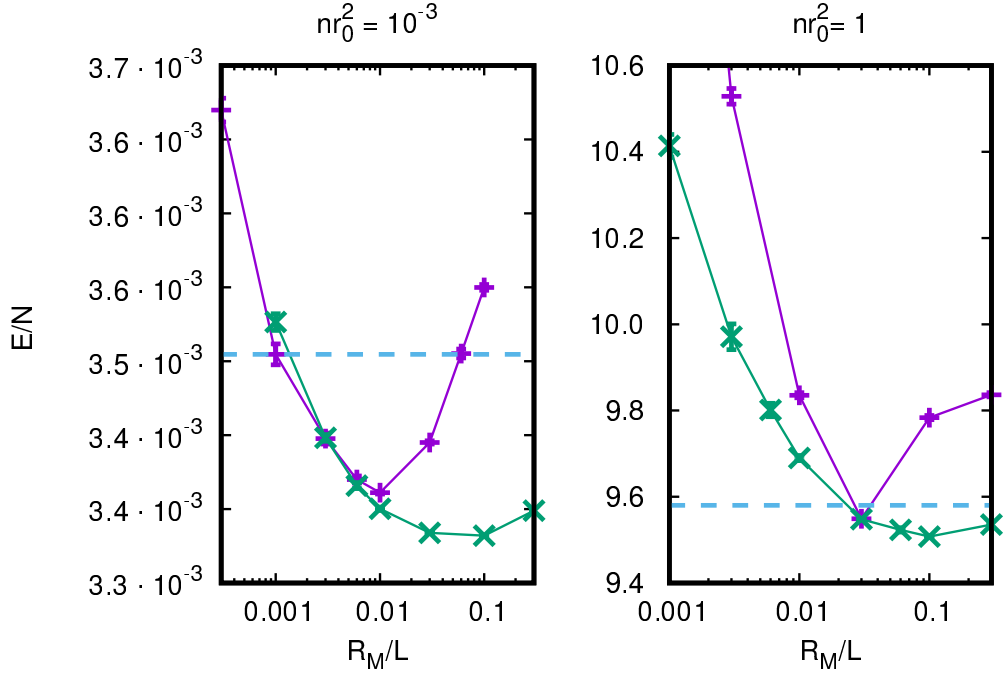

Fig. 4.6 Variational optimization of the parameter $R_M$ for two different densities $nr_0^2 = 10^{-3}$ (left) and $nr_0^2 = 1$ (right) in the dipolar system with an impurity. Purple points correspond to optimizing the wave function of Eq. (4.5) with the same unique variational parameter $R_M$. Green points represent the case in which two variational parameters are used $R_M^{\uparrow\uparrow}$ and $R_M^{\uparrow\downarrow}$. The horizontal dashed blue line correspond to the result obtained with the optimal variational parameter for the ferromagnetic phase $R_M^{FM}$ ($R_M^{\uparrow\uparrow} = R_M^{\uparrow\downarrow} = R_M^{FM}$).

### 4.5.3 Results

The QMC results that appear in this section are compared with two approximate theories to benchmark them. This is also useful to study the regime in which the system becomes universal in terms of the gas parameter. As a first approximation, we compare our energies with the prediction that mean-field theory offers for the system [220]. On the other hand, we also compare our results with a T-matrix study of the repulsive Fermi polaron [251]. The authors of Ref. [251] considered the ultra-dilute limit of spin-up impurities immersed in an spin-down bath, which is treated as an ideal Fermi gas. Quasi-particle properties (effective mass and quasi-particle residue) were then evaluated both for the attractive and the repulsive branches of a system where the impurity interacts via a short-range potential of scattering length $a_s$ with the bath, treated as an ideal Fermi gas. Due to the similarity of the repulsive branch studied in that model with the hard-disk system described in appendix B, it is worthy to compare both models with our results for the dipolar system.

### 4.5.4 The polaron energy

The energy of the polaron is an important and experimentally accessible observable. It is defined as the energy difference between the pure system of $N_\uparrow$ particles and the same system with an added impurity, at fixed volume. Making use of this definition it can be directly evaluated in QMC simulations as the chemical potential of the impurity. This has been already discussed in section 2.6.6.1, where an expression to estimate the polaron energy in DMC is given (*cf.* Eq (2.164)).

In mean-field theory the polaron energy in 2D reads

$$\varepsilon_{\mathrm{MF}} = \frac{4\pi\hbar^2 n}{m \ln(c_0 n a_s^2)} \ .$$
(4.23)

The dependence of the mean-field prediction (4.23) on a free parameter $c_0$ is a peculiarity of 2D systems that is related to the features of scattering theory in 2D [221]. This free parameter is related to a characteristic energy scale of the system [190, 94]. In the present work, we set it to the value $c_0 = e^{2\gamma}\pi/2 \simeq 4.98$, corresponding to using an energy scale equal to the Fermi energy $E_\mathrm{F} = 2\hbar^2\pi n/m$.

In Fig. 4.8, we show our QMC results compared to those of the mean-field approach (Eq. (4.23)). We plot the polaron energy in units of the mean-field energy, so that deviations from the mean-field prediction are enhanced. Although being a good approx-imation, mean-field fails to accurately reproduce even the lower densities considered in this work, which is a well known fact in two-dimensional gases [84]. As the density is increased, the mean-field prediction has a logarithmic divergence and thus it does not stand as a good energy scale for values of $na_s^2 > 10^{-3}$. For this reason, in the inset of Fig. 4.8 we plot the polaron energy, for the highest gas parameters, in units of the Fermi energy, $E_F$. The error bars that appear in Fig. 4.8 include both statistical and systematic errors, the latter being the largest contribution. In the low density regime, the systematic error is dominated by the finite value of the imaginary-time step $\delta\tau$, while for the higher densities the main source of error comes from finite-size effects. Concerning this latter issue, calculations have been done using 61 bath particles for all the dipolar system, while, for the hard-disk model, the exclusion of volume caused by the impurity makes it necessary to include 121 particles in the bath to maintain finite-size effects under control when the gas parameter is higher than $na_s^2 \geq 10^{-2}$. In the case of hard-disk interaction, systematic errors for the polaron energy are of the order of 0.5%, while for dipolar systems they grow up to 1%.

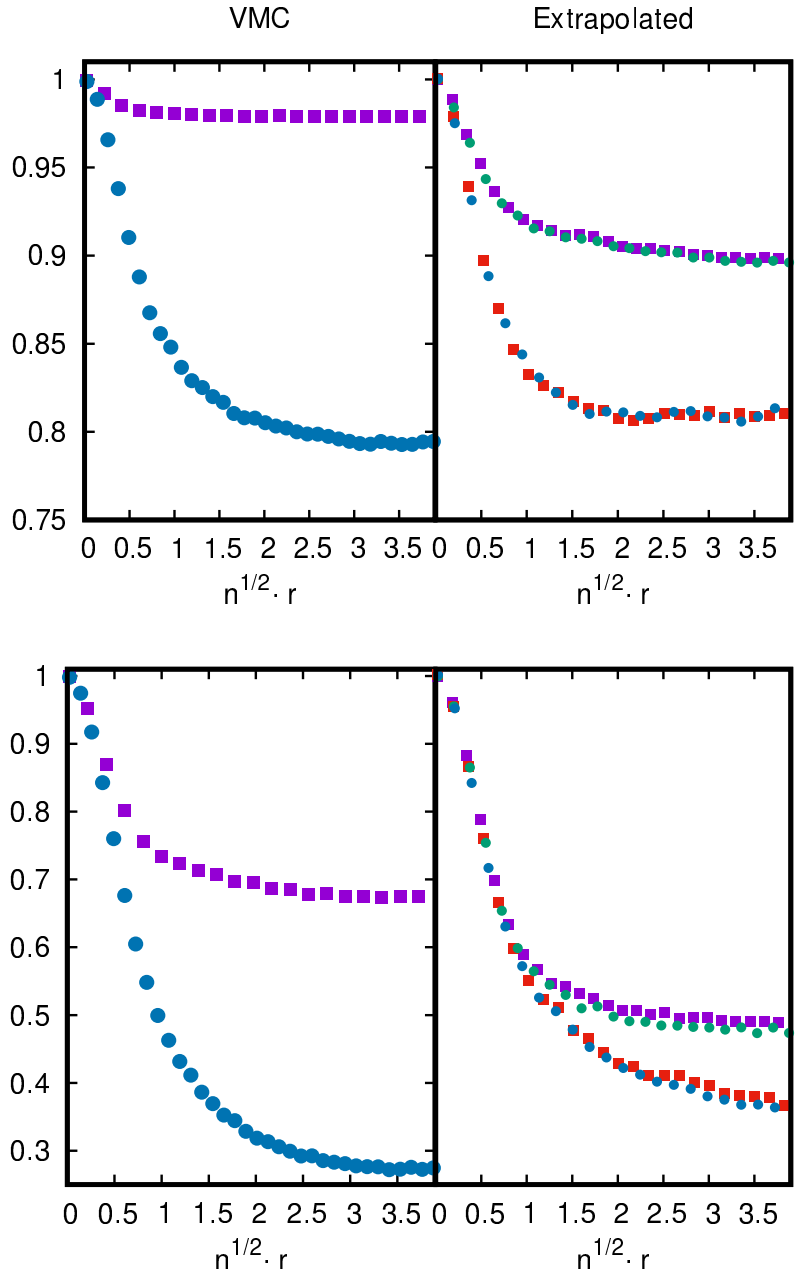

Fig. 4.7 Quasi-particle residue at different densities $nr_0^2 = 10^{-2}$ (top) and $nr_0^2 = 1$ (bottom) . Left panel: VMC results when the wave function employed has only one variational parameter $R_M = R_M^{\uparrow\uparrow} = R_M^{\uparrow\downarrow}$ (purple squares) and when $R_M^{\uparrow\uparrow}$ and $R_M^{\uparrow\downarrow}$ are optimized independently (blue dots). Right panel: Purple and green points correspond to extrapolated results when only one parameter is used while blue and red points are obtained with different values of $R_M^{\uparrow\uparrow}$ and $R_M^{\uparrow\downarrow}$. The extrapolations values are evaluated with the two different extrapolation methods of Eqs. (2.72) (Squares) and Eq (2.73) (circles).

### 4.5.5 Pair distribution functions

The presence of the impurity affects the local properties of the bath. This effect can be analyzed by looking at the pair distribution function between the background and the impurity $g^{\uparrow\downarrow}(r)$, sometimes referred to as the density profile of the bath around the impurity. In DMC simulations, we can evaluate both this distribution function and the one involving only bath particles, $g^{\uparrow\uparrow}(r)$, as it was pointed out in section 2.6.2.

Figure 4.9 shows $g^{\uparrow\downarrow}(r)$, as a function of the dimensionless quantity $r\sqrt{n}$, for different gas parameter values and for the two models considered in this work. The plot indicates that the hole around the impurity, arising from repulsive correlations between the impurity and bath particles, grows when the gas parameter is increased. We also notice that, at the lowest value of the gas parameter shown for the dipolar model ($na_s^2 \simeq 10^{-4}$), the distribution function closely resembles the one of the hard-disk model (except for distances that are comparable to the core radius $R = a_s$) indicating the approaching to the low-density universal regime, similar to what one finds when comparing the polaron energies for the two models. For the dipolar model, the radial distributions have been evaluated using the pure estimators technique [121] whilst the hard-disk model results correspond to the extrapolation of DMC results as it was explained in the method chapter (*cf.* Eq. (2.72)). This also applies for the data in Fig. 4.10, and in both cases error bars are chosen to cover systematic errors.

The dipolar model maintains its physical meaning in the high-density regime (as it was studied in section 4.4) where $g^{\uparrow\downarrow}(r)$ features Friedel oscillations, indicating the formation of shells of particles around the impurity. On the contrary, the radius of the hard-disk model starts to approach the mean inter-particle distance as the gas parameter approaches $na_s^2 \simeq 1$, and the model ceases to capture the physics of the repulsive branch with short-range interactions. It is worth mentioning that all the radial distributions shown in Fig. 4.9 are evaluated in a system containing 61 bath particles except for the two highest densities shown for the hard-disk model ($na_s^2 = 10^{-2}$ and $10^{-1}$). In these latter cases, the large amount of volume excluded by the impurity enhances the finite-size effects and the use of 121 bath particles is needed to keep them under control.

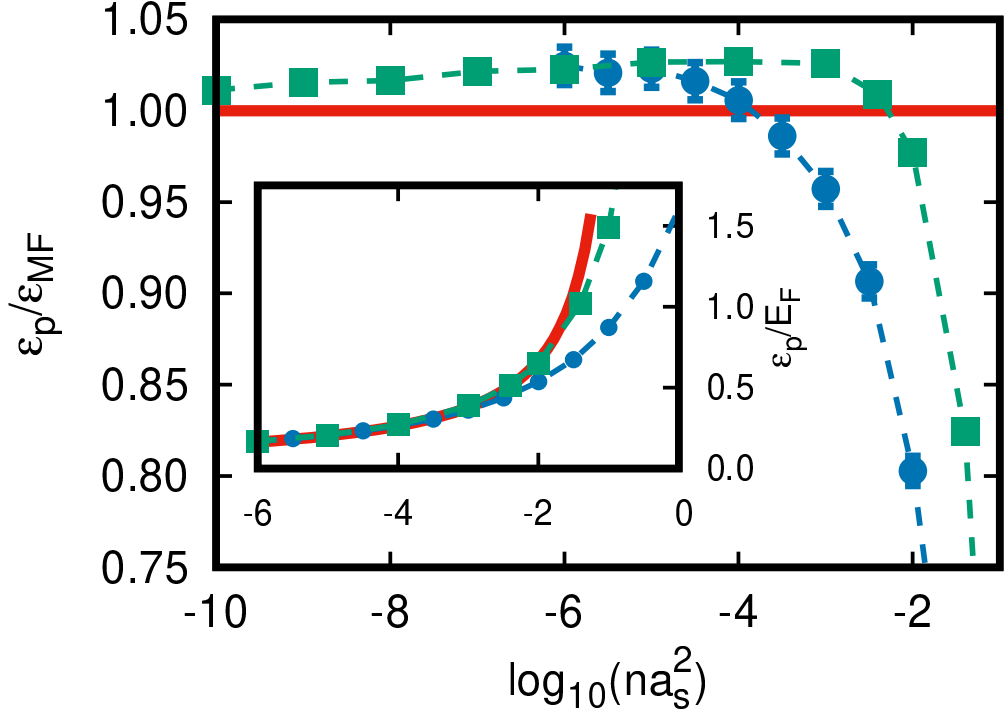

Fig. 4.8 Energy of the polaron in units of the mean-field energy in Eq. (4.23). The red line is the mean-field prediction, while green and blue symbols are DMC results for hard-disk and dipolar models, respectively. Dashed lines are guides to the eye. For large values of the gas parameter, the mean-field energy is not a good energy scale due to the logarithmic divergence of Eq. (4.23). Inset: polaron energy, in units of the bath Fermi energy $E_F$, plotted for larger values of $na_s^2$.

Due to the interaction between the impurity and the medium as well as the statistics of the particles in the bath, the volume occupied by the impurity is different from the one of any of the bath particles. A useful observable to measure this effect is the excess of volume parameter $\alpha$, that was already introduced in section 2.6.6.4 (see Eq. (2.169)). An estimation of $\alpha$ can be obtained from the $k = 0$ value of the static structure factor $S^{\uparrow\downarrow}(k)$ correlating the impurity and the bath particles [156, 157] (cf. Eq. (2.170)).

The sign of $\alpha$ carries information on whether there is an excess or deficit of volume induced by the inclusion of the impurity particle in the bath: $\alpha > 0$ ($\alpha < 0$) indicates that the impurity occupies more (less) volume than a given bath particle. This quantity has been evaluated in condensed-matter systems, for example for an $^3$He atom in bulk $^4$He. There, it was shown that the $^3$He atom occupies near 30% more volume than the average volume occupied by the particles of the $^4$He bath [151]. In that case, the increase of volume can be qualitatively explained in terms of the different zero-point motion that the two isotopes have, stemming from the mass difference.

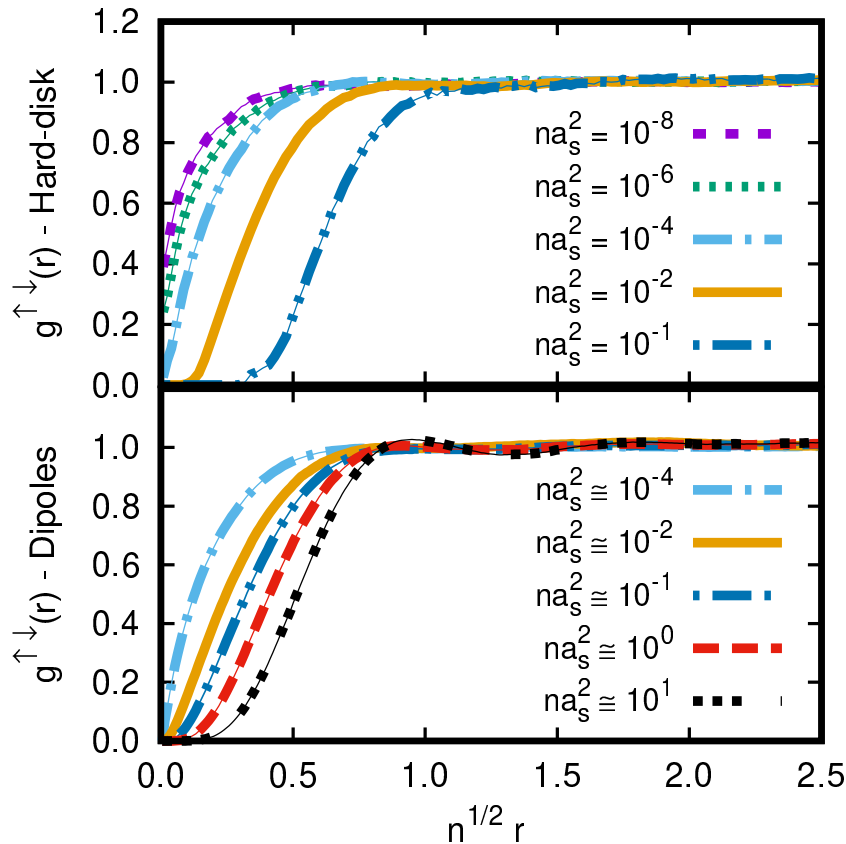

Fig. 4.9 Monte Carlo results for the pair distribution function $g^{\uparrow\downarrow}(r)$ between the impurity and the bath, evaluated for different values of the gas parameter $na_s^2$ for the hard-disk model (top panel) and for the dipolar one (bottom panel).

For a system where all atoms have the same mass and the same inter-particle interaction but where the species are distinguished by their spin component, as it is the case of our dipolar system, a decrease of volume would arise because of Fermi statistics. In order to quantify this reduction, we evaluate the impurity-bath static structure factor $S^{\uparrow\downarrow}(k)$ for our system of dipoles at different densities (see bottom panel of Fig. 4.10). For this model, the volume coefficient $\alpha$ is negative for all the range of densities that we analyze, meaning that the impurity occupies less volume than one of the bath particles, since these are pushed further apart from each other due to the Fermi repulsion. We see from our results that $\alpha$ decreases in magnitude with increasing density, that is, when the potential contributions to the energy start to be important compared to the Fermi repulsion. If one keeps increasing the density of the system up to the crystallization point ($nr_0^2 \sim 50$ [93]), the volume coefficient would approach zero ($\alpha \to 0$), as it would be the case for an impurity which is barely distinguishable from the bath atoms.

For the sake of comparison, we also show results for the excess of volume evaluated in a hard-disk model (see top panel of Fig. 4.10). In this case, however, the physics is different from the dipolar model, where the only difference between the two species comes from Fermi statistics. In this model one has also to consider that the only interaction present in the system is that of the impurity with the ideal Fermi bath. As a result, two effects compete and dominate over each other in different regimes. For low values of the gas parameter, where the hard-core radius is small compared to the mean inter-particle distance, one expects that all the deficit of volume would be caused by Fermi statistics, similar to the dipolar case. This is what can be seen when comparing the QMC results the two models in Fig. 4.10: up to values of $na_s^2 \leq 10^{-4}$, the two interactions potentials give the same $\alpha$ parameter. On the contrary, as the gas parameter increases and the system leaves the universal regime, the radius of the hard-core becomes similar to the inter-particle distance and $\alpha$ is greater than that from the equivalent dipolar system. It is worth noticing that, for the highest gas parameter considered for this model, $na_s^2 = 10^{-1}$, the excess volume coefficient becomes positive, meaning that the impurity, in this regime, occupies a bigger volume than an average particle in the ideal Fermi bath.

### 4.5.6 The quasi-particle picture

In the weakly-interacting regime, one can assume that the wave function $\phi$ describing the state of the bath plus the impurity system has an important overlap with the state $\Phi^{\text{NI}}$ in which interactions between the impurity and the bath are absent. The latter is a state representing a system containing a non-interacting impurity with momentum $k = 0$, immersed in an unperturbed single-component bath. This definition of the quasi-particle residue $Z$ was introduced in Eq. (2.167), and here make it explicit again

$$Z = \left| \langle \Phi^{\text{NI}} | \phi \rangle \right|^2 . \tag{4.24}$$

For the system with hard-disk interaction with which we compare, where the bath is an ideal Fermi gas, $|\Phi^{\text{NI}}\rangle$ reduces to $|\text{FS} + 1\rangle$, which stands for a Fermi sea with an added non-interacting impurity at zero momentum. In our dipolar model, in contrast, bath particles interact with each other, so that $\Phi^{\text{NI}}$ is the state of the interacting bath with the addition of a non-interacting impurity at zero momentum. The quasi-particle

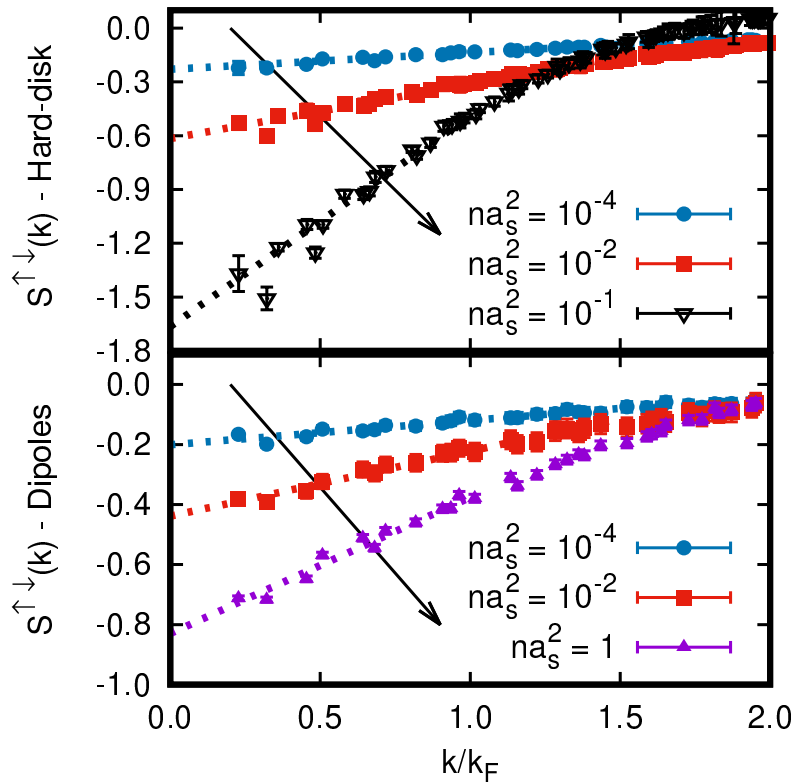

Fig. 4.10 Static structure factor $S^{\uparrow\downarrow}(k)$ involving correlations between the impurity and the bath particles, for small values of $k/k_F$ with $k_F = \sqrt{4\pi n}$ the Fermi momentum. Top (bottom) panel: results correspond to the hard-disk (dipolar) system for different values of the gas parameter. Same color and symbols are used to emphasize when the two models are evaluated at the same gas parameter. Dashed lines correspond to a linear extrapolation to $k \to 0$. The arrows indicate increasing density.

residue in Eq. (4.24) also represents the probability of free propagation of the impurity in the medium.

A discussion about the quasi-particle residue and its estimator was already introduced in section 2.6.6.3, so here we focus on the analysis of the QMC results. Since the DMC estimator of Eq. (2.168) is non-diagonal, the result is generally biased due to the choice of the trial wave function. Our estimation is based on the extrapolated estimator of Eq. (2.72) which we expect to be accurate enough due to the quality of the trial wave function, especially at low densities. In the top panel of Fig. 4.11, we show our results for the residue $Z$, following the prescription of Eq. (2.168), both for hard disks and dipoles. We find that a universal regime can be identified for gas parameters values lower than $na_s^2 < 10^{-3}$, up to where relative differences between the quasi-particle residues evaluated for the two models remain below 5%. These relative deviations are comparable to the ones reported for the polaron energy at that same gas parameter values, see Sec. 4.5.4. However, in the regime $na_s^2 > 10^{-3}$, clear differences between the two models appear: for the dipolar model the quasi-particle residue features values higher than 0.6 in all the interval of $na_s^2$ considered here. On the contrary, for the hard-disk model, $Z$ is highly suppressed as the gas parameter is increased. This fact reflects that the interaction radius begins to be comparable to the inter-particle distance, making it difficult for the impurity to perform a free displacement. Noticeably, for the largest value of the gas parameter ($na_s^2 = 4 \cdot 10^{-1}$) the residue almost vanishes, suggesting the tendency of the impurity to get localized as the interaction strength becomes very large. In the same plot, we include the T-matrix results from Ref. [251], corresponding to the quasi-particle residue of the repulsive branch of the 2D Fermi impurity problem with short-range interactions. These results are in reasonable agreement with the hard-disk impurity model, up to a regime where the excited repulsive polaron loses its identity.

Another relevant quantity in the study of the quasi-particle nature of the polaron is its effective mass, that is the mass of the quasi-particle formed by the impurity "dressed" by the medium. In a DMC simulation, the effective mass $m^*$ is obtained from the asymptotic diffusion coefficient of the impurity throughout the bath in imaginary time [151, 152], as it was advanced in Sec. 2.6.6.2 (*cf.* Eq. (2.165)).

In the bottom panel of Fig. 4.11 we report DMC results for the dipolar system, which show that interaction effects increase the effective mass of the polaron by roughly 30% as the gas parameter increases up to $na_s^2 \sim 1$. When compared to the data for short-range interactions from Ref. [251] (not shown), the effective mass of the dipolar model appears to be less affected by interactions and remains closer to its non-interacting limit ($m^* = m$), in analogy with what observed for the quasi-particle residue.

Finally, it is worth noticing that, through the knowledge of the effective mass, we can also access the excitation spectrum of the polaron at low momenta, as is made explicit in Eq. (2.165).

### 4.5.7 Polaron at high density

To close this section about the dipolar polaron, we study it in the high density regime, near the crystallization transition, $(nr_0^2)_c \approx 50$. Studying the polaron in this regime is of interest because it gives some insight into the stability of a possible ferromagnetic phase against one single spin flip. If the ferromagnetic state ($P = 1$) is the ground state, then any excitation must lead to a state with higher energy. If the true ground state has polarization $P < 1$, on the contrary, a single spin flip on top of a fully polarized state may decrease its energy, signaling the instability of the ferromagnetic state.

In DMC we can compute both the polaron energy $\epsilon_p^{(N^\uparrow)}$ and the chemical potential $\mu^{(N^\uparrow)}$ related to adding another spin-down (spin-up) particle on a bath of fully polarized

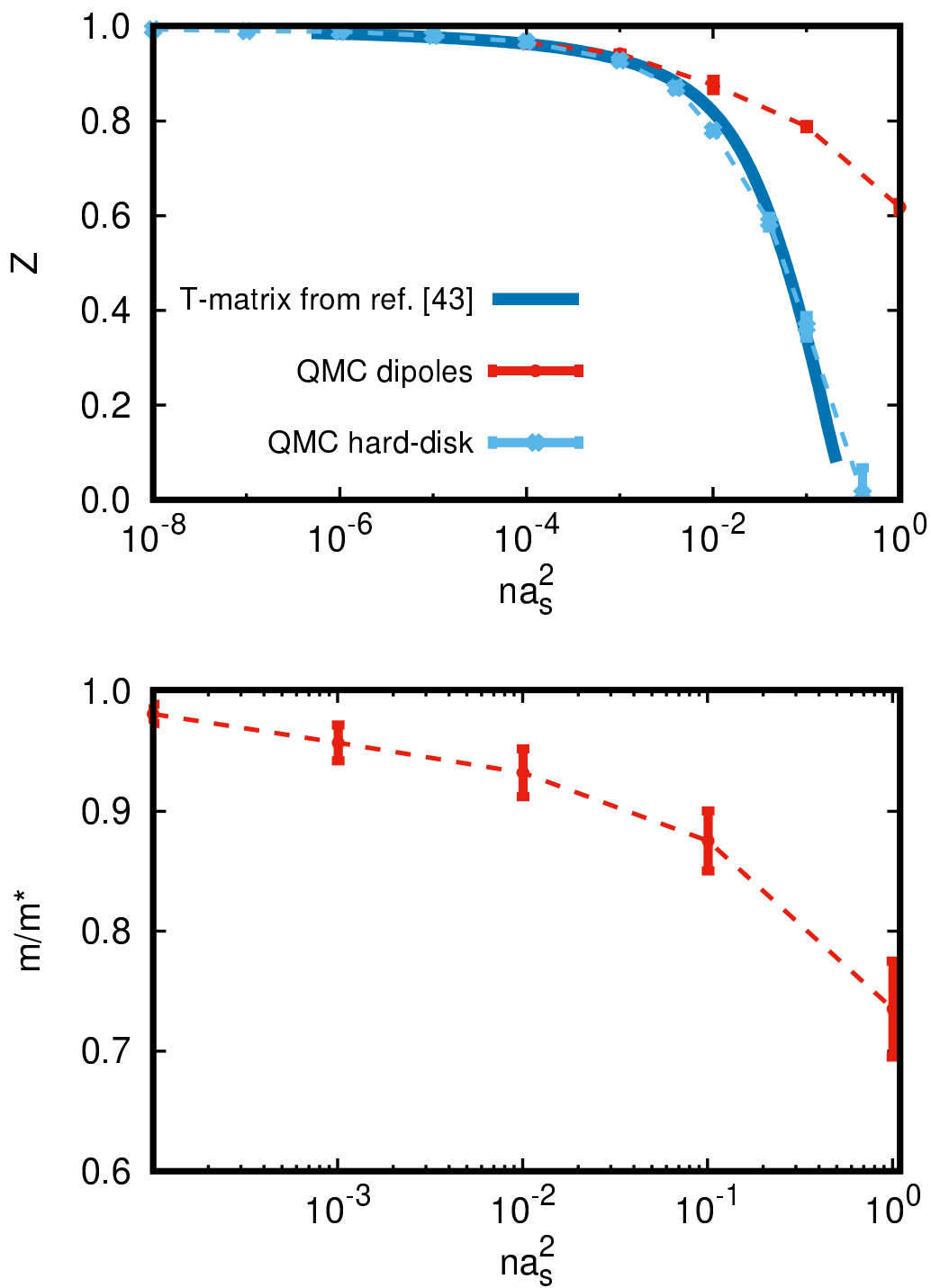

Fig. 4.11 Quasi-particle residue $Z$ (top panel) and effective mass of the polaron (bottom panel) as a function of the gas parameter $na_s^2$. Red symbols correspond to he dipolar system, blue ones to the hard-core impurity, and the solid blue line shows the many-body T-matrix theory results of Ref. [251]. Dashed lines are guides to the eye.

$N^\uparrow$ atoms. These two quantities read

$$\epsilon_p^{(N^\uparrow)} = \left[ E(N^\uparrow, 1) - E(N^\uparrow, 0) \right]_V \tag{4.25}$$

$$\mu^{(N^\uparrow)} = \left[ E(N^\uparrow + 1, 0) - E(N^\uparrow, 0) \right]_V \tag{4.26}$$

where $E$ stands total energy of the system. The subscript $V$ indicate that all the quantities are computed in the same fixed surface of area $L^2$, which leads to a density difference between the system with $N$ and $N + 1$ particles that only vanishes in the thermodynamic limit $N \to \infty$. In a large system, a single spin flip on top of a fully polarized state induces a change in the total energy which is

$$\Delta E_{\text{FLIP}} = \epsilon_p^{(\infty)} - \mu^{(\infty)} \tag{4.27}$$

If $\Delta E_{\text{FLIP}} > 0$, then the $P = 1$ state is robust against a single spin-flip excitation and might be the true ground state, while for a negative $E_{\text{FLIP}}$ such a state is unstable.

To evaluate in the region where the ground state may have polarization $P = 1$, we consider the density $n_\uparrow r_0^2 = 40$ (*cf.* Fig. 4.3, and notice that $n = n_\uparrow$ in the thermodynamic limit), where we perform calculations with the $\Psi_{\text{JS}}$ trial wave function. The polaron energy has a strong dependence on the system size, and so a finite-size-scaling study is necessary. We find that the DMC results are in reasonable agreement with a linear scaling, $\varepsilon_p^{(N^\uparrow)} = \varepsilon_p^{(\infty)} + \beta/N_\uparrow$ (see Fig. 4.12). The best-fit result is $\varepsilon_p^{(\infty)}/(\hbar^2 n_\uparrow/m) = 97.7(3)$.

The computation of both the polaron energy and the chemical potential as in Eqs. (4.25) and (4.26) is seriously affected by finite-size effects and statistical noise, as is evaluated from the difference of two energies of order $N$, while their difference is of order $1/N$. For this reason we directly compute the chemical potential from the equation of state of the Ferromagnetic phase that was reported in section 4.4 (*cf.* Table 4.2). The chemical potential computed from the equation of state reads

$$\mu^{(N_\uparrow)} = \left( 1 + n_\uparrow \frac{\partial}{\partial n_\uparrow} \right) E_{(N_\uparrow, 0)}. \tag{4.28}$$

To obtain the chemical potential using the above expression we fit the energy per particle to a curve of the form $A_{1/2} n_\uparrow^{1/2} + A_1 n_\uparrow + A_{5/4} n_\uparrow^{5/4} + A_{3/2} n_\uparrow^{3/2}$ [91], from which we obtain $\mu_{\text{JS}}^{(\infty)}/(\hbar^2 n_\uparrow/m) = 97.2(1)$, for $n_\uparrow r_0^2 = 40$. The error bar includes the statistical uncertainties and the systematic error due to finite-size effects.

In conclusion, we find that $\Delta E_{\text{FLIP}} > 0$ at density $n r_0^2 = 40$, pointing towards the stability of the $P = 1$ state. This is in agreement with the result that is found in section 4.4 from a nodal surface of the Jastrow-Slater type. However, and as it was already commented, an improvement of the nodal surface model used makes the ferromagnetic transition disappear. Finally, it is worth to comment that the large uncertainties on $\Delta E_{\text{flip}}$ make the results of this last section less reliable than those obtained by direct comparison of the energies in the two phases.

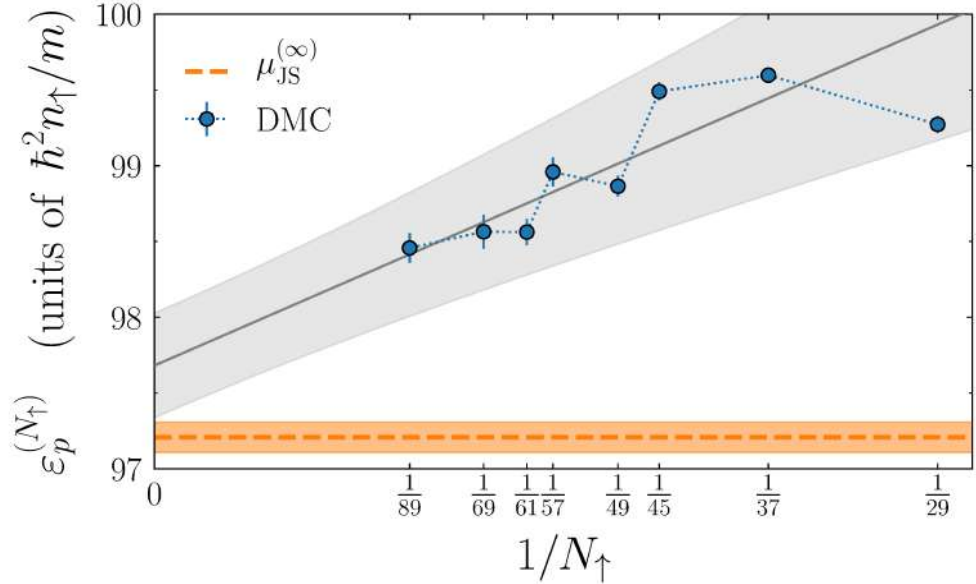

Fig. 4.12 Size scaling of the DMC polaron energy at a density $n_{\uparrow}r_0^2 = 40$, in units of $\hbar^2 n_{\uparrow}/m$ (blue circles – blue dotted line is a guide to the eye). For the linear fit (gray solid line), the shaded band includes both the statistical uncertainty and the systematic error due to the choice of the fit range. The $P = 1$ chemical potential $\mu_{\text{JS}}^{(\infty)}$ (orange horizontal dashed line, with shaded area representing its uncertainty) is extracted from the equation of state – see text. All calculations are based on the $\Psi_{\text{JS}}$ trial wave function.

## 4.6 Summary

By means of Diffusion Monte Carlo, we have studied the ground state properties of a two-component fermionic system of dipoles in two-dimensions. The system has been studied in different conditions of density an polarization. The determination of the equation of state of the system (in the paramagnetic phase $P = 0$) has allowed to show that the small deviations from the mean-field prediction persist even at the lowest densities considered in this work. Moreover the comparison of the EOS of the dipolar system to a hard-disk model permits to put an upper limit to the universal regime.

In the low-density regime, it is clear that the ground state of the system corresponds to the paramagnetic phase. However as density is increased, the interplay between correlations an inter-particle interactions become important, and it is not obvious to decide whether the ground state is polarized or not. The later would constitute an example of the *itinerant ferromagnetism* phenomena. We have shown that the usual backflow correlations are not accurate enough to answer this question, and that an improvement of the nodal surface is needed in order to obtain reliable results. Calculations performed with the best nodal surface available here discard the possibility of having a fully polarized phase before crystallization point. This result is in agrement with the one provided by the zero variance extrapolation method, and also with similar two-dimensional studies performed in liquid $^3$He [119] and the electron gas [201]. Regarding three-dimensional systems, no-polarized helium phase is found as the ground state [225], however, for the electron gas, and although for some time it has been accepted that its existence is possible in a narrow region of the phase diagram, this issue is still under debate [199, 253].

However, it is worth to remark that all the nodal surfaces employed so far rely on the backflow scheme, which is known to be mainly a s-wave correction, and thus, has only a minor effect on the $P = 1$ state due to the Pauli exclusion principle [224, 230].

In the last part of this chapter, we have studied the dipolar *Fermi polaron*, corresponding to the limit of an ultra-dilute concentration of impurities in a fully polarized Fermi bath. The computation of the polaron energy and its comparison to the one obtained with a hard-disk model allows us to determine the regime of universality of this problem. Indeed, non-universality effects start to appear at much lower densities compared with the pure paramagnetic phase. Moreover, our calculations show that some recent experiments are already in the regime of non-universality (*cf.* [207] for an experiment with $^{173}$Yb in the regime $na_s^2 \in [10^{-2}, 10^{-1}]$).

# Chapter 5

# Dysprosium liquid Droplets

In this chapter, we study the formation of quantum dipolar droplets of Dysprosium atoms. With the aim of understanding the properties of this system, we perform some Path Integral Ground State (PIGS) simulations – see section 2.5 for details about this method. This allows to evaluate relevant properties of the system in an exact way, and discuss the appearance of deviations from the mean-field scheme due to non-universal effects. In section 5.3.1, we evaluate the critical atom number for the dipolar Dysprosium droplets, whose comparison both with e-GPE prediction and experimental measurements constitutes the main result of this chapter. We also discuss results for other observables in order to get a better understanding of the differences between the PIGS and the extended Gross-Pitaevskii equation (e-GPE) predictions.

## 5.1 Introduction

In recent years, the observation of quantum droplets has attracted much attention in the ultracold gases community. These droplets are the result of the competition between attraction, repulsion, and quantum correlations. Despite their ultra-dilute density, (with their central density being orders of magnitude below the one of conventional liquids), they constitute an example of quantum self-bound objects. Up to now, these droplets have been observed in two different kind of experiments. The first one corresponds to Bose-Bose mixtures with attractive inter-species interaction and repulsive intra-species interaction. The second example are dipolar systems, where the inter-particle potential itself has attractive and repulsive contributions to the energy.

D. S. Petrov [62], in 2015, was the first to put forward the idea of employing Bose-Bose ultra-dilute mixtures with repulsive intra-species and attractive inter-species interaction to obtain self-bound systems. For a certain regime of experimental parameters, these droplets have been realized in different experiments [68, 69], within only some years of difference since their prediction. Motivated by this, a certain amount of theoretical work has been carried out in order to understand the underlying physics.

After the first mean-field studies, Quantum Monte Carlo (QMC) calculations have been carried out [70], showing that a precise description of these droplets requires going beyond the usual e-GPE description. The later relies on a universal approximation, where all the physics of the system can be described in terms of the gas parameter. However as density is increased, a more precise description would require the inclusion of at least, finite-range effects [70].

On the other hand, experiments carried out with ultracold dipolar atoms have also postulated themselves as good candidates for the study of ultra-dilute droplets [48, 51–53]. A lot of attention has been paid recently on arrays of droplets, as they are possible candidates for the supersolid phase predicted a few decades ago [54]. The idea is that, in an array of droplets, where translational invariance is broken, it can be possible to find phase coherence between the different droplets. Inspired by the theoretical work of Ref. [57], in which dipolar atoms were proposed to be confined in a cylindrical geometry, several experimental groups have studied the superfluidity of dipolar systems in this cylindrical configuration, where dipolar clusters are formed [52, 51, 53]. Indeed, for a certain combination of experimental parameters, phase coherent between the different clusters has been reported. Recently, the excitation spectrum of such systems has also been measured, characterizing the roton that appears in the spectrum [58, 60, 61]. Similarly to what happens with the case of droplets in Bose-Bose mixtures, the e-GPE, in which the Lee-Huang-Yang correction is included, captures qualitatively the physics of these systems. However quantitative deviations are found in some observable such as the roton spectrum [254] or the critical atom number [59].

The mechanism proposed by Petrov to form self-bound droplets has been generalized also to low-dimensional liquids [255]. In one-dimensional systems this has been studied both for mixtures of bosons [82] and for dipolar atoms [83]. Regarding two-dimensional configurations, droplet formation is predicted for different physical systems: for weakly interacting bosons [256], for Bose-Bose mixtures [255, 257, 258], and also for dipolar atoms [259].

One remarkable property of droplet systems, that distinguish them from the usual gaseous BEC ones is that, being self-bound objects, their density is larger, making their theoretical description more challenging. In particular the inclusion of beyond-mean-field effects in the form of the LHY correction [260] has some limitations when applied to droplet systems [261]. In particular, for the case of dipolar systems, the dipolar Lee-Huang-Yang (d-LHY) term includes in some cases a non-zero imaginary contribution to the energy that has to be removed *ad-hoc*, which constitutes an uncontrollable approximation [262, 263].

Here, we study quantum dipolar droplets of Dysprosium by means of the PIGS method, which allows to obtain exact results for the ground state of the system. By employing different potential models, we estimate the importance of non-universal effects. A systematic deviation from the e-GPE prediction is clearly found in several observables, such as the critical atom number needed to form a self-bound droplet, and the depletion of the condensate.

## 5.2 The system

We study a three-dimensional system of Dysprosium atoms with their magnetic moments fully polarized along the $Z$-direction. In analogy with experiments, the atoms are initially introduced into a trap, that in our model is characterized by an harmonic potential $V_{trap}$, which is removed after the droplets are formed.

The main ingredient required to perform a PIGS simulation is a good knowledge about the Hamiltonian. However, for ultracold gases this is not always possible, due to the lack of an accurate knowledge of the inter-particle potential. Here, we use a model that includes both the dipolar interaction and an effective short-range potential $V_{HC}$ with a repulsive core that prevents the system from collapsing. Assuming that all the dipoles are polarized along the $Z$ axis, the $N$-particle Hamiltonian reads

$$\hat{H} = -\frac{\hbar^2}{2m} \sum_{i=1}^{N} \nabla_i^2 + \frac{C_{dd}}{4\pi} \sum_{i<j}^{N} \frac{1 - 3\cos^2 \theta_{i,j}}{r_{i,j}^3} + \sum_{i<j}^{N} V_{HC}(r_{ij}) + \sum_{i}^{N} V_{trap}(r_i), \quad (5.1)$$

where $r_{i,j}$ and $\theta_{i,j}$ are the relative polar coordinates between the atoms, $m$ is the atomic mass, and $C_{dd} = \mu_0 \mu^2$ sets the strength of the dipolar interaction, with $\mu = 9.93\,\mu_B$ the magnetic dipole moment of $^{162}$Dy. In analogy to what was done in chapters 3 and 4, we use dipolar units, obtained from the characteristic dipolar length $r_0 = mC_{dd}/(4\pi\hbar^2)$ and the dipolar scale of energy $\epsilon_0 = \frac{\hbar^2}{mr_0^2}$, that allow to write the dipolar part of the Hamiltonian in a dimensionless way. In order to study whether there are universal properties in the system at the given conditions, we numerically solve the Hamiltonian for three different $V_{HC}$ models

$$
\begin{aligned}
V_{HC}^{(1)}(r) &= \frac{C_{12}}{r^{12}} - \frac{C_6}{r^6} \\
V_{HC}^{(2)}(r) &= \frac{C_9}{r^9} - \frac{C_6}{r^6} \\
V_{HC}^{(3)}(r) &= \frac{C_{12}}{r^{12}}.
\end{aligned}
\quad (5.2)
$$

The coefficient $C_6$ is known for Dysprosium [264] ($C_6 \approx 2.86 \cdot 10^{-2}$ in dipolar units). The other coefficients, $C_9$ and $C_{12}$, are fixed such that the complete interaction ($V_{HC}$ plus dipolar interaction) has the desired $s$-wave scattering length. This is accomplished by solving the low momentum limit of the scattering T-matrix, as is briefly described in the next section.

## Calculation of the s-wave scattering length for a two-body potential[1]

The $s$-wave scattering length of the combined two-body plus dipole-dipole interaction is obtained from the on-shell $T$-matrix, in the limit of vanishing momentum transfer. The $T$-matrix can be obtained by solving the Lippmann-Schwinger equation projected on a basis of free-particle eigenstates of definite momentum, according to the expression

$$
\begin{aligned}
T^{l,m}_{l',m'}(k',k) &= V^{l,m}_{l',m'}(k',k) \\
&+ \frac{\hbar^2}{M} \sum_{l_2,m_2} \int \frac{V^{l_2,m_2}_{l',m'}(k',q) T^{l,m}_{l_2,m_2}(q,k)}{\left( \frac{\hbar^2 k^2}{2M} - \frac{\hbar^2 q^2}{2M} + i\epsilon \right)} q \, dq \ ,
\end{aligned}
\tag{5.3}
$$

with $V^{l,m}_{l',m'}$ the matrix elements of the complete interaction, and $M$ the reduced mass of two atoms. Due to the anisotropy of the dipolar potential, the matrix elements of $T$, for different values of the quantum number $l$ and $l'$, are coupled. Moreover, the long-range character of the combined potential makes all partial waves to contribute significantly, even at low scattering energies [37]. Due to the nature of the dipolar interaction, different scattering lengths corresponding to different (coupled) channels appear and read

$$
a^{l,m}_{l',m} \equiv \lim_{k \to 0} \frac{\pi T^{l,m}_{l',m}(k,k)}{k},
\tag{5.4}
$$

with $l' = |l \pm 2|$. Still, the dominant one is the $s$-wave scattering length, corresponding to $l = l' = m = 0$. In practice, the low-momentum matrix elements $T^{l,m}_{l',m}(k,k)$ can be efficiently determined using the Johnson algorithm [265], which solves the Schrödinger equation and finds the logarithmic derivative of the wave function. Table 5.1 shows the $C_\alpha$ ($\alpha = 9, 12$) parameter of the hard core potentials used in this work for different scattering lengths expressed in Bohr radius $a_B$. These values have been chosen such that the resulting interactions do not have any two-body bound state.

| $a^{0,0}_{0,0}$ | $V^{(1)}_{HC}$ | $V^{(2)}_{HC}$ | $V^{(3)}_{HC}$ |
|---|---|---|---|
| $60 a_B$ | $4.83 \cdot 10^{-4}$ | $7.00 \cdot 10^{-3}$ | $2.07 \cdot 10^{-4}$ |
| $70 a_B$ | $7.10 \cdot 10^{-4}$ | $9.10 \cdot 10^{-3}$ | $3.47 \cdot 10^{-4}$ |
| $80 a_B$ | $1.06 \cdot 10^{-3}$ | $1.19 \cdot 10^{-2}$ | $5.79 \cdot 10^{-4}$ |
| $90 a_B$ | $1.61 \cdot 10^{-3}$ | $1.62 \cdot 10^{-2}$ | $9.6 \cdot 10^{-4}$ |

Table 5.1 Values for the parameter $C_\alpha$ ($\alpha = 9, 12$) of the potentials in Eq. (5.2), in dipolar units, for different scattering lengths in Bohr radius $a_B$.

---

[1]The parameters for the potential that reproduce the scattering length that are listed here have been derived by Juan Sanchez-Baena.

### 5.2.1 Extended Gross-Pitaevskii equation

In the following we compare our PIGS results to those from the extended mean field theory, (see Refs [49, 50, 266, 59], that is, the Gross-Pitaevskii equation with the LHY correction (e-GPE)

$$
\begin{aligned}
i\hbar\partial_t\Psi(\vec{r},t) \quad = \quad & \left[ -\frac{\hbar^2\nabla^2}{2m} + V_{\text{ext}} + g\,|\Psi|^2 - i\,\frac{\hbar L_3}{2}\,|\Psi|^4 \right. \\
& + \int \frac{C_{dd}}{4\pi}\,\frac{1 - 3\cos^2\theta}{\vec{r}^3}(\vec{r} - \vec{r'})\,|\Psi(\vec{r'})|^2\,d\vec{r'} \\
& \left. + \frac{32\,g\,\sqrt{a_s^3}}{3\sqrt{\pi}}\,\mathcal{Q}_5(\varepsilon_{\text{dd}})\,|\Psi|^3 \right]\,\Psi(\vec{r},t),
\end{aligned}
\tag{5.5}
$$

with $g = 4\pi\hbar^2\,a_s/m$ the contact interaction parameter, and the factor $\mathcal{Q}_l$ appearing in the dipolar Lee-Huang-Yang (d-LHY) correction is obtained from the following integral

$$
\mathcal{Q}_l = \frac{1}{2}\int_0^\pi d\alpha\,\sin\alpha\,\left[ 1 + \varepsilon_{dd}(3\cos^2\alpha - 1) \right]^{l/2},
\tag{5.6}
$$

with $\varepsilon_{dd} = \frac{a_{dd}}{a_s}$, $a_{dd} = \frac{1}{3}r_0 = \frac{mC_{dd}}{12\pi\hbar^2}$, and $L_3$ the three-body loss coefficient (for Dysprosium $L_3 \sim 10^{-41} - 10^{-40}$ m$^6$/s). It is important to notice that in the mean-field droplet regime $\varepsilon_{dd} > 1$ [49, 50], $\mathcal{Q}_l$ may have imaginary contributions. Although it constitutes an uncontrolled approximation, this imaginary term is usually neglected assuming that it is small.

In order to get the critical atom number curve (*cf.* Fig. 5.1) two different methods are used, leading to the same result. For both cases $V_{\text{ext}} = 0$ is chosen and the simulation is started with $N > N_{\text{crit}}$, initially prepared with an elongated Gaussian density distribution. Then the ground state is found by imaginary time evolution of the e-GPE. Next one can either simulate atom losses like in the experiment, or repeat this process of finding the ground state with lower atom number, until a stable solution cannot be found anymore. In the second method a certain uncertainty due to the step size that is chosen for the atom number is present. In the first method real-time evolution of the e-GPE is performed in order to simulate the dynamics of three-body losses. Due to the losses, the density and the effective two-body attraction reduces with time, until $N = N_{\text{crit}}$ where the energy is essentially zero. This leads to the evaporation of the droplet into the gaseous phase.

## 5.3 Results for the droplets

In this section we present our PIGS results for the dipolar Dysprosium droplets. We start evaluating the critical atom number, whose comparison with the experimental one and with the e-GPE prediction constitutes the main result of this chapter. After

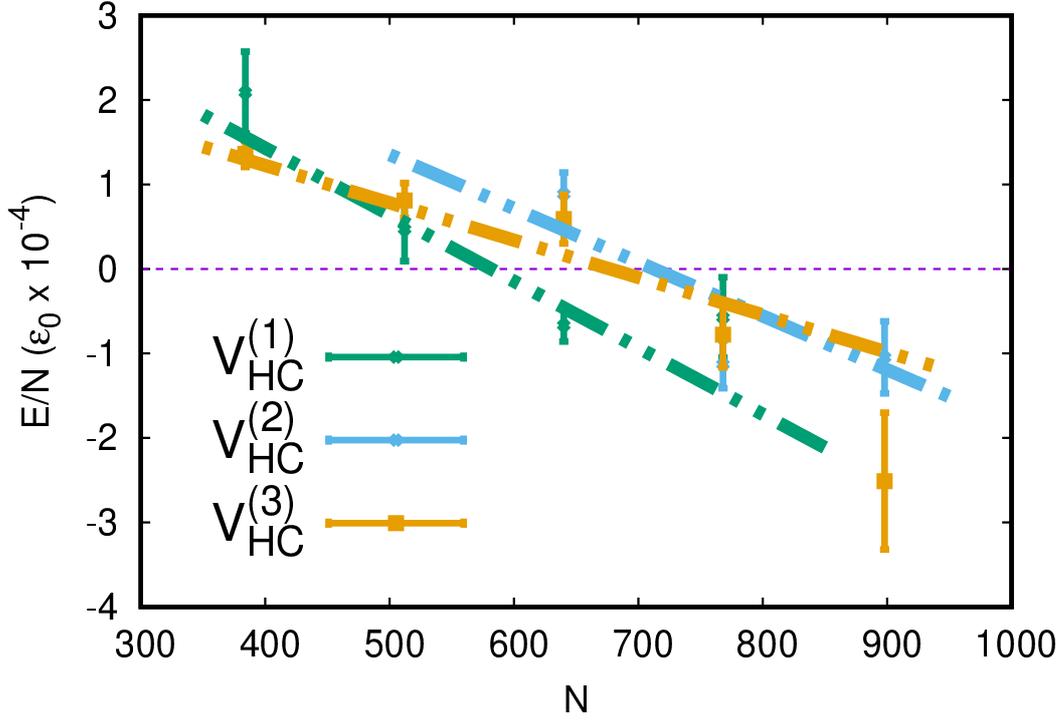

Fig. 5.1 Energy per particle as a function of the number of particles in units of $\hbar^2/mr_0^2$ for the dipolar system confined in a trap, evaluated with the three interaction potentials of Eq. (5.2) and for the $s$-wave scattering length $a_s = 60a_B$. The lines represent a fit to the data, and the intersection with the $E = 0$ axis defines the critical number of the model at this scattering length value.

that, we compare the PIGS density profiles of the droplets to those obtained with the e-GPE.

### 5.3.1   The critical atom number

One of the fundamental quantities that can be obtained from the PIGS simulations is the ground state energy, which is negative for a self-bound droplet. It is a well known fact that there is a critical number $N_c$ below which the system ceases to be self-bound ($E/N > 0$). Fig. 5.1 shows, for $a_s = 60a_B$, the ground-state energy obtained for the Hamiltonian in Eq. (5.1) with the three different $V_{HC}$ models of Eq. (5.2), as a function of the total number of particles. We also include a linear fit (near $E/N \sim 0$) that helps us to determine the point $N_c$ where the energy is zero. As it can be seen, different models lead to slightly different predictions (breaking the universality in the gas parameter), which adds an additional uncertainty to the evaluation of the critical number.

The resulting critical atom numbers for several values of the scattering length are shown in Fig. 5.2: Red symbols correspond to PIGS results with error bars that take into account the effect of the non-universality, according to the analyzed model potentials, as well as the statistical errors. The PIGS critical atom numbers are systematically lower than the e-GPE predictions (solid green line), and in good agreement with the experimental measurements[2]. The improvement of the PIGS predictions with respect to the e-GPE results points to the relevance of finite-range effects, similar to what is found in dilute Bose mixtures [67, 70]. As the scattering length increases, $N_c$ also increases and, unfortunately, since the computational cost of the simulation grows very rapidly with the number of particles, we can not reliably determine $N_c$ for scattering lengths larger than $a_s = 90a_B$.

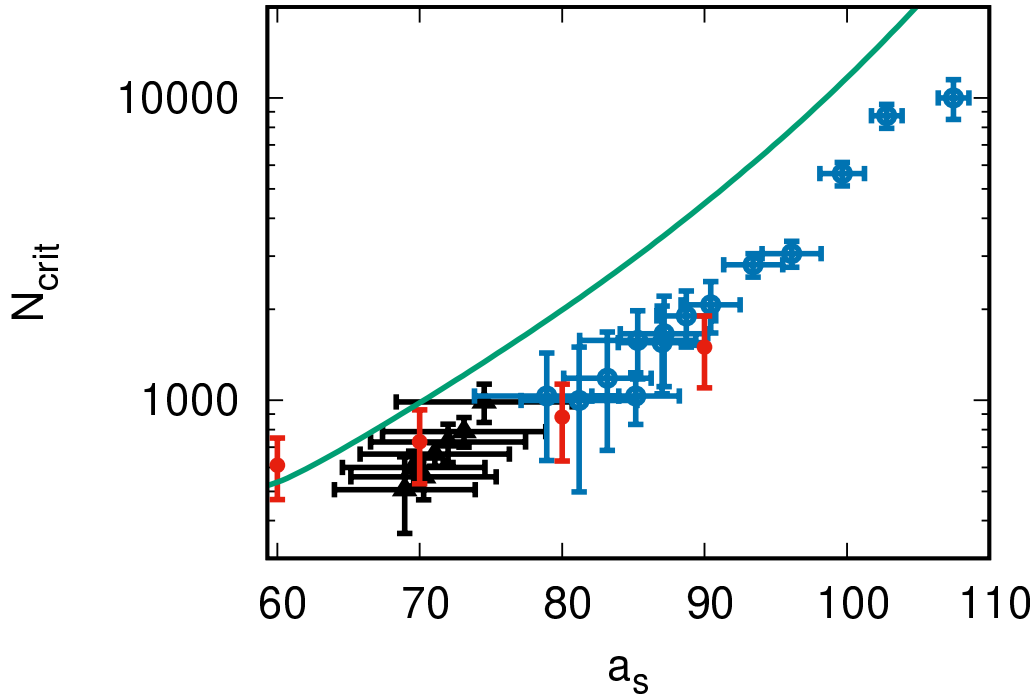

Fig. 5.2 Critical atom number $N_{crit}$ for a self-bound dipolar quantum droplet of $^{162}$Dy (black symbols) and $^{164}$Dy (blue symbols) atoms. Solid green line corresponds to the e-GPE prediction and red symbols are results from PIGS.

### 5.3.2 Density profiles

The droplets obtained by PIGS differ from those obtained in the e-GPE approximation, not only in the critical number, but also on the density profiles. Figure 5.3 shows the integrated density profiles along the axial directions of the droplet, obtained from both methods, for 1000 (left panel) and 2000 (right panel) atoms, and for a scattering length of $60a_B$. As it can be seen, for these (low) particle numbers, the profiles are quite different, with the PIGS one broader and with a lower central density. Still, the difference reduces when the number of atoms grows from 1000 to 2000. Increasing the atom number even more, we expect the differences to be reduced.

[2]We thanks Fabian Böttcher and Matthias Wenzel for providing the experimental results and the e-GPE prediction for the critical atom number shown in Fig. 5.1 as well as the mean-field density profiles of Fig. 5.3. For details about the experimental measurement of the critical number see Ref. [59], and for the mean-field density profiles of Fig. 5.3 – see the Phd. Thesis of Matthias Wenzel.

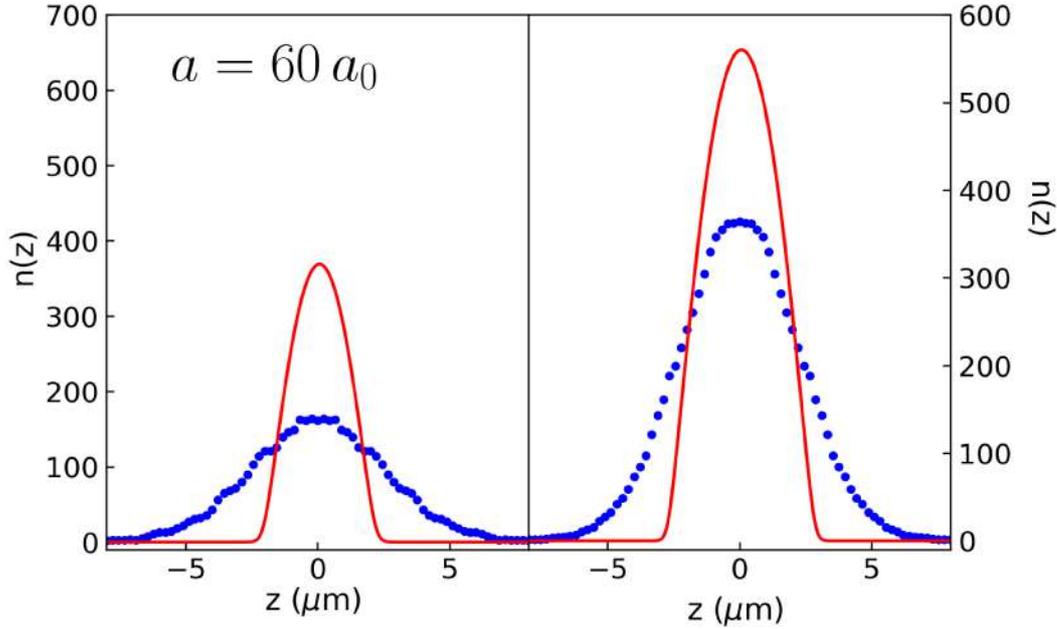

Fig. 5.3 Density profiles along the $Z$ direction in the e-GPE (red solid line) and PIGS (blue dots) approximations for a scattering length $a = 60a_B$. The left and right panels show the e-GPE results for $N = 1000$ and $N = 2000$ atoms, compared with the PIGS results for $N = 1024$ and $N = 2048$ atoms, respectively. Each profile has been properly normalized to its corresponding particle number.

Further improvement on the experimental side, in order to obtain high resolution images of the smaller droplets that are accessible with PIGS ($N \approx 2000$), would help to check our theoretical predictions. On the theoretical side, unfortunately, due to the big size of experimentally accessible droplets ($N \approx 20000$) it is not possible to perform direct PIGS calculations of them. An alternative approach, that has been already used in Bose-Bose droplets [70] consists in using the Local Density Approximation (LDA) based on the equation of state (EOS) computed with PIGS. The equation of state for the bulk has a lower computational cost in Monte Carlo, as it can be computed in a box with periodic boundary conditions with a reduced number of particles (*cf.* section 5.4.1).

## 5.4   Results for the bulk

In order to understand where the differences between the mean-field approach and the PIGS prediction come from, it is worth to have a look at other observables that are easily computed in the bulk system. We start computing the EOS of the system and the short-inter-particle distance structure, that is accessible through the radial distribution functions. Finally, we evaluate the depletion of the condensate, to analyze the validity of employing the e-GPE approach.

### 5.4.1   Equation of state

The Hamiltonian of Eq. (5.1) can also be used to study the homogeneous fluid system. In order to reproduce the uniform infinite system, we set $V_{trap} = 0$ and perform simulations in a box of length $L = \sqrt[3]{N/n}$, with $N$ the total number of particles in the box and $n$ the density. Computing the EOS of such a Hamiltonian allows to determine the equilibrium density for a given scattering length, which should equal the central density of a saturated quantum droplet.

To calculate energies we use the PIGS estimator that was introduce in Eq. (2.141). The top panel of figure 5.4 shows results for the EOS for different scattering lengths. These calculations are performed with 512 particles and using $V_{HC}^{(1)}$ (see Eq. (5.2)), notice that the dipolar potential is always included) as a repulsive potential. For each potential and once the s-wave scattering length is fixed, we determine the equilibrium density ($a_s \in [60a_B, 140a_B]$, that are on the experimentally available region). It is worth to notice that an equilibrium density is found even for the largest scattering length, for which $\varepsilon_{dd} = \frac{a_{dd}}{a_s} < 1$, and where in principle the e-GPE does not predict neither the formation of a liquid state nor droplets [50, 49]. In the bottom panel of the same figure, we show calculations in which the scattering length is fixed to the value $a_s = 90a_B$. Blue and red circles correspond to calculations performed with 512 particles using different model potential ($V_{HC}^{(1)}$ and $V_{HC}^{(2)}$ respectively) what allows to determine an upper limit to the universal regime for densities $n < 10^{21}$. In that regime, the system can be studied in terms of the gas parameter $na_s^3$. In the same plot we show the EOS evaluated with the $V_{HC}^{(1)}$ potential with 128, 256 and 512 particles to estimate the relevance of finite size effects in this calculation. The potentials $V_{HC}$ are short-ranged, and therefore the main contributions to finite size effects are expected to come from the dipolar part. In order to understand the small finite-size effects that the data on the bottom panel of Fig. 5.4 hints, we evaluate the tail of the potential for

distances $r > L/2$:

$$
\frac{E_{tail}^{dip}(n, L)}{N} = \frac{1}{2} \int_{|\mathbf{r}|=L/2}^{\infty} V_{dip}(\mathbf{r}) g(\mathbf{r}) \rho d\mathbf{r}
$$

$$
= \frac{1}{2} \int_{L/2}^{\infty} \int_{0}^{\pi} \frac{1 - 3\cos^2\theta}{r^3} \rho 2\pi r^2 \sin\theta dr = 0, \quad (5.7)
$$

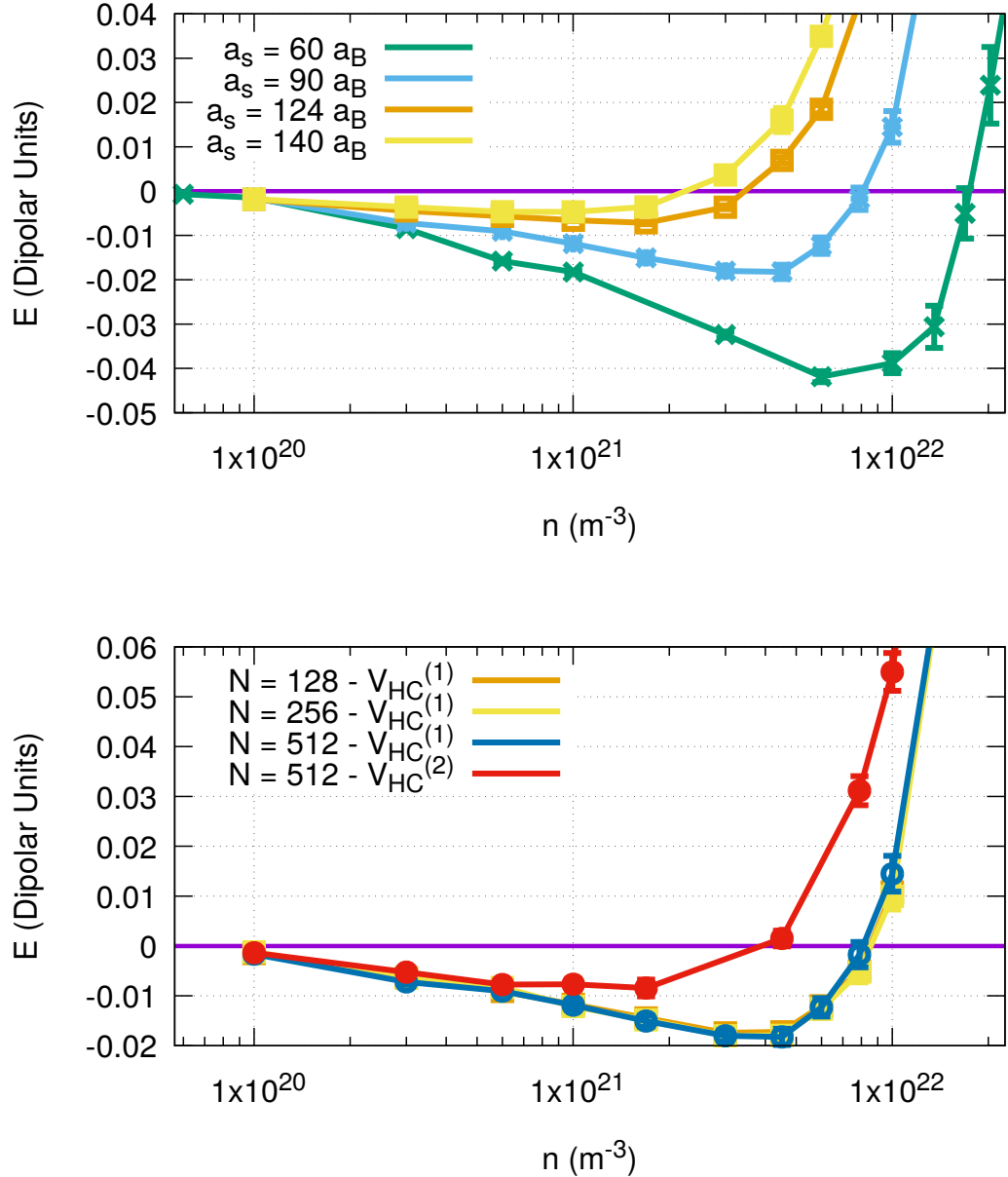

Fig. 5.4 Equation of state of the bulk dipolar system. Top panel: EOS for the $V_{HC}^{(1)}$ potential and for different scattering lengths. Bottom panel: EOS evaluated for a fixed scattering length $a_s = 90a_B$ and different particle numbers for the $V_{HC}^{(1)}$ and $V_{HC}^{(2)}$ potential.

the above integral vanish once the angular part is integrated assuming that the density $\rho$ is constant and that $g(r) \approx 1$ for $r > L/2$. The first assumption is justified as long as no clusters (or filaments) are formed inside the box, the second one is justified in next section – see Fig. 5.5. On the other hand, the contribution to the potential tail coming from the short range potential of Eq. (5.2) is small but finite, and is evaluated in a similar way

$$\frac{E_{HC,tail}^{(1)}(n,L)}{N} = \frac{1}{2}\int_{|\mathbf{r}|=L/2}^{\infty} V_{HC}^{(1)}(\mathbf{r})g(\mathbf{r})\rho d\mathbf{r} \approx \frac{2\pi\rho}{3}\left(\frac{C_{12}}{3(L/2)^9} - \frac{C_6}{(L/2)^3}\right) \quad (5.8)$$

$$\frac{E_{HC,tail}^{(2)}(n,L)}{N} = \frac{1}{2}\int_{|\mathbf{r}|=L/2}^{\infty} V_{HC}^{(1)}(\mathbf{r})g(\mathbf{r})\rho d\mathbf{r} \approx \frac{2\pi\rho}{3}\left(\frac{C_9}{2(L/2)^6} - \frac{C_6}{(L/2)^3}\right) \quad (5.9)$$

$$\frac{E_{HC,tail}^{(3)}(n,L)}{N} = \frac{1}{2}\int_{|\mathbf{r}|=L/2}^{\infty} V_{HC}^{(1)}(\mathbf{r})g(\mathbf{r})\rho d\mathbf{r} \approx \frac{2\pi\rho}{9}\frac{C_{12}}{(L/2)^9} \quad (5.10)$$

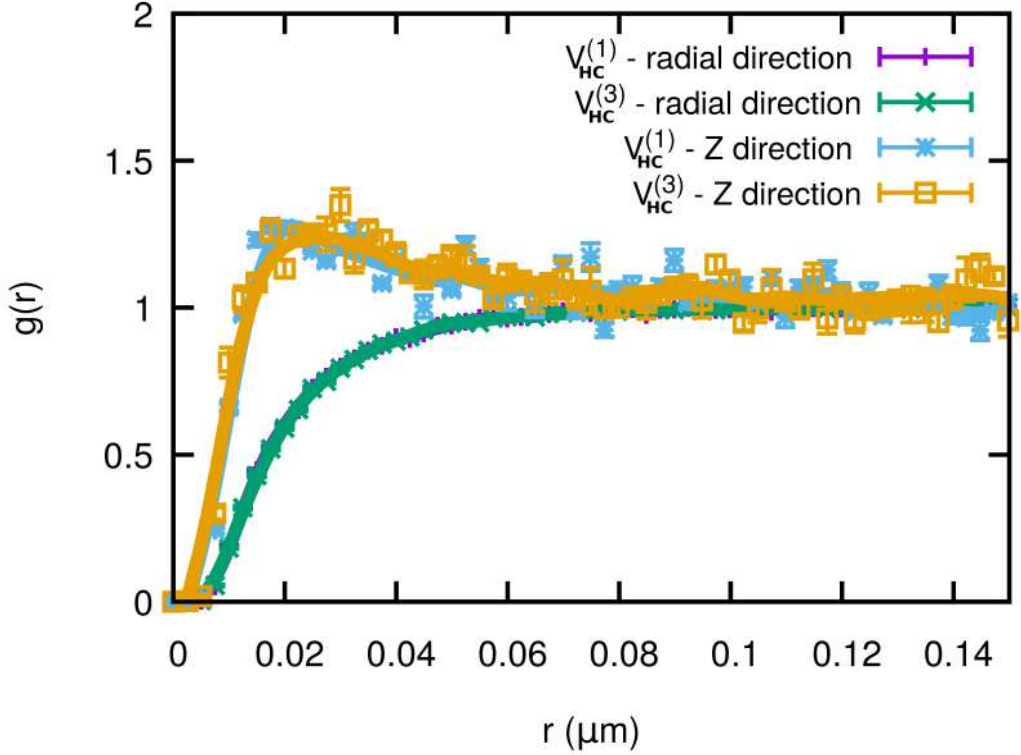

Fig. 5.5 Pair distribution function $g(\mathbf{r})$ for the bulk system at a density of $n = 5.88 \times 10^{21}\,\mathrm{m}^{-3}$, corresponding to the central density of a saturated quantum droplet at $a_s = 60\,a_B$ evaluated with the two model potentials $V_{HC}^{(1)}$ and $V_{HC}^{(3)}$ of Eq (5.2).

### 5.4.2 Radial distribution functions

To understand the role of correlations in this system, we evaluate the two-body radial distribution function for the equilibrium density $\rho = 5.88 \cdot 10^{21}\,\mathrm{m}^{-3}$ corresponding to the EOS evaluated at the scattering length of $a_s = 60 a_B$. We calculate it along the direction in which the dipole moments are aligned ($Z$) and also in the orthogonal (radial) direction. In Fig. 5.5 we show these quantities evaluated with the two different potential models ($V_{HC}^{(1)}$ and $V_{HC}^{(3)}$ of Eq. (5.2)), and including the dipolar interaction. The comparison of the results shows that the g(r) is not seriously affected by the model potential. In the radial direction, the pair correlation function is a monotonic function of the distance that resembles the one of a weakly interacting system. On the other hand, along the polarization direction the $g(r)$ shows signatures of local ordering, as it is highlighted by the broad peak at short distances. The hole at short distances is caused by the repulsive core of the two-body model potential.

### 5.4.3 Depletion of the condensate

An important quantity that gives an idea of the validity of employing the e-GPE equation is the depletion of the condensate $\frac{n_{dep}}{n}$, or its complementary quantity, the condensate fraction $\frac{n_c}{n}$

$$\frac{n_{dep}}{n} = 1 - \frac{n_c}{n}. \tag{5.11}$$

In principle, one expects that the e-GPE would give a good description of the system when almost all the system remains in the condensate. In the Bogoliubov approximation, for a contact interacting gas, it is possible to estimate the depletion of the condensate according to [260]

$$\left(\frac{n_{dep}}{n}\right)_{Bog} = \frac{8}{3\sqrt{\pi}}\sqrt{na_s^3}. \tag{5.12}$$

When dipolar inter-particle interactions are present in the system, the above expression is modified in a similar way as the LHY term in the e-GPE (see Eq. (5.5)) Taking it into account, the depletion of the condensate reads [262, 263]

$$\left(\frac{n_{dep}}{n}\right)_{Dip} = \frac{8}{3\sqrt{\pi}}\sqrt{na_s^3}\,\mathcal{Q}_3(\epsilon_{dd}), \tag{5.13}$$

with $\mathcal{Q}_3$ defined in Eq. (5.6).

In PIGS, the condensate fraction can be computed from the long-distance asymptotic behavior of the One-body Density Matrix (OBDM). In our algorithm, this is achieved by employing the *worm* algorithm and evaluating the estimator that appear in Eq. (2.157). In Fig. 5.6 we compare the predictions of the Bogoliubov theory, both for a fluid without (LHY, red dashed line – see Eq. (5.12)) and with dipolar interaction (d-LHY, orange dashed line – see Eq. (5.13)), with our PIGS results (blue symbols – dashed blue line is a guide to the eye) for a scattering length of $a_s = 60 a_B$. First, it is worth noticing that, in the Bogoliubov approximation, at the equilibrium density that is calculated with the $V_{HC}^{(1)}$ potential in section 5.4.1 for this value of $a_s$ ($\rho_{eq} \sim 6 \cdot 10^{21} \mathrm{m}^{-3}$), the correction to the condensate fraction due to the presence of dipolar interaction is roughly twice the value of that corresponding to a gas with no dipolar terms. For densities $\rho \leq 10^{-21} \mathrm{m}^{-3}$, our PIGS results coincide with the d-LHY prediction, which is in agreement with a possible universal behavior below that density, as discussed in section 5.4.1. On the other hand, as the density is increased and approaches to the equilibrium density, clear deviations from the d-LHY correction appear, although the condensate fraction still has large values $\frac{n_c}{n} \sim 90\%$, which are much larger than those observed in more correlated systems such as $^4$He, where $\frac{n_c}{n} \sim 8\%$ – see, for example, Ref. [7].

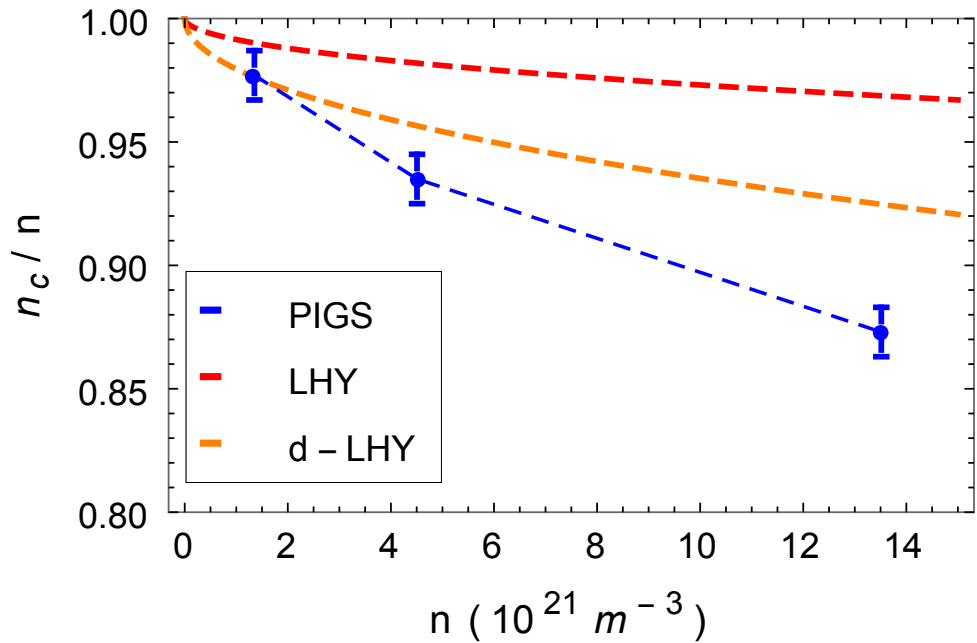

Fig. 5.6 Condensate depletion as predicted by the PIGS calculations and the Bogoliubov theory without (LHY, see Eq. (5.12)) and with dipolar interaction (d-LHY, see Eq. (5.13)), for a scattering length of $a_s = 60\,a_B$.

## 5.5  Summary

In this chapter we have studied the ground state properties of quantum dipolar droplets of Dysprosium atoms by means of the PIGS method. We have evaluated the minimum critical atom number that is needed to obtain a self-bound droplet as a function of the scattering length. Being a measurable quantity in current state-of-the-art experiments, it can be used to benchmark different theoretical predictions. In particular we have compared our results with those coming from a e-GPE approximation, showing that the critical atom number predicted by PIGS is systematically lower than the one predicted by the mean-field approach. However, our PIGS results for different model potentials agree well with the experimental measurements done with $^{162}$Dy and $^{164}$Dy experiments in the range of scattering lengths that we can study $a_s \in [70a_B, 90a_B]$. Unfortunately, PIGS calculations become too computationally demanding as the number of particles is increased, being inaccessible for atom numbers larger than 2000. We have also shown that, by performing simulations with different model potentials, non-universal effects appear in this system: we have included the uncertainty coming from the lack of knowledge of the inter-atomic potential by including the non-universality effects into the error bars of the critical number prediction.

As direct simulations of saturated droplets are not possible due to the huge number of particles needed ($N > 10^4$), we have performed simulations of the infinite bulk system, whose conditions at the equilibrium density should be comparable to those at the center of a saturated droplet. Calculations performed with different model potentials situate the equilibrium density outside the universal regime. Finally, we have evaluated the condensate fraction and compared it to the Bogoliubov prediction for the dipolar gas. The e-GPE prediction coincides with the PIGS result for the lower densities considered, that would correspond to the densities of non-saturated droplets. However, the e-GPE overestimates this quantity near the equilibrium density. This deviation questions the accuracy of employing e-GPE to describe saturated droplets.

# Chapter 6

# Conclusions.

In this Thesis we have performed numerical Monte Carlo simulations of different dipolar systems. When dealing with bosonic systems, the employment of these techniques yields exact results. This is true both at zero temperature, when using the Diffusion Monte Carlo (DMC) or the Path Integral Ground State (PIMC) algorithms, and at finite temperature with the Path Integral Monte Carlo (PIMC) method. The capability that PIMC offers to obtain exact results at finite temperature is remarkable, making this method extremely productive. On the other hand, for fermionic systems, the *sign problem* emerges, which in general makes the signal-to-noise ratio of the calculation unacceptable. To tackle with this problem we use the Fixed Node (FN) approximation, which provides controlled but otherwise variational solutions to the problem at zero temperature.

Regarding the experimental realization of the two-dimensional (2D) systems studied in this Thesis, two assumptions about the theoretical model need to be discussed, in view of a possible connection with experiments of ultracold dipolar atoms or molecules: The reduced dimensionality and the shape of the inter-particle potential. The experimental realization of a two-dimensional system is carried out by imposing a tight confinement along the transverse direction, characterized by the harmonic-oscillator length $a_z$ and by its typical energy scale $\hbar\omega_z$. At zero temperature, the condition to be in the two-dimensional regime reads $\mu \ll \hbar\omega_z$, where $\mu$ is the chemical potential, which is of the order of $E_{\text{IFG}}$. Such condition corresponds to $nr_0^2 \ll (r_0/a_z)^2$, showing that the maximum allowed value of $nr_0^2$ depends on the ratio $a_z/r_0$. As an example, the dipolar length for Dysprosium atoms is $r_0 \approx 20$ nm and a realistic value for the trapping potential is $a_z \approx 500$ nm. Then, the confined system can be described by a two-dimensional model up to $nr_0^2 \approx 10^{-3}$.

The second issue is that our 2D model neglects the presence of an additional contact interaction, on top of the dipolar repulsion. On the one hand, this is partially justified by the fact that the two-dimensional scattering length for a three-dimensional contact interaction with scattering length $a_{3D}$ scales as $\exp(-\sqrt{\frac{\pi}{2}}a_z/a_{3D})$, in presence of transverse confinement [221]. Thus it is strongly suppressed when $a_{3D} \ll a_z$, which is the typical case away from Feshbach resonances. On the other hand, the two-dimensional scattering length of the dipolar potential is of the order of $r_0$.

In what follows, we summarize the main results of this Thesis.

## Superfluid properties of bosonic dipolar system in two-dimensions

In chapter 3, we have characterized the superfluid properties of the different phases appearing in the phase diagram of the two-dimensional (2D) dipolar system. It is a well known fact that two-dimensional systems can only support a condensate in the limit of zero temperature. However the existence of a quasi-condensate is possible, showing up as algebraic decay of the correlations. In much the same way, 2D systems can be superfluid at finite temperature. The transition from a superfluid to a non-superfluid phase in this geometry is driven by the appearance of topological defects as temperature is increased, being of the Berenzinskii-Kosterlitz-Thoulesss (BKT) type. From a phenomenological point of view, the main difference between the superfluid transition in two-dimensional systems with respect to the three dimensional case is that the superfluid density performs a jump at the critical temperature, instead of having a smooth decay as the temperature is increased. In section 3.3.2 we have studied the system at zero temperature ($T = 0$), computing the superfluid and condensate fractions to find that, both in the gas and in the stripe phase, these quantities are finite. The study is completed in section 3.4 by extending it to finite temperature. This is achieved by employing PIMC, what allows to determine the critical temperature at which the BKT transition occurs, both in the gas and in the stripe phase. Therefore, we suggest that the dipolar stripe phase is a good candidate for the supersolid state of matter.

## Two-dimensional properties of the dipolar stripe phase

Also in chapter 3 (see section 3.5), and as a check to validate the results obtained for the BKT transition in the stripe phase, we have analyzed whether it is possible to understand this phase as an ensemble of one-dimensional systems. With this aim in mind, we have compared our PIMC results for the superfluid density with the predictions that Luttinger Theory (LL) offers for one-dimensional systems. The LL theory predicts a scaling law for the superfluid fraction as a function of the length of the system and the temperature. Indeed, in 1D systems superfluidity can only arise as a finite size effect, vanishing for infinite large systems. Our results show that not only the scaling predicted by the LL theory do not apply to the dipolar stripe phase, but also that the superfluid signal in this phase presents large values for conditions of temperature and system size for which the superfluidity in a 1D system would be zero or almost zero.

# Equation of state of the fermionic dipolar system in two-dimensions

A study of the unpolarized phase of the two-component mixture of dipolar fermions, both at low and high density is presented in chapter 4 – see section 4.3. There, both the Equation Of State (EOS) and the radial distribution functions are reported. These calculations have been done with fixed-node DMC calculations employing plane waves determinant as the nodal surface, which is accurate enough for the low densities considered in this section. The comparison of our results for the dipolar system with those provided by a hard-disks model and with the mean-field prediction, allows to establish a regime of universality for values of gas parameter $na_s^2 \ll 10^{-2}$.

# Absence of itinerant ferromagnetism in two component dipolar Fermi system

Although in principle it is away from the limits of current state-of-the-art experiments (see discussion at the beginning of the this chapter), it is of theoretical interest whether it could exist a polarized state as the ground state of the system. This phenomena is referred in the literature as *itinerant ferromagnetism* and has been a long-standing topic in the condensed matter community, as its solution is extremely sensitive to quantum correlations. When performing DMC calculations for fermionic systems with the FN approximation, only upper bounds to the exact ground state energy can be obtained, whose quality depends on the nodal surface employed in the trial wave function. We have discussed this problem for the 2D dipolar system, to show that the usual backflow-corrected wave function, is not enough to give a reliable answer to this problem. This can be attributed to the extremely small energy difference that is found between the ferromagnetic and the paramagnetic phases. The most accurate calculations, performed with the best known nodal surface, discard the possibility of having an itinerant ferromagnetic phase in the 2D dipolar system. Similar results have been found both for liquid ${}^3$He and the electron gas in two-dimensions [119, 201]. However, as all the trial wave functions employed in these calculations rely on a backflow approximation (that constitutes a correction mainly in the s-wave channel), it barely corrects the fully polarized state [224, 230]. For this reason, a final answer to this problem may need a more accurate description of the many-body wave function, enhancing p-wave and higher partial waves corrections to the nodal surface.

## The dipolar Fermi polaron

In this thesis we have also studied the repulsive dipolar Fermi polaron, consisting on a atomic impurity immersed in a bath of fully polarized identical dipoles. In the particular case that we study, the impurity has the same mass and dipolar moment as the rest of particles of the bath, and thus, the only difference between the impurity and the bath relies on the Fermi statistics of the later. By means of fixed-node DMC we compute the polaron energies in the weakly interacting regime. The comparison of the results for the dipolar system with those obtained with a model in which an impurity interacts by a hard-disk potential with an ideal Fermi gas, allows us to find a regime where both models essentially coincide with the mean-field prediction. We find that the polaron problem is more challenging than the evaluation of the EOS of the unpolarized system. This is reflected in the appearance of non-universal effects at lower values of the gas parameter $na_s^2 \gg 10^{-5}$ . This constitutes an exciting finding because it situates recent experiments out of the universal regime (*cf.* Ref. [207] for an experiment with $^{173}$Yb in the range $na_s^2 \in [10^{-2}, 10^{-1}]$), making effective-range effects achievable in present and future ultracold atoms experiments. Precisely in this regime it is where we find that the quasi-particle picture starts to fail, with values of the quasi-particle residue below 80%. These results point out that more efforts should be put on the theoretical side in order to correctly describe this problem.

## Dipolar Dysprosium Droplets

Finally, in chapter 5, we have focused on the study of quantum dipolar droplets of Dysprosium atoms by means of the Path Integral Ground State (PIGS). The main result obtained is the evaluation of the critical atom number, which is the minimum number of atoms that is needed to obtain a self-bound droplet. Our Monte Carlo results are compared to the extended-Gross-Pitaevskii equation (e-GPE) ones, showing a systematic overestimation of the critical atom number $N_c$ when using the e-GPE. On the other hand, the PIGS results are in good agreement with the available experimental measurements for this quantity, corresponding to $^{164}$Dy and $^{162}$Dy in the range where we are able to evaluate it with PIGS ($N_c < 2000$). Calculations performed with different model potentials show that effects of non-universality can be observed in the range of densities that are spanned in the experiments, which is reflected, for example, in a non-universal equilibrium density (that we have determined by studying the bulk system for different model potentials). This density would be the one in the interior of a saturated droplet. The sensitivity of the density profiles to the details of the many-body approach employed to evaluate it, make them to be good candidates as a testbed for the different theoretical approaches. To this aim, new experimental data regarding droplets of different sizes would be useful. Finally, and also for the bulk system, we have evaluated the Bose-Einstein condensate fraction for dipolar matter. A comparison between the PIGS results with the Bogoliubov (LHY) prediction for this quantity shows that for low values of the density, both approaches are in agreement. However, as density is increased the LHY prediction underestimates the depletion of the condensate. High values of this quantity (of about 10% or more) limit the validity of the e-GPE framework to describe quantum droplets, as it relies on the assumption that almost all the system is in the condensate.

# Appendix A

# Related publications

Here, we present a list with the papers in which the work that is discussed in this Thesis has been published.

# Appendix B

# Details about the Hard-Disks model employed to compare with the dipolar system in Chapter 4

Here we present a description of the Hard-Disks (HD) model that is employed for the comparison with the dipolar system in chapter 4. This model is used in section 4.3 to discuss the equation of state of a two-dimensional fermionic system in the low-density regime and again in section 4.5 where the repulsive Fermi polaron is presented. In both cases, the comparison between the dipolar and the HD model, allows to determine a regime of universality. Results regarding this model have been originally published in Refs. [190, 56]

## Low density equation of state

The model that we consider to study the low density equation of state of the two-dimensional repulsive Fermi system, composed of $N^\uparrow$ spin-up and $N^\downarrow$ spin-down atoms, is described by the Hamiltonian

$$H = -\frac{\hbar^2}{2m} \sum_{i=1}^{N^\uparrow} \nabla_i^2 - \frac{\hbar^2}{2m} \sum_{i'=1}^{N^\downarrow} \nabla_{i'}^2 + \sum_{i,i'} V(r_{ii'}), \tag{B.1}$$

where simple indexes $i$ refer to spin-up ($\uparrow$) particles and primed ones to spin-down ($\downarrow$) particles. To describe the unpolarized phase we consider the total number of particles $N = N^\uparrow + N^\downarrow$ with $N^\uparrow = N^\downarrow$. And $V(r_{ii'})$ is the HD inter-particle potential, that acts only between pairs of particles of different spin component. It reads

$$V^{\text{int}}(r) = \begin{cases} \infty & r \leq R \\ \\ 0 & r > R. \end{cases} \tag{B.2}$$

An important difference between this model and the dipolar one introduced in chapter 4 is that, in the present one, same spin particles do not interact between each other. It is important to recall that, in 2D, the scattering amplitude depends logarithmically on momentum, so that the definition of the scattering length $a_s$ involves an arbitrary constant. Two alternative conventions are typically used. In the first one, $a_s$ is defined to fulfill $a_s = R$ for a hard-core potential, so that the two-body scattering wave function vanishes at $r = a_s$ [216] in analogy with the 3D case. This is the convention that we use in this work. With such definition, the two-body binding energy for an attractive contact interaction is $|\epsilon_b| = 4\hbar^2/(ma_s^2 e^{2\gamma})$, with $\gamma \simeq 0.577$ Euler's constant [203, 190]. Another definition of the 2D scattering length (now indicated by $b$) aims at maintaining a simple relation with the binding energy $|\epsilon_b| = \hbar^2/(mb^2)$, in analogy with the 3D attractive problem [267, 251]. The relation between the two conventions is $b = a_s e^{\gamma}/2$.

For the HD model, all the physics in the system is condensed into the gas parameter $na_s^2$, and thus, this is the parameter that we use to compare different models. We also notice that the closer $na_s^2$ is to unity, the less this model is expected to faithfully describe the repulsive branch of the polaron, since coupling to molecular states is completely ignored.

## The Fermi polaron

When the repulsive Fermi polaron is studied in section 4.5, an analog to the previous model can be used. In this case a single spin down impurity is immersed in a fully polarized bath of spin-up particles. The Hamiltonian describing this system, is the equivalent to the one introduced in Eq. (4.21), that in this particular case reads

$$\hat{H} = -\frac{\hbar^2}{2m}\nabla_\downarrow^2 - \frac{\hbar^2}{2m}\sum_{i=1}^{N_\uparrow}\nabla_i^2 + \sum_i^{N^\uparrow} V(r_{i\downarrow}), \tag{B.3}$$

with $V(r)$ a potential of the form of Eq. (B.2) accounting for interactions between the impurity and the bath. In this model the bath is considered to be non-interacting.

For both the models in Eqs. (B.1) and (B.3) the importance sampling technique (*cf.* section 2.3.2 for details) is employed when performing the DMC simulations. In this framework, Jastrow correlations are implemented only between different spin-component pairs of particles since intra-species interaction is neglected in the HD model ($f_{\uparrow\uparrow}(r) = f_{\downarrow\downarrow}(r) = 1$). Then, in this implementation, only one variational parameter has to be optimized $r_{HD} \leq L/2$, corresponding to the distance at which the conditions $f_{\uparrow\downarrow}(r_{HD}) = 1$, $f'_{\uparrow\downarrow}(r_{HD}) = 0$ are imposed.

# Appendix C

# Iterated Backflow procedure.

To benchmark our DMC calculations and test their validity, we compare our results with those obtained with an independent implementation of DMC algorithm by Markus Holzmann [118–120][1]. In this appendix, we summarize the nodal surfaces that have been used to construct the trial wave functions of this implementation. Results are summarized in Table C.1.

Wave functions labeled as JS[MH] and JS3 correspond to calculations performed with the usual Jastrow-Slater form including two and three-body Jastrow correlations respectively[2]. In this implementation, the Jastrow wave functions are parameterized via a locally Hermite interpolation (splines) [268, 269]. As the nodal surface is the usual plane wave determinant, its DMC result for the energy has to coincide to the ones of our JS implementation, although their energy and variance in a VMC implementations have to be reduced. This can be check by having a look at Table 4.2 and Table C.1, for densities $nr_0^2 = 40$ and $nr_0^2 = 48$.

The Backflow wave functions (BF[MH]) includes two-body backflow correlations as in Eq. (4.18). The difference is that in this case, $f_{\mathrm{BF}}(r)$ is evaluated through Hermite interpolants, like the Jastrow factors in the wave functions JS[MH] and JS3[MH]. This choice is different that the one that we chose in chapter 4, where $f_{\mathrm{BF}}(r)$ was chosen to have Gaussian form, this changes the nodal surface and in the end DMC backflow energies of Table 4.2 are slightly larger than the ones for appearing on Table C.1 for the BF wave function. It is important to remark that up to this level in the accuracy of the nodal surface, conclusions regarding the existence of a possible itinerant ferromagnetic phase are the same in both implementations.

Calculations with an explicit three-body backflow correlation are also shown in Table C.1, as it was done in Ref. [118], it are labeled as BF3.

---

[1]Calculations performed with the trial wave functions listed at this point have been performed by Markus Holzmann and have been originally published in Ref. [94]

[2]Although all the calculations presented in this appendix have been done by Markus Holzmann, the superscript "MH" is employed explicitly when differentiation from the results presented in chapter 4 is needed

Results in which the iterative backflow procedure is used are also shown in Table C.1. It iteratively constructs wave functions that include backflow correlations $ITN$, that constructed from the previous iterated level times $IT(N-1)$ The starting point of this method chose $IT0$, as the usual Jastrow-Slater wave function with two-body correlations. At each iteration, the new backflow coordinates $q_i^\alpha$ are determined as:

$$\mathbf{q}_i^{(\alpha)} \equiv \mathbf{q}_i^{(\alpha-1)} + \sum_{j \neq i} \left( \mathbf{q}_i^{(\alpha-1)} - \mathbf{q}_j^{(\alpha-1)} \right) f_{\mathrm{BF}}^{(\alpha-1)} \left( |\mathbf{q}_i^{(\alpha-1)} - \mathbf{q}_j^{(\alpha-1)}| \right), \qquad \text{(C.1)}$$

where $\mathbf{q}_i^{(0)} = \mathbf{x}_i$ are the particle coordinates. The iterative-backflow functions $f_{\mathrm{BF}}^{(\alpha)}(r)$, are chosen to be Gaussians, depending of three parameters, and thus depend on three parameter such as in Eq. (4.18). When optimizing the Gaussian parameters it is important to be sure that $\mathbf{q}_i^{(0)}$ and its derivatives vanish at distances $r = L/2$.

Finally BF3T1 calculations are performed by a combination of IT1 and BF3: an iterated two-body Jastrow and backflow potential are used together with a non-iterated three body backflow correlations.

| $\Psi_T$ | $P=1$ | | | $P=0$ | | | $\Delta E_T$ | $\Delta E$ |
|---|---|---|---|---|---|---|---|---|
| | $\sigma_T^2$ | $E_T$ | $E$ | $\sigma_T^2$ | $E_T$ | $E$ | | |
| JS$^{\mathrm{MH}}$ | 102(8) | 1615.98(5) | 1607.92(1) | 311(10) | 1636.56(8) | 1610.85(5) | 20.6(1) | 2.93(6) |
| JS3 | 47(2) | 1610.90(3) | - | 185(6) | 1625.21(6) | - | 14.31(6) | - |
| BF$^{\mathrm{MH}}$ | 40(2) | 1609.28(3) | 1607.09(2) | 98(2) | 1613.94(17) | 1606.91(17) | 4.7(2) | -0.2(2) |
| BF3 | 28.5(8) | 1608.59(7) | 1606.94(4) | 96(2) | 1612.63(7) | 1606.29(6) | 4.0(1) | -0.65(7) |
| IT1 | 32.6(7) | 1608.40(7) | 1606.91(3) | 78(1) | 1610.78(11) | 1605.69(9) | 2.38(13) | -1.22(1) |
| BF3-IT1 | 26.8(5) | 1608.15(7) | 1606.87(5) | 75(2) | 1609.65(11) | 1605.57(6) | 1.50(13) | -1.31(8) |
| IT2 | 30.6(7) | 1608.02(6) | 1606.91(6) | 74.8(9) | 1609.54(7) | 1605.31(13) | 1.52(9) | -1.59(14) |
| VMC$_{\mathrm{ext}}$ | | 1605.1(8) | | | 1600.5(12) | | | |

| $\Psi_T$ | $P=1$ | | | $P=0$ | | | $\Delta E_T$ | $\Delta E$ |
|---|---|---|---|---|---|---|---|---|
| | $\sigma_T^2$ | $E_T$ | $E$ | $\sigma_T^2$ | $E_T$ | $E$ | | |
| JS$^{\mathrm{MH}}$ | 167(6) | 2094.2(2) | 2083.05(8) | 481(23) | 2119.6(3) | 2087.2(3) | 25.40(8) | 4.1(4) |
| BF$^{\mathrm{MH}}$ | 60.0(9) | 2084.78(9) | 2082.09(7) | 151(2) | 2091.4(2) | 2082.15(8) | 6.6(1) | 0.0(2) |
| BF3 | 45(1) | 2083.97(10) | 2081.86(5) | 140(20) | 2089.44(6) | 2081.5(2) | 5.5(1) | -0.4(2) |
| IT1 | 54(2) | 2083.89(9) | 2081.85(8) | 159(4) | 2088.11(14) | 2081.0(2) | 4.(1) | -0.9(2) |
| BF3-IT1 | 44(1) | 2083.59(8) | 2081.69(5) | 114(3) | 2086.82(14) | 2080.5(2) | 3.23(8) | -1.2(2) |
| VMC$_{\mathrm{ext}}$ | | 2079.9(5) | | | 2075(3) | | | |

Table C.1 Energies per particle and variances at density $nr_0^2 = 40$ and $nr_0^2 = 48$ (up and down tables respectively), in units of $\varepsilon_0$, for different trial wave functions $\Psi_T$. The variational energy per particle $E_T$ and the DMC result $E$ are reported, and for $E_T$ we also report the variance $\sigma_T^2$. The last two columns report the energy difference between the unpolarized and polarized states, for variational results [$\Delta E_T \equiv E_{T,P=0} - E_{T,P=1}$] and for DMC energies [$\Delta E$, *cf.* Eq. (4.20)]. The line "VMC$_{\mathrm{ext}}$" is obtained through a linear extrapolation of the $E_T$ values – see text. Data for $P = 1$ ($P = 0$) are obtained with $N = 121$ ($N = 122$) particles. For the details about the wave functions, and for a comparison with Table 4.2. Note that the JS and JS3 wave functions have the same nodal structure, so that their DMC energies should be the same within statistical uncertainty.